\documentclass{article}

\usepackage{arxiv}

\usepackage[utf8]{inputenc} 
\usepackage[T1]{fontenc}    
\usepackage{hyperref}       
\usepackage{url}            
\usepackage{booktabs}       
\usepackage{amsfonts}       
\usepackage{nicefrac}       
\usepackage{microtype}      
\usepackage{lipsum}
\usepackage{graphicx}
\graphicspath{ {./images/} }
\usepackage{cite}
\usepackage{amsmath,amssymb,amsfonts}
\usepackage{algorithmic}
\usepackage{graphicx}
\usepackage{algorithm,algorithmic}
\usepackage{hyperref}
\usepackage{textcomp}
\newtheorem{theorem}{Theorem}[section]
\newtheorem{problem}[theorem]{Problem}
\newtheorem{definition}[theorem]{Definition}
\newtheorem{assumption}[theorem]{Assumption}
\newtheorem{proposition}[theorem]{Proposition}

\newtheorem{example}[theorem]{Example}
\newtheorem{remark}[theorem]{Remark}
\usepackage{tikz,xcolor,pgfplots}
\usepackage{url}
\usetikzlibrary{shapes,arrows,calc,positioning,arrows.meta,graphs}
\usepackage{lipsum}
\usepackage{amsfonts,graphicx,epstopdf,algorithmic}
\usepackage[outline]{contour} 
\contourlength{1.2pt}
\usetikzlibrary{positioning,calc}
\usetikzlibrary{backgrounds}

\ifpdf
  \DeclareGraphicsExtensions{.eps,.pdf,.png,.jpg}
\else

  \DeclareGraphicsExtensions{.eps}
\fi

\pgfmathdeclarefunction{gauss}{3}{%
  \pgfmathparse{0.25*1/(#3*sqrt(2*pi))*exp(-((#1-#2)^2)/(2*#3^2))}%
}
\pgfmathdeclarefunction{cdf}{3}{%
  \pgfmathparse{0.3*1/(1+exp(-0.07056*((#1-#2)/#3)^3 - 1.5976*(#1-#2)/#3))}%
}
\pgfmathdeclarefunction{fq}{3}{%
  \pgfmathparse{0.3*1/(sqrt(2*pi*#1))*exp(-(sqrt(#1)-#2/#3)^2/2)}%
}
\pgfmathdeclarefunction{fq0}{1}{%
  \pgfmathparse{0.3*1/(sqrt(2*pi*#1))*exp(-#1/2))}%
}
\pgfmathdeclarefunction{fq1}{1}{%
  \pgfmathparse{0.05*(8*(#1-16)*(#1-12)*(#1-8)*(#1-4)/((-16) * (-12) *(-8) * (-4))+ 1*(#1-16)*(#1-12)*(#1-8)*#1/((-12) * (-8) * (-4) * 4)+ 6*(#1-16)*(#1-12)*(#1-4)*#1/ ((-8) * (-4)* 4 * 8)+ 3*(#1-16)*(#1-8)*(#1-4)*#1/ ((-4)* 4 * 8 * 12)+ 8*(#1-12)*(#1-8)*(#1-4)*#1/ (4 * 8 * 12 * 16))}
  }

\colorlet{mydarkblue}{blue!30!black}

\usepgfplotslibrary{fillbetween}
\usetikzlibrary{patterns}
\pgfplotsset{compat=1.12} 

\usepackage{enumitem}
\setlist[enumerate]{leftmargin=.5in}
\setlist[itemize]{leftmargin=.5in}

\newcommand{\doa}{\mbox{${\mathrm {DoA}}$}}
\newcommand{\spec}{\mbox{$\mathrm{spec}$}}
\newcommand{\sign}{\mbox{$\mathrm{sign}$}}

\newcommand{\realpart}{\mbox{$\mathrm{Re}$}}
\usepackage{amsopn}
\DeclareMathOperator{\diag}{diag}
\begin{document}
\title{
\vspace{8mm}
Control of the Power Flows \\
of a Stochastic Power System}
\author{Zhen Wang
%
\thanks{
Zhen Wang is grateful to the Chinese Scholarship Council 
for financial support (grant number 202106220104) for his stay at Delft University of Technology
in Delft, The Netherlands.
Zhen Wang is with the School of Mathematics, Shandong University, 
Jinan, 250100, Shandong Province, China and
visited the Dept. of Applied Mathematics, Delft University of Technology, 
Delft, The Netherlands from November 2021 to February 2024
(email: wangzhen17@mail.sdu.edu.cn, wangzhen17.sdu@vip.163.com).
}~,
\and
Kaihua Xi\footnotemark[2]~,\and
Aijie Cheng
\thanks{
Kaihua Xi and Aijie Cheng are with 
the School of Mathematics, Shandong University, Jinan, 
250100, Shandong Province, China 
(email: kxi@sdu.edu.cn, aijie@sdu.edu.cn).
}~,
\and Hai Xiang Lin \footnotemark[3]~,\and
Jan H. van Schuppen
\thanks{
Hai Xiang Lin and Jan H. van Schuppen are with the
Dept. of Applied Mathematics, Delft University of Technology, 
Delft, The Netherlands
(email: H.X.Lin@tudelft.nl, J.H.vanSchuppen@tudelft.nl).
}~.
}
%
%
\maketitle
\begin{abstract}
How to determine the vector of power supplies
of a stochastic power system for the next short horizon,
such that the probability is less than a prespecified value
that any phase-angle difference of a power line of the power network 
exits from a safe set?
The power system is modelled 
such that the differential equation of each frequency
is affected by a Brownian motion process. 
A safe set can be selected to be any subset of the interval $(-\pi/2, ~ +\pi/2)$,
which is a sufficient condition for not losing synchronization.
That the controlled system has an improved performance is shown by
numerical results of three academic examples including 
a particular eight-node academic network,
a twelve-node ring network, and
a Manhattan-grid network.
\end{abstract}
\section{Introduction}\label{sec:intro}
The aim of this paper is 
to present the solution of a control problem 
for a power system which is subject to stochastic disturbances
and which may be in danger of losing transient stability. 
\paragraph{Motivation}  
The motivation of this paper is the fact that current power systems
experience fluctuations in power lines
and that such fluctuations are expected to increase 
in intensity in the coming decades.
The fluctuations of the power flows
are due to the power sources,
in particular to the renewable power sources including
wind turbines, wind parks, solar panels, photovoltaic panels, 
biomass generators, tidal energy, 
all without online $CO_2$ production. 
The power supplied by wind turbines and photovoltaic panels 
will not be steady during either short or long time scales, 
but will fluctuate with the weather and other factors.  
The fluctuations of power demand
are expected to increase in intensity
due to an increasing variety of power loads.
The fluctuations of power flows in the power network
endanger the transient stability of the power system,
for which a form of control is needed,
\cite{haehne:schmietendorf:tamrakur:peinke:ketteman:2019,johnson:rhodes:webber:2020}.
\paragraph{Problem Introduction}  
The overall control objective of this paper
is to maintain transient stability of a stochastic power system.
\par 
Control is the focus of this paper.
The control objective of transient stability of the stochastic power system 
is strengthened to include the additional control objective
that the probability is less than a prespecified value
that any of the phase-angle differences of a power line
exits from the safe set $(-\pi/2, ~ +\pi/2)$ 
or from a strict subset of that set. 
The restriction to a safe set of the phase-angle differences
rather than to a domain of attraction of the state set,
is an important step,
which was used in
\cite{wu:xi:cheng:lin:schuppen:2023:chaos,xianwu:kaihuaxi:etal:2023:arxiv}.
The time index set is discretized into a sequence
of short horizons of approximately 3 to 15 minutes.
%
\paragraph{Problem}
Determine a power supply vector
such that the controlled power system 
has a probability less than a prescribed value
that the phase-angle difference over any power line 
exits from the interval $(-\pi/2, ~  +\pi/2)$.
Instead of the described set,
an engineer may select a strictly smaller subset of $(-\pi/2, ~ +\pi/2)$.
The vector of power supplies has to be based on a prediction
of the power demand vector over the next future short horizon,
so that equality of the sum of power supplies and the sum of power demands 
is obtained.
The computed control input will be the power supply vector
which is held constant for a period equal to the length of the short horizon.
%
\paragraph{Literature Review}
\par
Since the voltage is assumed to be constant in the short horizon, we first refer to \cite{perninge2010risk} for the computation of the first exit-time to voltage instability of the power system with uncertainties in future loading, which reference would be useful for the voltage stability problem.
Our paper investigates the control of power flows 
by using a sequential optimization approach on power dispatch 
in a secondary or tertiary control framework 
that may be extended to the Security-Constraint Optimal Power Flow (SC-OPF) 
methodology in the future.
 An SC-OPF problem involves determining power dispatch while considering constraints such as generator capabilities and voltage limits, with optimization criteria including generator costs and transmission losses.
Here, we first review the literature related to the SC-OPF 
so as to better address the literature of our problem. 
A comprehensive survey about the Security-Constrained Optimal Power Flow 
can be found in \cite{bhaskar2011security}.
\par
Security-Constraint Optimal Power Flow (SC-OPF)  with constraints.
\cite{8822460}, N. Ngaa et al. 
put the frequency deviation as a nonlinear constraint 
in the optimal power flow network and 
then used the genetic algorithm to solve the problem.
They clarify the importance of considering this constraint.
However, the dispatch cost will be higher than before 
caused by a preference for choosing a generator with higher inertia. 
Note that this paper considers both active power and reactive power.
In \cite{gutierrez2010neural}, 
a differentiable function on power system variables is considered, 
which is extracted from the Neural network representation of the secure boundary.
This approach can better approximate the secure boundary than a typical method, 
which uses as a constraint the OPF model for optimal dispatch.
\par 
\cite{Abhyankar2017solution,alejandro2017directional} 
consider transient stability as a constraint for power dispatch. 
\cite{Abhyankar2017solution} presents several techniques 
for the Tansient Stability Constained-Optimal Power Flow (TSC-OPF) problems. 
\cite{alejandro2017directional} proposed a directional derivative-based method 
to make the angle deviation 
from the center of inertia decrease 
in the steepest direction to determine a transiently-stable power dispatch, 
by this method only two constraints are added 
to the conventional optimal power flow framework. 
Two numerical experiments 
show that the increased cost of the proposed procedure 
from the conventional OPF analysis is quite minor.
\par 
The economic issue is addressed in the following references.
\cite{6882841} discuss using power router (PR) control 
in the security-constrained optimal power flow 
as a real-time power dispatch setpoint, 
where the PR phase-angle injection 
is included in the objective function and constraints. 
The economic cost is minimized while the constraints are still satisfied. 
In this literature, the AC power flow equation is used. 
Furthermore, the proposed algorithm is easy to be extended 
by the conventional SC-OPF algorithms.
\cite{4335095} considers system corrective capabilities 
into an economic dispatch problem with security constraints. 
By a decomposition technique which allows separate contingency analysis 
with generator rescheduling and an iterative procedure, 
the same security level of the conventional security-constrained dispatch 
can be achieved, while the economic cost is lower. 
Note that this work deals with AC power flows, not with linearized power flow. 
Load forecast and network configuration 
can be taken into account in their proposed framework, 
while these two items are strengthened in our paper. 
\par 
A probabilistic framework is investigated in the following references.
\cite{vrakopoulou2013probabilistic} extend the SC-OPF 
when incorporating renewable energy sources to a probabilistic robust one. 
They use simulation results to justify their choice 
by comparing the OPF, the SC-OPF, and the probabilistic robust SC-OPF. 
Furthermore, the performances of adopting AC and DC power flow 
are also compared, 
while the AC power flow model respects the violation level and 
the DC power flow does not. 
Moreover, the cost of the proposed corrective scheme 
(policy based Automatic Voltage Regulation (AVR) set-point) 
is lower than the constant AVR set-point. 
Furthermore, in this paper, a convex relaxation is used. 
\par
From a centralized manner to a distributed manner.
\cite{mohammadi2016agent} 
put forward two multi-agent distributed approaches 
to solve the DC Security Constraint Optimal Power Flow problem 
due to the urgent need for distributing integrated renewable sources 
like solar panels in a decentralized manner.
\par
However, few papers have investigated analytically 
the stochastic fluctuations of power systems and their control. 
Even fewer papers presented concrete procedures 
to suppress or to mitigate the fluctuations of a power system 
with integrated renewable sources. 
In our paper, an AC power flow model is adopted.
\par
Secondly, we review literature related to the $H_2$ norm. 
This is done because we investigate the stochastic power system 
in the manner of the invariant distribution 
of the stochastic linearized power system.
\cite{marris2008upgrading} 
introduced a management with power supply and demand, and 
how the future power system would be like, 
from which one learns that renewable sources will replace 
the traditional generation gradually in the future and 
therefore the inertia will dramatically drop down 
to zero after the penetration of these renewable sources, 
\cite{optimal_inertia_placement} put forward an algorithm 
on where to place virtual inertia 
aiming to optimize a $H_2$ norm which 
is a global metric containing both phase-angle difference and frequency deviation.
\cite{H2norm}, $H_2$ norm is also used to evaluate the resistive power losses 
in terms of the design of future power systems 
where more generators and transmission lines should be accommodated. Note that in \cite{optimal_inertia_placement,H2norm}, a linear DC model is used. 
\par
However, as already pointed out in \cite{WANG2023110884} and 
later in Section \ref{Invariantdistribution}, 
the variance of the power flow at one line 
or of a particular node can also be calculated 
by the $H_2$ norm of the input-output LTI system, 
setting the output as only the phase-angle difference 
at one line or as only the frequency at one node. 
If one wants to control the most vulnerable line 
or the most vulnerable node, 
one needs to solve many Lyapunov equations. 
Thus, the $H_2$ norm is not used in this paper.  
%
\paragraph{Contributions of this Paper} 
(1) A detailed derivation of the probability and of an upper bound on that probability
that the phase-angle difference across any power line exits from a safe set 
according to an invariant probability distribution.
(2) The numerical results for the optimal power supply vectors and 
of the phase-angle differences of the power line flows 
for three academic examples, including 
a particular eight-node example, 
a twelve-node ring network, and 
a small Manhattan-grid example.
These three examples are chosen so as to investigate 
the influence of different network structures on the probabilities.
%
\paragraph{Paper Organization} 
Section~\ref{sec:intropowersystem} introduces a deterministic power system and 
a stochastic linearized  power system. 
The control objective function is introduced and defined 
in Section~\ref{sec:problem}.
The performance of the controlled power systems
for three illustrative examples,
based on their numerical computations,
are summarized and displayed in Section~\ref{sec:examples}. 
Section~\ref{conclusionfurtherinvestigation} states conclusions and 
describes open research issues. 
The reader may find results on a related optimization problem
in the companion paper
\cite{zhenwang:reportthree:2023}.
\section{The Power System}\label{sec:intropowersystem} 
\subsection{Notation}
The set of the integers is denoted by $\mathbb{Z}$ and 
that of the positive integers by
$\mathbb{Z}_+ = \{ 1, ~ 2, ~ \ldots \}$. 
The natural numbers are denoted by $\mathbb{N} = \{ 0, ~ 1, ~ 2, ~ \ldots \}$.
For any positive integer $k\in \mathbb{Z}_+$ denote the finite sets
$\mathbb{Z}_k = \{ 1, ~ 2, ~ \ldots, ~ k \}$ and
$\mathbb{N}_k = \{ 0, ~ 1, ~ 2, ~ \ldots, ~ k \}$.
The real numbers, the positive real numbers, and the strictly positive
real numbers are respectively denoted by
$\mathbb{R}$, $\mathbb{R}_+=[0,~\infty)$, and $\mathbb{R}_{s+}=(0,~\infty)$.
The complex numbers are denoted by $\mathbb{C}$.
The open left part of the complex plane 
is denoted by $\mathbb{C}_{o}^-=\left\{c\in\mathbb{C}~|~\realpart{(c)}<0\right\}$.
Denote the {\em sign} function as,
$\sign(x) = +1, ~~\text{if} ~ x > 0, ~ = -1 ~\text{if} ~ x < 0, ~ = 0, ~ 
\text{otherwise.}$
\par
The vector space of $n \in \mathbb{Z}_+$ tuples 
of the real numbers is denoted by $\mathbb{R}^n$.
Denote
by $e_k$ the $k^{th}$ unit vector 
whose $k^{th}$ component equals one 
while the other components equal zero.
Denote by $1_n\in \mathbb{R}^n$, 
the n-dimensional vector whose components are all equal to $1$.
The set of matrices of size $n \times m$
with entries in the real numbers,
is denoted by $\mathbb{R}^{n \times m}$.
The n-th column of matrix $A$ is denoted by $A_n$,
and the n-th row of matrix $A$ is denoted by $A(n)$.
The set of matrices of size $n \times n$ with
elements in the real numbers whose off-diagonal elements are all zeros
is denoted by $\mathbb{R}_{diag}^{n \times n}$.
Denote a diagonal matrix,
with on the diagonal the elements of the vector $v\in \mathbb{R}^n$,
by $\diag(v)\in \mathbb{R}^{n \times n}$.
A matrix $Q \in \mathbb{R}^{n \times n}$
is called {\em symmetric and positive definite}
if for all $v \in \mathbb{R}^n$, $v^T Q v \geq 0$
and denote the set by $\mathbb{R}_{pds}^{n \times n}$.
Such a matrix is called {\em strictly positive definite} 
if, for all $v \in \mathbb{R}^n$ with $v \neq 0$, $v^T Q v > 0$
and denote the set of such matrices by $\mathbb{R}_{spds}^{n \times n}$.
\par
The spectrum of the square matrix $A \in \mathbb{R}^{n \times n}$
is denoted by $\spec(A)$,
which set is defined as the set of eigenvalues of that matrix.
Denote the {\em spectral index} of matrix
$A \in \mathbb{R}^{n \times n}$
as the tuple
$n_{si}(A) = \{ n_-, ~ n_0, ~ n_+ \}$
where 
$n_- \in \mathbb{N}$ is the number of eigenvalues
with strictly negative real part, thus in
$C_o^- = \{ c \in \mathbb{C} |~ \realpart(c) < 0 \}$,
$n_0 \in \mathbb{N}$ is the number of eigenvalues
with real part equal to zero, and
$n_+ \in \mathbb{N}$ is the number of eigenvalues
with strictly positive real part, thus in
$C_o^+ = \{ c \in \mathbb{C} |~ \realpart(c) > 0 \}$.
%
%
\par
The symbol $x\in G(m_x,Q_x)$ denotes
that the random variable $x:\Omega\to \mathbb{R}^n$ 
has a Gaussian probability distribution function 
with mean $m_x\in \mathbb{R}^n$ and 
variance $Q_x\in \mathbb{R}_{pds}^{n\times n}$. 
\subsection{The Deterministic Nonlinear Power System}\label{sec:powersystem}
The 
{\em nonlinear power system}
is defined by a set of nonlinear differential equations 
driven by the input signal $p$, 
\cite{Kundur1994},
\cite{ilic2000dynamics},
\cite{arapostathis1982global}, and
\cite{Dorfler2012}. 
The {\em nominal frequency}
is defined to be the frequency of a rotating frame
with respect to which the actual power system will be defined.
The value of the nominal frequency equals 50 Hz in Europe and in Asia,
and equals 60 Hz in North America.
\par 
The graph of the power network is described by the tuple
$(\mathcal{V},\mathcal{E})$, 
where $\mathcal{V}$ is the set of nodes and 
$\mathcal{E}$ is the set of lines. 
Denote by $n_\mathcal{E}$ the number of lines and 
by $n_\mathcal{V}$ the number of nodes.
It is assumed that the undirected graph of the power network
is connected, meaning that for every tuple
$(i, ~ j) \in \mathcal{E}$,
there exists a path from $i$ to $j$.
\par 
The power system dynamics is specified by the following equations,
 \begin{small}
\begin{align}
        \frac{d \theta(t)}{dt}
    & = \omega(t), ~
          \theta(0) = \theta_0, \\
        M \frac{d \omega(t)}{dt}
    & = - D \omega (t)
          - B W f_{d}(\theta(t)) 
          + p(t), ~ \omega(0) = \omega_0; 
                          \end{align}
          \begin{align}
     t\in T 
    & = [0, ~ +\infty), \mbox{where}\nonumber   \\   
                    f_d(\theta)
    & = \left(
            \sin(\theta_{i_k} - \theta_{j_k}), ~ 
            \forall ~ k = (i_k, ~ j_k) \in \mathcal{E}
          \right) \\
    & = \sin(B^\top \theta)\in \mathbb{R}^{n_\mathcal{E}}; ~ 
          \mbox{which satisfies that,} \nonumber \\
    &   \forall ~ \theta\in \mathbb{R}^{n_\mathcal{V}}, ~
          \forall ~ \psi \in \mathbb{R},  \nonumber \\
              f_d(\theta)
    & = f_d(\theta + \psi \times 1_{n_\mathcal{V}} ); 
          \label{eq:fdshiftsangles} \\
      &   M \in \mathbb{R}_{\diag}^{n_\mathcal{V} \times n_\mathcal{V}}, ~
          \forall ~ i \in \mathbb{Z}_{n_\mathcal{V}}, ~ M_{i,i} > 0; ~  
          \nonumber \\
    &   D \in \mathbb{R}_{\diag}^{n_\mathcal{V} \times n_\mathcal{V}}, ~
          \forall ~ i \in \mathbb{Z}_{n_\mathcal{V}}, ~ D_{i,i} \geq 0; ~ 
          \nonumber \\
    &   B \in \mathbb{R}^{n_\mathcal{V} \times n_\mathcal{E}}, ~
          L \in \mathbb{R}^{n_\mathcal{V} \times n_\mathcal{V}}, ~
          W \in \mathbb{R}_{\diag}^{n_\mathcal{E} \times n_\mathcal{E}}, 
          \nonumber \\
        B_{i,k}
    & = \left\{
          \begin{array}{ll}
            +1, & \mbox{if} ~ k = (i, j_k) \in \mathbb{Z}_{n_\mathcal{E}},~ \nonumber \\
            -1, & \mbox{if} ~ k = (j_k, i) \in \mathbb{Z}_{n_\mathcal{E}}, ~ \nonumber\\
            0,  & \mbox{else;} 
          \end{array}
          \right.  \\
        L_{i,j}
    & = \left\{
          \begin{array}{ll}
             b_{i,j} ~ \hat{V}_i \hat{V}_j, & \mbox{if} ~ (i,j) \in \mathcal{E}, ~ \\
             0,                             & \mbox{else;} 
          \end{array}
          \right. \nonumber \\ 
        W_{k,k}
    & = \begin{array}{ll}
            L_{i_k,j_k} > 0, & \mbox{if} ~ k = (i_k, j_k) \in \mathcal{E}; 
          \end{array}
          ~ \nonumber
\end{align}
 \end{small}
where
$L_{i,j}=b_{ij}\hat{V}_i \hat{V}_j$ is 
the effective capacity of line $(i, ~ j) \in \mathcal{E}$, 
in which $\hat{V}_i$ is the voltage at node $i$ and 
$b_{ij}$ is the susceptance of line $(i, ~ j)$, 
the dynamics of the voltage is neglected in the short horizon, and
$L_{i,j}$ is considered as a constant on this horizon.
\par
Call
$\theta: T \rightarrow \mathbb{R}^{n_\mathcal{V}}$
the {\em phase deviation vector},
$\omega: T \rightarrow \mathbb{R}^{n_\mathcal{V}}$
the {\em frequency deviation vector} 
($\theta, \omega$ are relative to the nominal frequency),
$p=[p^+,-p^-]^\top: T \rightarrow \mathbb{R}^{n_\mathcal{V}}$ 
the {\em power vector},
$p^+: T \rightarrow \mathbb{R}^{n^+}$ 
the {\em vector of power supplies},
$p^-: T \rightarrow \mathbb{R}^{n_\mathcal{V}-n^+}$ 
the {\em vector of power loads},
$M \in \mathbb{R}_{s+}^{n_\mathcal{V} \times n_\mathcal{V}}$
the {\em matrix of inertias},
$D \in \mathbb{R}_+^{n_\mathcal{V} \times n_\mathcal{V}}$ 
the {\em matrix of damping constants}, 
$L \in \mathbb{R}_+^{n_\mathcal{V} \times n_\mathcal{V}}$, 
the {\em matrix of power line capacities}, 
$k$ is the index of line $(i_k, j_k)$,
$B \in \mathbb{R}^{n_\mathcal{V} \times n_\mathcal{E}}$, 
the {\em incidence matrix}, where the directions of the lines are consistently specified,
and
call $f_{d} \in \mathbb{R}^{n_\mathcal{E}}$ 
the {\em vector of sines of the phase-angle differences across power lines}.
It is assumed that,
for a sufficiently rich subset of initial conditions (not specified in the paper),
there exists a unique solution 
of the differential equation of the power system.
Denote by $(\theta(t; ~ 0, ~ \theta_0), ~ \omega(t; ~ 0, ~ \omega_0))$ 
for all times $t \in T$,
the solution of the differential equation
where the state at time $t \in T$ is denoted by
$(\theta(t), ~ \omega(t) )^T$.
\par
For any power line indexed by $k = (i_k, ~ j_k) \in \mathcal{E}$
denote the {\em phase-angle difference} of that power line and
the {\em power flow} through that power line, both used below,
respectively by,
\begin{align*}
      & ( \theta_{i_k} - \theta_{j_k} ): T \rightarrow \mathbb{R}, \\
      & L_{i_k, j_k} 
        \sin
        \left(
          \theta_{i_k} - \theta_{j_k} 
        \right): T \rightarrow \mathbb{R}.
\end{align*}
The flow of a power line is bounded by the capacity $L_{i_k, j_k}$
of the line because the absolute value of the sine function is bounded 
by the real number one.
%
\subsection{The Synchronous State}\label{sec:sychronousstate}
Consider the 
nonlinear
power system 
described in Section~\ref{sec:powersystem}.
A short while after the start of the short horizon,
the electric machines will run steady. 
One calls such a state a synchronous state rather than a steady state. 
In the power network, we suppose a subset of the nodes provide only power supply and a complementary subset have only power demand. 
The nodes are labeled such that:
\begin{align*}
        p(t)
    & = \begin{bmatrix}
            p^+(t)\\
            -p^-(t)
          \end{bmatrix} ~ \Leftrightarrow ~
         p^+(t) = 
         \begin{bmatrix}
            p_1^+(t) & \cdots & p_{n^+}^+(t) 
         \end{bmatrix}^\top, \\
        p^-(t)
    & = \begin{bmatrix}
           p_{n^++1}^-(t) & \cdots & p_{n_{\mathcal{V}}}^-(t)
          \end{bmatrix}^\top,
\end{align*}
which means that 
the nodes $1\to n^+$ have only power supply and 
the nodes $n^++1\to n_{\mathcal{V}}$ have only power demand. 
\par
Consider a steady power vector 
$p_{sp}\in \mathbb{R}^{ n_\mathcal{V}}$.
Recall that $\theta$ and $\omega$ are defined 
with respect to a frame rotating at the nominal frequency.
A {\em synchronous state} is defined as a tuple 
$(\theta_s, ~ \omega_s 1_{n_{\mathcal{V}}}) 
  \in \mathbb{R}^{2 n_{\mathcal{V}}}$ 
with $\omega_s \in \mathbb{R}$,
which satisfy the {\em synchronous state equations},
\begin{align}
        0
    & = - D 1_{n_{\mathcal{V}}} \omega_s
        - B W f_d(\theta_s) 
        + p_{sp}, ~ \omega_s 1_{n_{\mathcal{V}}} = 0 ~
         \Leftrightarrow  
         \label{eq:synchronousstateequation} \\
          0
	& = - D_{(i,i)} \omega_s
          - \sum_{j=1}^{n_\mathcal{V}} ~ 
              L_{i,j} \sin(\theta_{s,i} - \theta_{s,j}) + p_{sp,i}, ~
          0 = \omega_s.  \label{eq:synchronousstateequationcomponent} 
\end{align}
\begin{remark}
For the purposes of this paper,
it is essential that the nonlinear character of the power system is used.
Without this nonlinear character,
the understanding of the effects
of the stochastic fluctuations on the stability of the power system is lost.
For a linear direct current (DC) analysis,
the readers are referred to 
\cite[Remark 4.1]{wu:xi:cheng:lin:schuppen:2023:chaos}, 
where it is explained that the linear DC model,
in which the $\sin$ function is absent from (\ref{eq:synchronousstateequationcomponent}),
is insufficient for the analysis of the variance of the invariant distribution.
\end{remark}
%
\subsection{The Existence of a Strictly-Stable Synchronous State}
\label{sec:existenceofastable}
\begin{assumption}\label{assumption:existencestablesynchronousstate}
It is assumed that,
for the nonlinear power system
,
there exists a synchronous state
$(\theta_s, ~ \omega_s 1_{n_{\mathcal{V}}})$,
which satisfies the condition that
$(\theta_{s,i_k} - \theta_{s,j_k}) \in (-\pi/2, ~ +\pi/2)$,
for all $(i_k, ~ j_k) \in \mathcal{E}$ and $\omega_s=0$.
That state will be referred to as a 
{\em strictly-stable synchronous state}.
If a strictly-stable synchronous state exists 
then it is unique, \cite{skar_uniqueness_equilibrium}.
\end{assumption}
%
\par
The synchronous state is a solution of the
synchronous state equation
(\ref{eq:synchronousstateequation}).
The solvability of this equation has received much attention
in the literature, 
\cite{baillieul1982geometric},
\cite{Jafarpour},
\cite{DorflerPNAS},
\cite{skar_uniqueness_equilibrium}. 
Below we use a solution procedure from the literature.
\begin{theorem}
\label{criterion}
\cite{DorflerPNAS}.
Consider the power system specified in \ref{sec:powersystem}.
There exists a unique strictly-stable synchronous state
if there exists a $\gamma \in (0, ~ \pi/2)$ such that,
\begin{align}
    &   \| B^{\top} ( B W B^\top)^{\dagger} p_{sp} \|_{\infty}
        \leq \sin(\gamma).
\end{align}
\end{theorem}
\par
Note that $B^{\top} ( B W B^\top)^{\dagger} p_{s}$ 
has been detailed in Appendix \ref{computingAb}
as a formula $A~p_{s} + b$ 
where $A$ and $b$ can be computed from the given line capacities, 
network incidence matrix and 
the predicted power demand of next short horizon, and 
$p_{s}$ is the decision vector 
which will be explained in detail later in Section \ref{sec:domain}.
\begin{remark}\label{bifurcation}
Assumption \ref{assumption:existencestablesynchronousstate} is related to the existence of bifurcations of power systems.
The bifurcation scenario of the RTS 96 network 
is introduced on page 13 of the supporting information of \cite{DorflerPNAS}. 
In a recent publication \cite{jafarpour2022flow},
the authors localize the solutions in each winding cell, and 
they claimed and proved that there is at most one solution in each wind cell. 
However, there still exist cases where no solution can be found. 
Hence the above assumption is motivated.
\end{remark}
%
\subsection{The Domain of the Power Supply Vector}\label{sec:domain}
\begin{definition}\label{def:domain}
{\em The domain of the power supply vector}.
\par
Consider the 
maximal  available
power supply and the power demand
for the next short horizon, denoted respectively by
$p^{+,max} \in \mathbb{R}_+^{n^+}$ and 
$p^- \in \mathbb{R}_{s+}^{n^-}$.
Define then
$p_{sum}^{+,max} = \sum_{i=1}^{n^+} ~ p_i^{+,max}$ and
$p_{sum}^- = \sum_{i=1}^{n^-} ~ p_i^{-}$.
%
It is assumed that $p_{sum}^{+,max} \geq p_{sum}^{-}$.
%
Note that $p_{sum}^-$ is strictly positive.
\par
Construct a power supply vector $p^+ \in \mathbb{R}^{n^+}$
according to the following steps,
\begin{align*}
        (1)
    &   ~~ \mbox{choose} ~ p_{sum}^+ \in \mathbb{R} ~
          \mbox{such that} ~ p_{sum}^+ = p_{sum}^-, \\
        (2)
    &   ~~ \mbox{choose} ~ 
          p_i^+ \in [0, ~ p_i^{+,max}], ~
          \forall ~ i = 1, ~ 2, ~ \ldots, ~ n^+-1, \\
    &   ~~ \mbox{and set} ~
          p_{n^+}^+ 
          = p_{sum}^+ - \sum_{i=1}^{n^+-1} ~ p_i^+;  \\
	(3) 
    &   ~~  p^+ ~ \mbox{has to satisfy that} ~
          \forall ~ j \in \mathbb{Z}_{n^+}, ~ \\
    &   ~~ p_{sum}^+ - \sum_{i=1, i \neq j}^{n^+} ~ p_{i}^+ 
	 \in [0, ~ p_{j}^{+,max}].
\end{align*}
The latter conditions denote that:
(1) the power supply of $p^+$ equals the power demand; 
(2) the power supply components 
$p_1^+, ~ p_2^+, ~ \ldots, ~ p_{n^+-1}^+$ are feasible;
(3) the power supply vector $p^+$ satisfies these conditions.
\par
Define the {\em decision vector} as a subvector of the power supply vector,
by the formula\\
$p_s= \begin{bmatrix} p_1^+ & p_2^+ & \ldots & p_{n^+-1}^+ \end{bmatrix} 
  \in \mathbb{R}^{n^+-1}$. 
\par           
Define the {\em domain of the feasible vector} as the set $P^+$,
where the matrices $A_1,~ b_1,~ b_2$ 
are provided in \ref{ExpressionOfMatricesA1b1b2},
\begin{align*}
    P^+ 
      & = P^+(p^{+,max}, ~ p^-)  \\
      & = \left\{
          \begin{array}{l}
            p_s \in \mathbb{R}_+^{n^+-1} |~ 
            \mbox{conditions (1), (2), (3) above} \\
            \mbox{all hold, then we can derive} ~
            b_1 \leq A_1 p_s, ~ p_s \leq b_2
          \end{array}
          \right\}. 
\end{align*}
\end{definition}
%
%
%
%
\par
Condition (1) that power supply $p_{sum}^+$ equals power demand $p_{sum}^-$ and
that the synchronous state equation \eqref{eq:synchronousstateequation} holds,
imply that the synchronous frequency $\omega_s$ equals zero.
\begin{proposition}\label{thmdomain}
The domain $P^+$ is compact and convex. Moreover, it is a polytope.
\end{proposition}
\subsection{Linearize the Nonlinear Power System} 
\label{sec:linearpowersystem}
The procedure to linearize 
the deterministic nonlinear power control system
at a strictly-stable synchronous state
may be found in the references
\cite{zab2} and \cite{motter2013spontaneous}.
\begin{align}
        \frac{dz(t)}{dt}
    & = J(\theta_s, 0) z(t) 
          +
          \begin{bmatrix}
            0 \\ I
          \end{bmatrix} \Delta p(t), \label{SystemJ}  \\
        z(0) 
    & = \begin{bmatrix}
            \theta(0) -\theta_s \\\omega(0) -0
          \end{bmatrix}, ~
          \Delta p(t) = p(t) - p_{sp}, \nonumber \\
        z(t)
    & = \begin{bmatrix}
            \Delta \theta(t) \\ \Delta \omega(t)
          \end{bmatrix}
          =
          \begin{bmatrix}
            \theta(t) - \theta_s\\
            \omega(t) - \omega_s 
          \end{bmatrix}, ~ \nonumber  \\
         J(\theta_s, ~ 0)
    & = \begin{bmatrix}
            0                    & I_{n_\mathcal{V}} \\
            - M^{-1} B W F(\theta_s) & - M^{-1} D 
          \end{bmatrix}, ~
          \label{eq:jacobianmatrix} \\
        F(\theta_s)
   &  = \left(
            \frac{df_{d}(\theta)}{d\theta}
          \right)|_{\theta = \theta_s} 
          \in \mathbb{R}^{n_\mathcal{V} \times n_\mathcal{V}}.
          \nonumber
\end{align}
where the matrix of equation \eqref{eq:jacobianmatrix}
is called the {\em Jacobian matrix} and 
the product $BWF(\theta_s)$ is called 
the {\em Laplacian matrix} of the power network. 
The Laplacian matrix depends on the synchronous state considered.
The reader finds 
in Appendix~\ref{Jacobian}
the Laplacian matrix of a complete power network.
In general, if the synchronous state is such that
the phase-angle difference is closer to either $+\pi/2$ or to $-\pi/2$,
the closer the modulus of the eigenvalues of $J(\theta_s, ~ 0)$
will be to zero. 
\par
The Laplacian matrix of the power network is in general a singular matrix,
\cite{WANG2023110884}.
For all of the examples of Section~\ref{sec:examples}
the Laplacian matrix has precisely one eigenvalue at zero.
The eigenvalues of the Jacobian matrix of a power system 
for a {\em complete}
power network,
are discussed in 
\cite{skar_uniqueness_equilibrium,zaborszky:huang:leung:zheng:1985}.
%
\begin{figure}
\begin{center}
\begin{tikzpicture}
 \message{Expected sensitivty & p-value^^J}
   \def\N{50}
  \def\q{4.8}
  \def\Bs{2.6}
  \def\S{\q}
  \def\Ss{1.4}
  \def\xmax{16}
  \def\ymin{-0.15}
  \def\ymax{0.4}
   \begin{axis}[every axis plot post/.append style={
               mark=none,domain=0:{\xmax},samples=\N,smooth},
               xmin={-0.08*(\xmax)}, xmax=\xmax,
               ymin=\ymin, ymax=\ymax,
               axis lines=middle,
               axis line style=ultra thick,
               enlargelimits=upper, 
               ticks=none,
               every axis x label/.style={at={(current axis.right of origin)},anchor=north west},
               y=240pt,
               domain=0:(\xmax)
              ]
\addplot[name path=S22,line width=0.7mm,blue!90!red ] {fq1(x)};
 \node[below=0.05pt,black] at (2.8,0){$x_{s,a}$} ;
   \addplot[black,dashed,line width=0.7mm]
      coordinates {(8.38,-0.1) (8.38, {fq1(8.38)})}
       node[above =1pt] {$x_{s,c}$};
  \node[below=13.8pt,black] at (4,0) {$\doa(x_{s,a})$};
  \draw [stealth-stealth,line width=0.4mm](0,-0.07) -- (8.38, -0.07);
  \draw [-stealth,line width=0.4mm]({0.5},{fq1(0.5)+0.06}) -- (1, {fq1(1)+0.1});
    \draw [-stealth,line width=0.4mm]({1.5},{fq1(1.5)+0.06}) -- (2, {fq1(2)+0.05});
     \draw [-stealth,line width=0.4mm]({5},{fq1(5)+0.06}) -- (4, {fq1(4)+0.05});
     \draw [-stealth,line width=0.4mm]({7},{fq1(7)+0.03}) -- (6, {fq1(6)+0.05});
     \draw [-,line width=0.4mm] (2.8,-0.02)--(2.8,0.02);
  
\end{axis}
   \begin{axis}[every axis plot post/.append style={
               mark=none,domain=8.4:{\xmax},samples=\N,smooth},
               xmin={-0.08*(\xmax)}, xmax=\xmax,
               ymin=\ymin, ymax=\ymax,
               axis lines=middle,
               axis line style=ultra thick,
               enlargelimits=upper, 
               ticks=none,
               every axis x label/.style={at={(current axis.right of origin)},anchor=north west},
               y=240pt,
               domain=8.4:(\xmax)
              ]
\addplot[name path=S23,line width=0.7mm,red!90!blue ] {fq1(x)};
\addplot[black,dashed,line width=0.7mm]
      coordinates {({\xmax+0.06},-0.1) ({\xmax+0.06}, {fq1(\xmax)})};
        \draw [stealth-stealth,line width=0.4mm](8.38,-0.07) -- (\xmax, -0.07);
\addplot[black,dashed,line width=0.7mm]
      coordinates {(13.4,0) (13.4, {fq1(13.4)})};
  \node[below=13.8pt,black] at (13.1, 0) {$\doa(x_{s,b})$};
   \node[below=0.05pt,black] at (13.4,0){$x_{s,b}$} ;
     \draw [-stealth,line width=0.4mm]({9.5},{fq1(9.5)+0.02}) -- (11, {fq1(11)+0.03});
      \draw [-stealth,line width=0.4mm]({12},{fq1(12)+0.02}) -- (13, {fq1(13)+0.01});
        \draw [-stealth,line width=0.4mm]({15.7},{fq1(15.7)+0.08}) -- (15, {fq1(15)+0.07});
        \draw [-stealth,line width=0.4mm]({14.7},{fq1(14.7)+0.04}) -- (13.7, {fq1(13.7)+0.01});

\end{axis}
\end{tikzpicture}
\caption{Two stable points $x_{s,a}$ and $x_{s,b}$ with their domain of attraction and a saddle point $x_{s,c}$}
\label{fig:domainofattraction}
\end{center}
\end{figure}
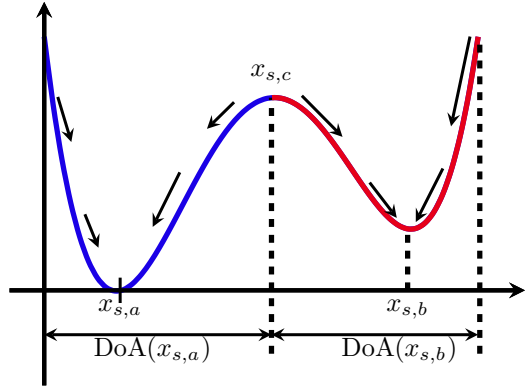
\subsection{The Stability of a Nonlinear Power System}
\label{subsec:dof}
The reader finds in this subsection the concepts 
of the domain of attraction of a synchronous state and
the definition of transient stability of a power system.
For concepts and a classification 
of power system stability, see 
\cite{Kundur1994,kundur:etal:2004}.
The main basis for this subsection are 
the papers on stability of nonlinear power systems,
\cite{zab2} and \cite{chiang1988stability}.
\begin{definition}\label{def:domainoa}
Define the {\em domain of attraction} 
of a strictly-stable synchronous state, 
$(\theta_s, ~ \omega_s 1_{n_{\mathcal{V}}}) \in \mathbb{R}^{2 n_{\mathcal{V}}}$,
of a deterministic nonlinear power system 
as a subset of the state set
such that the state trajectory of the power system, 
starting at any state of this subset,
will converge to the considered synchronous state;
in terms of mathematical notation,
\begin{align*}
        \lefteqn{
          \doa(\theta_s, ~ \omega_s)
        } \\
    & = \left\{
          \begin{array}{l}
            (\theta_0, ~ \omega_0) \in \mathbb{R}^{2 n_\mathcal{V}} |~ \\
            \lim_{t \rightarrow +\infty} ~
            (\theta(t; ~ 0, ~ (\theta_0, ~ \omega_0)), ~ 
             \omega(t; ~ 0, ~ (\theta_0, ~ \omega_0))
            ) \\
            = (\theta_s, ~ (\omega_s \times 1_{n_\mathcal{V}}))
          \end{array}
        \right\}.
\end{align*}
\par
Call the dynamic behavior of the power system
at the strictly-stable synchronous state $x_s$ 
{\em transient stable} if
(1) $( \spec (J(\theta_s, ~ 0)) \backslash \{ 0 \} ) \subset \mathbb{C}^-$ and
(2) the state trajectory, when started inside the open domain,
remains inside the domain during the entire horizon.
\end{definition}
\par
An example with domains of attraction is displayed in
Fig.~\ref{fig:domainofattraction}. 
The figure displays an abstract energy function as a function
of a one-dimensional state.
Displayed are the open domains of attraction of the two steady states 
$x_{s,a}$ and $x_{s,b}$ and the steady state $x_{s,c}$
whose domain of attraction is only the steady state itself.
\par
In the literature, there are necessary and sufficient conditions
for stability of the synchronous state of a deterministic nonlinear power system,
which is often condition (1) of transient stability.
A reference for these conditions is the paper
of J. Zaborszky et al., \cite{zaborszky:huang:leung:zheng:1985}
in which 
also
monitoring of a power system is described.
\par
Which concept of stability is needed 
for a {\em deterministic nonlinear power system} 
when it is subjected to 
either a single disturbance 
or a sequence of deterministic disturbances,
not specified in the nonlinear power system?
Then the state trajectory of such a system
may return to the synchronous state 
or will fluctuate in the domain of attraction near the synchronous state.
However, due to particular disturbances,
the state trajectory may leave the domain of attraction
and move into the domain of another synchronous state
where its dynamic behavior may be different.
There are no concepts for this form of instability,
caused by deterministic disturbances of power systems.
The reader is referred to the papers
\cite{zab2},
\cite{chiang1988stability} for this dynamic behavior.
A departure from the domain of a synchronous state 
is best avoided by a form of control. 
\par
In the remainder of the paper
a necessary condition for the transient stability of a power system 
will be used.
\begin{definition}
The deterministic nonlinear power system will be said to be
{\em phase-angle-difference stable}
if the phase-angle differences of all power lines
remain in the {\em safe subset} 
$(-\pi/2, ~ +\pi/2) \subset \mathbb{R}$ for all times. 
If a power system is {\em no longer} phase-angle-difference stable
then one says that the power system has lost {\em synchronization}.
\end{definition}
\par
The domain of attraction of a state set is difficult to compute.
Hence the use of the subset $(-\pi/2, ~ +\pi/2) \subset \mathbb{R}$ 
for the phase-angle differences of any power line.
The literature on the relation 
of a safe subset of the phase-angle differences
to that of the domain of attraction of a synchronous state,
including stability of a deterministic power system,
includes the following:
\cite{skar_uniqueness_equilibrium},
\cite{zaborszky:whang:prasad:katz:1981},
\cite{zaborszky:huang:leung:zheng:1985}
and \cite[Lemma 5.2]{tsolas1985structure}.
\par
There is no proof in the literature
that the phase-angle-difference stability implies
that the state of the power system remains in the domain of attraction
of a synchronous state.
\par
The phase-angle differences between buses 
have been widely used as a measure of 
the transient stability of power systems in the literature. 
See  \cite{xianwu:kaihuaxi:etal:2023:arxiv} 
for the escape probability of the phase-angle differences 
from the set $(-\pi/2,\pi/2)$ 
as a metric for transient stability of a stochastic power system. 
A phase plane of high-order derivatives of angle difference 
for a Single Machine Infinite Bus System 
is plotted in \cite{li2014transient} 
through which the unstable equilibrium can be found, and 
the transient stability can thus be evaluated. 
In \cite{song2017network}, 
the line whose phase-angle difference 
is inside the domain $(\pi/2,3\pi/2)$ 
is used as a critical line for the investigation 
of the small-disturbance rotor stability of power systems. 
%
\subsection{The Stochastic Linearized Power System}\label{sec:linearsto}
A stochastic power system is formulated
based on 
the linearization of the deterministic nonlinear power system.
The deterministic functions $(\theta, ~ \omega)$
become stochastic processes in the stochastic system defined below.
The process $\Delta p$ 
is below defined as a Brownian motion process.
It denotes the difference of the power supply and the power demand,
each of which as a difference
from their values at the synchronous state on a short horizon. Here, we refer to \cite{dong2012numerical}, where the uncertainty of the load is modeled as a Brownian motion.
The 
{\em stochastic linearized power system}
is then defined by the linear stochastic differential equation, 
\cite{I.Karatzas},
\cite{H.Kwakernaak},
\begin{align}
        dx(t)
    & = J(\theta_s, 0)~ x(t)~dt
          +
         K ~ dv(t), ~ x(0), \label{eq:linsde}\\
        y(t)
    & = C x(t),~
          \label{eq:linsdeoutput} \\
    &   x: \Omega\times T\rightarrow \mathbb{R}^{n_{\mathcal{V}}},~
          y: \Omega\times T\rightarrow \mathbb{R}^{n_{\mathcal{E}}},
          \nonumber \\
        x(t) 
    & = \begin{bmatrix}
            \theta(t) \\\omega(t)
          \end{bmatrix}, ~
          K = 
          \begin{bmatrix}
            0 \\ K_2
          \end{bmatrix} \in 
          \mathbb{R}^{2n_{\mathcal{V}}\times n_{\mathcal{V}}}, 
          \nonumber \\
        C 
    & = \begin{bmatrix}
            B^\top & 0
          \end{bmatrix} 
          \in \mathbb{R}^{n_{\mathcal{E}}\times 2n_{\mathcal{V}}},
          \nonumber 
\end{align}
where $(\Omega, ~ F, ~ P)$ denotes a probability space consisting of 
a set $\Omega$,
a $\sigma$-algebra $F$, and 
a probability measure $P$;
$T=[0,\infty)$ denotes the time index set;
$x(0):\Omega\rightarrow \mathbb{R}^{2n_{\mathcal{V}}}$ 
denotes the initial state of the stochastic system 
which is assumed to have a Gaussian distribution 
with expectation $m_{x(0)} \in \mathbb{R}^{2n_{\mathcal{V}}}$ and 
a symmetric positive-definite variance matrix 
$Q_{x(0)}\in \mathbb{R}_{spds}^{2n_{\mathcal{V}}\times 2n_{\mathcal{V}}}$,      
hence $x(0)\in G(m_{x(0)},Q_{x(0)})$;
$v:\Omega\times T\rightarrow \mathbb{R}^{n_{\mathcal{V}}}$ 
denotes a Brownian motion process 
such that 
$\forall s,~t\in T, s<t,~v(t)-v(s)\in G(0,(t-s)\times I_{n_{\mathcal{V}}})$, 
while $F^{x(0)}, F_\infty^v$ are independent $\sigma$-algebras;
$ K_2\in \mathbb{R}_{diag}^{n_{\mathcal{V}}\times n_{\mathcal{V}}}$ 
denotes a matrix which multiplies the vector-valued Brownian motion, 
which is a diagonal matrix with 
$K_2(i,i)\geq 0,\forall ~i\in \mathbb{Z}_{n_{\mathcal{V}}}$.
\subsection{Stochastic Stability of a Stochastic Power System}
%
Stochastic stability of nonlinear stochastic systems
is extensively treated in the literature, 
\cite{hasminski:1980,kozin:1969,kunita:1976,kushner:1967,meyn:tweedie:1993}.
The main concept is the invariant probability distribution of the state 
of the system and its properties.
For the characterization of stochastic linearized power system considered,
(\ref{eq:linsde},~ \ref{eq:linsdeoutput}),
attention will be focussed on  
the variance of the vector of phase-angle differences.
%
\begin{definition}\label{def:safeset}
Consider the stochastic linearized power system 
(\ref{eq:linsde},~ \ref{eq:linsdeoutput}),
and consider a synchronous state 
$x_s = (\theta_s, ~ \omega_s 1_{n_{\mathcal{V}}})$
which satisfies
Assumption~\ref{assumption:existencestablesynchronousstate}.
Define the {\em safe set} $X_c$ 
 of the phase-angle differences
as a subset of 
$\mathbb{R}^{2 n_{\mathcal{V}}}$
by the 
formulas,
\begin{align*}
        \Theta_c
    & = \left\{ 
          \begin{array}{l}
            \theta \in \mathbb{R}^{n_{\mathcal{V}}} |~
		  \forall ~ k = (i_k, ~ j_k) \in \mathcal{\mathcal{E}}, \\
            ( \theta_{i_k} - \theta_{j_k} ) \in (-\pi/2, ~ +\pi/2) 
          \end{array}
          \right\}, \\
        X_c
    & = \Theta_c \times \mathbb{R}^{n_\mathcal{V}}
        \subset \mathbb{R}^{2 n_{\mathcal{V}}}.
\end{align*}
Call the phase-angle differences of a power system
at the strictly-stable synchronous state $x_s$ during a short horizon 
and for a prespecified value $\epsilon \in (0, ~ 1)$,
{\em (probabilistically) safe-stable} 
(1) if $\spec ( J(\theta_s, ~ 0) \backslash \{0\} ) \subset \mathbb{C}_o^-$, and
(2) if the stochastic power system is started at the synchronous state, 
then the probability is higher than $(1 - \epsilon) \in (0, ~ 1)$
that the 
phase-angle differences,
remain inside the safe set 
$X_c = \Theta_c \times \mathbb{R}^{n_\mathcal{V}}$
during the considered horizon.
\end{definition}
\par
In this paper attention is focused
on a stochastic 
 linearized
 power system 
and a strictly-stable synchronous state 
such that the phase-angle differences of all power lines are safe-stable. 
This condition implies that synchronization of the power system is maintained.
%
%
\subsection{The Invariant Distribution of the Stochastic Linearized Power System}\label{Invariantdistribution}
\par
It follows from \cite[Theorem 6.17]{I.Karatzas} and 
\cite[Theorem 1.52]{H.Kwakernaak} 
that the stochastic processes $x$ and $y$ 
of the stochastic linearized power system
(\ref{eq:linsde},~ \ref{eq:linsdeoutput})
are Gaussian processes with for all times $t \in T$,
$x(t)\in G(m_x(t),Q_x(t))$ and $y(t)\in G(m_y(t),Q_y(t))$. 
\par 
If the Jacobian matrix $J(\theta_s,0)$ is Hurwitz, 
hence $\spec\left(J(\theta_s,0)\right)\subseteq \mathbb{C}_o^-$, 
the following properties 
of the mean value and the variance matrix hold,
\begin{align*}
        0
    & = \lim\limits_{t\to \infty}{m_x(t)}, 
    & 0 = \lim\limits_{t\to \infty}{m_y(t)}, \\
        Q_x
    & = \lim\limits_{t\to \infty}{Q_x(t)}, ~
    & Q_y = \lim\limits_{t\to \infty}{Q_y(t)},
\end{align*}
and where $Q_x$ is 
the unique solution of the following Lyapunov equation and 
$Q_y$ satisfies,
\begin{align}
        0
    & = J(\theta_s,0)Q_x+Q_xJ(\theta_s,0)^{\top}+KK^{\top},\label{Lyapeqn}\\
        Q_y
    & = CQ_xC^{\top}.\label{Lyapouteqn}
\end{align}
But in fact, the system matrix $J(\theta_s,0)$ is not Hurwitz 
because the Laplacian matrix $BWF(\theta_s)$ has a zero eigenvalue. 
\par
The reader finds in \cite{WANG2023110884}
a discussion for the determination
of the variances of the power lines
based on an eigenvalue decomposition and a reduction process.
\par
In control engineering, 
the system operators want to determine that power line 
which is the most vulnerable for instability.
Hence there is a need to compute
the variances of the phase-angle differences of all power lines. 
The $H_2$ norm of the input-output LTI system is insufficient for the computation of the variance due to high computational complexity, see \cite{WANG2023110884}.
%
\begin{assumption}\label{assumption:supportabilitylinearizedpowersystem}
%
Consider a stochastic linearized power system 
(\ref{eq:linsde},~ \ref{eq:linsdeoutput}),
where the nonlinear power system has been linearized at a strictly-stable synchronous state.
\begin{enumerate}
\item[(1)]
Assume that the diagonal matrix $K$
has strictly-positive diagonal elements, 
hence, for all $ k \in \mathcal{\mathcal{V}}$, $K_{k,k} > 0$.
\item[(2)]
$(J(\theta_s, ~ 0), ~ K)$ is a controllable  pair.
\end{enumerate}
\end{assumption}
\par
Condition (1) 
of Assumption~\ref{assumption:supportabilitylinearizedpowersystem} and
the fact that, in the graph associated with the system
with nodes at states and inputs, 
there exists a path from any node representing a frequency
to a node with a disturbance,
imply that Condition (2) 
of Assumption~\ref{assumption:supportabilitylinearizedpowersystem} holds,
\cite{davison:1977}.
%
\begin{proposition}\label{prop:stdevstrictlypositive}
Consider a stochastic linearized power system 
(\ref{eq:linsde},~ \ref{eq:linsdeoutput}).
Assume that either 
Condition (1) or Condition (2) of
Assumption~\ref{assumption:supportabilitylinearizedpowersystem} holds.
Then, for any power line $k \in \mathbb{Z}_{n_\mathcal{E}}$ and any
fixed power supply vector $p_s \in P^+$,
the standard deviation $\sigma_k(p_s)$ is strictly positive;
in terms of notation,
$\sigma_k(p_s) = (Q_{y,(k,k))})^{1/2} > 0$.
\end{proposition}
\par
The proof of the above proposition may be found in 
Appendix~\ref{apsubsec:stdevstrictlypositive}.
\par
From the discussion above follows that the random variable 
of the phase-angle difference over power line $k$ of the power network,
$\theta_{i_k}(t) - \theta_{j_k}(t)$,
has the invariant Gaussian probability distribution,
$\forall ~ t \in T$ and 
$\forall ~ k = (i_k,j_k) \in \mathcal{\mathcal{E}}$, 
\begin{align*}
    &   \theta_{i_k}(t) - \theta_{j_k}(t) 
        \in G( m_k(p_s), ~ \sigma_k^2(p_s)); \\
    m_k(p_s) 
    & = \theta_{s, i_k}(p_s) - \theta_{s, j_k}(p_s), ~\\
    \sigma_k(p_s) 
    & = Q_{y,(k,k)}(p_s)^{1/2} > 0; ~
          \mbox{then,} \\
    &   \frac{ \theta_{i_k}(t) - \theta_{j_k}(t) - m_k(p_s)
          }{
            \sigma_k(p_s)
          } 
          \in G(0, ~ 1); \\
    \epsilon
    & = \int_{-\infty}^{r_{\epsilon}} ~ p_{G(0,1)}(w) ~ dw.
\end{align*}
See Table~\ref{table:gaussianpdfvalues}
for the relation between the threshold values and 
the values of the parameter, 
$(\epsilon,r_{\epsilon})$, of $G(0,1)$.
\begin{table}
 \begin{center}
\begin{tabular}{|l|l|}
\hline
$\epsilon$ & $r_{\epsilon}$ \\
\hline
0.050 & -1.65 \\
0.040 & -1.76 \\
0.030 & -1.89 \\
0.020 & -2.06 \\
0.010 &-2.33 \\
0.001 &-3.08 \\
\hline
\end{tabular}
\end{center}
\caption{Relation of probabilities and thresholds of a Gaussian
probability distribution with mean zero and variance one. 
Thus, if $x\in G(0,1)$ and $\epsilon\in (0,1)$ 
then $P(\left\{x\leq r_\epsilon\right\})\leq \epsilon$ and 
$r_\epsilon$ is an approximation of the maximal such real number.}
\label{table:gaussianpdfvalues}
\vspace{-4mm}
\end{table}
\subsection{An Upper Bound}
\label{subsec:towardscontrolobjectivefunction}
The probabilities that the maximum of a stationary Gaussian process
on a finite interval is larger than a particular threshold,
are available in the literature,
see for example
\cite[Section 6.5]{gikhman:skorokhod:1969}.
However,
even for a Brownian motion process
the probabilities are expressed as an infinite series
of integrals over Gaussian density functions.
Such expressions require approximations
which are computationally intensive.
Therefore it has been decided to
restrict attention to the probability at
a particular time that the process exits the safe set
according to the invariant probability distribution.
%
\begin{proposition}\label{prop:probabilitybound}
Consider the stochastic 
linearized
power system.\\
(a) The probability that the power flow of any power line exits the safe set 
according to the invariant distribution
of the vector of power line flows, satisfies,
\begin{align*}
       &~~~~~ \forall ~ k \in \mathbb{Z}_{n_\mathcal{E}}, ~
          \forall ~ p_s \in P^+,  \\
        \lefteqn{
          p_{out,k}(p_s)
        } \\
       & = P
           \left(
           \left\{
             \nu \in \Omega |~
               \theta_{i_k}(\nu, t) - \theta_{j_k}(\nu, t) 
               \leq - \pi/2
           \right\}
           \right)  + \\
       & ~~~  + P
           \left(
           \left\{
             \nu \in \Omega |~
               \theta_{i_k}(\nu, t) - \theta_{j_k}(\nu, t)
               \geq + \pi/2
           \right\}
           \right) \\
       & = f_{G(0,1)}(r_{a,k}(p_s)) + f_{G(0,1)}(-r_{b,k}(p_s)), \\
       & 
           \leq   2 f_{G(0,1)}(r_{a,b,k}(p_s)), 
         \\
     r_{a,k}(p_s)
       & = \frac{- \pi/2 - m_k(p_s)}{\sigma_k(p_s)}, \\
     - r_{b,k}(p_s)
       & = \frac{- \pi/2 + m_k(p_s)}{\sigma_k(p_s)};\\
     r_{a,b,k}(p_s)
       & = \max \{ r_{a,k}(p_s), ~ - r_{b,k}(p_s) \}, ~
	\mbox{then,} \\
     r_{a,b,k}(p_s)
       & = \frac{-\pi/2 + | m_k(p_s)| }{\sigma_k(p_s)}.
\end{align*}
(b) An upper bound.
If $r_{a,b,k}(p_s) \leq r_{\epsilon /2}$ 
then $p_{out,k}(p_s) \leq \epsilon$. 
\end{proposition}
\par
%
%
Infimization of the lower bound derived above
leads to the following alternative optimization criteria,
\begin{align*}
        (a)
    &   \inf_{p_s \in P^+} ~ \max_{k \in \mathbb{Z}_{n_\mathcal{E}}} ~
                r_{a,b,k}(p_s) ~ \\
    & = \inf_{p_s \in P^+} ~ \max_{k \in \mathbb{Z}_{n_\mathcal{E}}} ~
              \left(
                \frac{- \pi/2 + |m_k(p_s)|}{\sigma_k(p_s)}
              \right);\\
        (b)
    &   \inf_{p_s \in P^+} ~ \max_{k \in \mathbb{Z}_{n_\mathcal{E}}} ~
              \left(
                \frac{|m_k(p_s)|}{\sigma_k(p_s)} - r_{\epsilon} 
              \right)  \\
    & = \inf_{p_s \in P^+} ~ \max_{k \in \mathbb{Z}_{n_\mathcal{E}}} ~
              \left(
                \frac{|m_k(p_s)| - r_{\epsilon} \:\sigma_k(p_s)}{\sigma_k(p_s)} 
              \right); \\
           (c)
    &   f_d^* = \inf_{p_s \in P^+} ~ \max_{k \in \mathbb{Z}_{n_\mathcal{E}}} ~
              \left[ ~
                |m_k(p_s)| - r_{\epsilon} \:\sigma_k(p_s) ~
              \right], \\
    &   \mbox{where} ~ r_{\epsilon} \in (-\infty, ~ 0).
\end{align*}
Below optimization criterion (c) will be used.
The criteria (a) and (b) involve division by $\sigma_k$
which involves computational issues that are best avoided.
\par
Note that 
$f_d^* < \pi/2$
with a parameter $r_{\epsilon}$ 
implies that there exists
a supply vector $p_s \in P^+$ such that
$p_{out}(p_s) \leq  2 \:\epsilon$.
The proof of this proposition can be found in 
Appendix~\ref{proofofpropprobabilitybound}.
\subsection{A Sequence of Short Horizons}
Recall that the {\em time index set}, also called the {\em horizon},
is the set of the positive real numbers $T = [0, ~ +\infty)$.
This time index set is now partitioned into 
an infinite sequence of {\em short horizons} all of the same length, 
for example 3 or 5 or 15 minutes,
as is done in discretization of a continuous-time system.
\section{The Control Problem}\label{sec:problem}
\subsection{The Setting of the Control Problem}
\label{sec:introtoconpro}
%
Consider the nonlinear power system 
and the stochasic linearized power system
both defined in the previous section.
Consider the time index $T = \mathbb{R}_+$
partitioned into an infinite sequence of short horizons.
It is assumed that,
during any short horizon,
there is available to the controller
a prediction of the total power demand $\hat{p}_{sum}^-$ 
of the power system in the next short horizon
which is based on output feedback measurements.
\par
The problem to be investigated and formalized below is to compute,
during any short horizon,
an input for the next short horizon
in the form of a vector $p^+$ of power supplies for all nodes with power supply.
That constant input is then used during the entire next short horizon.
\par
The computation of the vector $p^+$ of power supplies proceeds 
as described in Subsection  \ref{sec:domain}. 
but with the power demand of the next short horizon 
replaced by a prediction of that power demand. 
Set first $p_{sum}^+ = \hat{p}_{sum}^-$, 
thus total power supply equals the prediction of total power demand.
Secondly, 
determine the vector $p_s = \{ p_1^+, ~ p_2^+, ~ \ldots, ~ p_{n^+-1}^+ \}$
according to the optimization problem {\em described below} and 
satisfying the conditions (1)-(3) of Subsection \ref{sec:domain}.
\par
The approach of control sketched above,
is a generalization of secondary frequency control,
\cite{EAGC,Dorfler2016,Dorfler2014,PIAC}.
%
%
In the quoted papers, one computes
the vector $p_s = \{ p_1^+, ~ p_2^+, ~ \ldots, ~ p_{n^+-1}^+ \}$
to {\em minimize a cost function based only on the economic cost}.
\par
The main difference of the approach proposed in this paper
compared with classical secondary frequency control,
is therefore the cost criterion for the determination
of the components of the vector of power supplies.
\par
The term {\em control problem} is preferred by the authors
over the term {\em optimization problem}.
It is stated by P. Kundur in his book,
\cite[Subsec. 11.1.6, p. 617]{Kundur1994},
that the controllers of a power system have available
very recent information of line flows, system frequencies,
and MW power loadings for automatic generation control.
Thus secondary frequency control is based on output feedback.
From the observations one has to compute
a prediction for the power demand in the next future horizon.
In this paper it is assumed that the control computer of a power system
can compute such a prediction,
hence the control is based on output feedback.
\par
The problem formulated is a problem of {\em stochastic control theory}.
The analytic form of the stochastic control problem is nonlinear
due to the optimization criterion of the probabilities
of exiting the safe subset.
In the research areas of communication networks and of motorway control,
there are treated stochastic control problems 
with a finite-capacity of network lines 
and a probability of exceeding a safe subset,
\cite{smulders1990control,walrand:varaiya:2000}.
\par
\begin{figure}\label{fig:controlcriterionptwo}
\begin{center}
\begin{tikzpicture}
 \message{Expected sensitivty & p-value^^J}
  \def\N{100}
  \def\q{0.4995}
  \def\Bs{2}
  \def\S{\q}
   \def\Ss{0.3302}
     \def\xmin{-4.5*\q}
  \def\xmax{3.2*\q}
  \def\ymin{{-0.16/\Bs}}
  \def\ymax{{0.58/\Bs}}
   \begin{axis}[every axis plot post/.append style={
               mark=none,domain=\xmin:{1.32*\xmax},samples=\N,smooth},
               xmin=\xmin, xmax=1.20*\xmax,
               ymin=\ymin, ymax=\ymax,
               axis lines=middle,
               axis line style=thick,
               enlargelimits=upper, 
               ticks=none,
               xlabel=$v$,
               every axis x label/.style={at={(current axis.right of origin)},anchor=north west},
               y=240pt,
              ]
   \addplot[name path=S,line width=0.35mm,blue!90!red ] {gauss(x,\S,\Ss)};
       \addplot[blue,dashed,line width=0.35mm]
      coordinates {(\q,{-0.04*exp(-\q/\Bs)/\Bs}) (\q, {1.08*gauss(\q,\S,\Ss)})}
       node[below=-2pt,pos=0] {$m$};
    \addplot[red,line width=0.35mm]
      coordinates {(1.57,{-0.04*exp(-1.57/\Bs)/\Bs}) (1.57, {1.38*gauss(1.57,\S,\Ss)})}
       node[below=-5pt,pos=0] {$\pi/2$};
          \addplot[red,line width=0.35mm]
      coordinates {(1.57, {0.1*gauss(1.57,\S,\Ss)}) (1.39*\xmax, {0.1*gauss(1.57,\S,\Ss)})};

       \addplot[red,line width=0.35mm]
      coordinates {(-1.57,{-0.009*exp(1.57/\Bs)/\Bs}) (-1.57, {10.5*gauss(-1.57,\S,\Ss)})}
       node[below=-4pt,pos=0] {$-\pi/2$};    
              \addplot[red,line width=0.35mm]
      coordinates {(\xmin,{2.5*gauss(-1.57,\S,\Ss)}) (-1.57, {2.5*gauss(-1.57,\S,\Ss)})};

    \path[name path=xaxis1]
     (0,0) -- (1.08*\xmax,0);
    \addplot[red!60!blue] fill between[of=xaxis1 and S, soft clip={domain=1.57:1.50*\xmax}];
       \path[name path=xaxis2]
     (-1.08*\xmax,0) -- (0,0);
    \addplot[red!60!blue] fill between[of=xaxis2 and S, soft clip={domain=-1.08*\xmax:-1.57}];
    \node[below left=2pt] at (0,0) {0};

    \node[above right=1.3pt,black!20!blue] at ( 0.45*\Bs,0.13) {$p_{\theta_i - \theta_j}(v)$};
    
  \end{axis}
  
   \end{tikzpicture}
\end{center}
\caption{The probability density function of the phase-angle difference 
$(\theta_{i_k}-\theta_{j_k})$ of the power flow
in power line  $k=(i_k,j_k)\in \mathcal{\mathcal{E}}$; and 
two red bars for the probabilities that the phase-angle difference 
is larger than $+\pi/2$ or less than $-\pi/2$. 
The parameters of the probability density function displayed
are $(m, ~ \sigma) = (0.4995, ~ 0.3344)$
which values are chosen identical to those of 
Fig.~\ref{fig:ringnetcontrolwithoutwith} for `without control'.}
\label{fig:gaussianprobdensityfunctionsafeset}
\vspace{-3mm}
\end{figure}
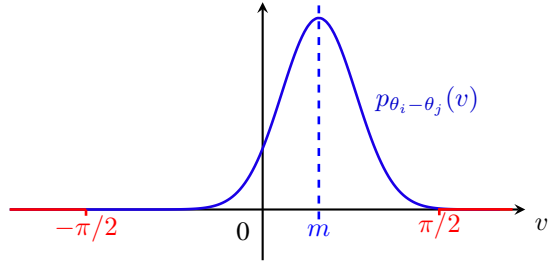

 
Fig.~\ref{fig:gaussianprobdensityfunctionsafeset}
illustrates the stochastic control problem.
The extremal probabilities of power line $k$
that $(\theta_{i_k}-\theta_{j_k})>+\pi/2$ and 
$(\theta_{i_k}-\theta_{j_k})<-\pi/2$
are to be made sufficiently small by a choice of the power supply vector.
\subsection{The Control Objective Function}\label{sec:conobj}
\begin{definition}\label{def:controlobjectivefunction}
{\em The control objective function}.
\par
Consider the stochastic linearized power system 
of Section~\ref{sec:linearsto}.
Define the functions,
\begin{align}
    &   \forall ~ k = (i_k, ~ j_k) \in \mathbb{Z}_{n_\mathcal{E}}, ~ 
          \forall ~ p_s \in P^+, \nonumber \\
        f_{as,k}(p_s)
    & = | \theta_{s,i_k}(p_s) - \theta_{s,j_k}(p_s) | \nonumber \\
    & = \arcsin( | (A~ p_s + b)_k | )=|m_k(p_s)|,~ 
      \label{eq:matricesAb} \\
           f_k(p_s)
    & = f_{as,k}(p_s) + r_\epsilon ~ \sigma_k(p_s) \in \mathbb{R}_+,  
      \label{eq:fkps} \\
          f(p_s)
    & = \| (f_k(p_s)) \|_{\infty} 
          = \max_{k \in \mathbb{Z}_{n_\mathcal{E}}} ~ f_k(p_s), 
      \label{eq:conobjectivefunction}   \\
    &   f_{as,k}: P^+ \rightarrow \mathbb{R}_+, ~
          f_{as}: P^+ \rightarrow \mathbb{R}^{n_\mathcal{E}}, 
          \nonumber \\
    &   f_{k}: P^+ \rightarrow \mathbb{R}_+, ~
          f: P^+ \rightarrow \mathbb{R}_+, ~ 
            r_\epsilon \in \mathbb{R}_{s+}.  \nonumber 
\end{align}
\end{definition}
 $f(p_s)$ is the {maximum over all power lines of an upper bound on the probability 
that the power flow 
of 
any power line is outside the safe set,
when the power supply vector is $p_s \in P^+$. 
$f$ can be considered to be the {\em control objective function}.
The parameter $r_\epsilon \in \mathbb{R}_{s+}$ which appears in \eqref{eq:fkps}
is a parameter of the control objective function. 
Appendix~\ref{computingAb} contains
the formulas of the matrices $A$ and $b$ of $f_{as,k}(p_s)$ in (\ref{eq:matricesAb}) .
%
\par
For a particular decision vector $p_s \in P^+$,
$f(p_s)$ can be computed by the following steps:
(1) For the given power network parameters, compute $A$ and $b$ of $f_{as,k}(p_s)$ according to Appendix~\ref{computingAb} which is also addressed in \ref{sec:existenceofastable} and determine the synchronous phase-angle differences vector $\theta_{s,i_k}-\theta_{s,j_k}=\arcsin(A~ p_s + b) \in (- \pi/2, ~ +\pi/2)^{n_{\mathcal{E}}}$.
Then compute
$| m_k(p_s) |$ for all $k \in \mathcal{\mathcal{E}}$.
(2) Compute $F(\theta_s, ~ 0)$ and 
the Jacobian matrix $J(\theta_s, ~ 0)$.
(3) Using the matrices $(J(\theta_s, ~ 0), ~ K, ~ C)$
and a reduction process,
solve a Lyapunov equation,
and then compute $Q_y$ and $\sigma_k(p_s)$
for all $k \in \mathbb{Z}_{n_{\mathcal{E}}}$; and
(4) compute $f(p_s)$ according to the equations
(\ref{eq:matricesAb},\ref{eq:fkps},\ref{eq:conobjectivefunction}).
\subsection{The Control Problem}\label{subsec:conproblem}
\begin{problem}\label{probleminfimizefunctionf}
{\em Control of the probability that any phase-angle difference
of a power line leaves the safe subset}. 
\par
Consider the stochastic linearized power system described by the equations
(\ref{eq:linsde}, \ref{eq:linsdeoutput}) and
a value $\epsilon \in (0, ~ 1)$.
\par
The control problem is 
to determine a vector of power supplies for the next short horizon,
such that the probability mentioned above,
is less than the value $\epsilon$. 
Furthermore, that probability is best minimized.
\par
In terms of a mathematical formula,
the problem is to solve the following optimization problem,
including to determine a {\em minimizer} $p_s^*\in P^+$
and a {\em value} $a \in \mathbb{R}$, according to,
\begin{align*}
        a
    & = f(p_s^*)= \inf_{p_s \in P^+} ~ f(p_s) \\
    & = \inf_{p_s \in P^+} ~
        \max_{k = (i_k, ~ j_k) \in \mathcal{E}}
          \left[ ~ 
          \begin{array}{l}
            |\theta_{s,i_k}(p_s)-\theta_{s,j_k}(p_s)| + \\
            + r_{\epsilon} ~ \sigma_k(p_s) ~
          \end{array}
          \right].
\end{align*}
If the value satisfies $a < \pi/2$ 
then there exists an input vector $p_s \in P^+$
such that the probability is less than $\epsilon$ that the phase-angle difference of any power flow
leaves the safe subset during a short horizon.
\end{problem}
\par
It is proven in a companion paper,
\cite{zhenwang:reportthree:2023},
that the control objective function $f$ defined above,
is nondifferentiable and nonconvex. The nondifferentiable property is caused by the absolute value operator in the $f_{as,k}$ in (\ref{eq:matricesAb}), and
the infinity norm in (\ref{eq:conobjectivefunction}). The nonconvexity is due to the variance $\sigma_k(p_s)$ in (\ref{eq:fkps}).
A minimizer of the control objective function 
over the set of power supply vectors exists and
an iterative procedure based on 
a generalized subgradient or first directional derivative 
produces an approximation sequence 
which is proven to converge 
to a rather good local minimizer in a finite number of steps. 
\par
After applying the proposed procedure, 
the determined power supply vector makes the power system relatively 
more stable
with as consequence that the most vulnerable line,
(the line with the highest exit probability from the safe set),
has a probability to leave the safe subset, below the preset bound.
Hence the transient stability of the power system is improved.
%
\section{The Performance of the Controlled Power System for Three Examples}
\label{sec:examples}
Example~\ref{ex:eightnodenet}
has been chosen because it has been used to illustrate the Braess paradox. 
This phenomenon concerns the power flows 
when scheduling or planning the operations of a power system.
Such functions influence the transient stability of the power system,
\cite{Witthaut2012}.
Example~\ref{ex:ringnettwelvenodes} of a ring network 
has been chosen because such rings often occur in power networks.
Example~\ref{ex:manhattengrid} of a Manhattan grid 
has been chosen because it was expected to be relatively stable.
For all three examples, 
the value of the parameter $r_\epsilon$ of the control objective function 
is chosen to be $3.08$, 
which leads to a probability of $99.8 \%$
of the power flow inside the safe set. 
\par
The reader finds the details of the computations of the examples
in a report,
\cite{zhenwang:2023:arxiv}.
\par
\par
Note that, if one assumes that the probabilities of leaving the safe set
are independent and that one is willing to accept a probability
of leaving the safe set once per year,
then the probability of leaving the safe set in a three minute period
may be approximated by 
$\epsilon = 6 \times 10^{-6} \approx ((60/3) \times 24 \times 365)^{-1}$.
\par
Of interest to control is the comparison of the performance of 
a power system {\em with control}, as proposed in this paper, and
a power system {\em without control}.
\par
Define the {\em proportional control law} for the power supply vector
by the following procedure for a short horizon.
Consider the vector of maximal available power supply
$p^{+,max}$ and the vector of power demands $p^-$
which satisfy that 
$\sum_{i=1}^{n^+} ~ p_i^{+,max} = p_{sum}^{+,max} 
\geq p_{sum}^- = \sum_{j=1}^{n^-} ~ p_j^-$.
Choose $p_{sum}^+ = p_{sum}^-$ hence power supply equals power demand.
Define the fraction 
$s = p_{sum}^+ / p_{sum}^{+,max} \in (0, ~ 1)$.
Define then the power vector $p^+ \in \mathbb{R}^{n^+}$,
$p_i^+ = s \times p_i^{+,max}$ for all $i \in \mathbb{Z}_{n^+}$.
Then $\sum_{i=1}^{n^+} ~ p_i^+ = p_{sum}^+ = p_{sum}^-$.
\par
The comparison between {\em without control} and {\em with control}
will then be focused on the three values of,
\begin{align*}
    & \left\{ 
        ( |\theta_{i_k} - \theta_{j_k}|, ~ \sigma_k, ~ 
          f_k(p_s) = |\theta_{i_k} - \theta_{j_k}| + r_\epsilon~ \sigma_k), ~
        \forall ~ k \in \mathcal{E}
      \right\}.
\end{align*}
The results of this comparison are displayed below
for the examples of a ring network and of a Manhattan grid.
\begin{example}\label{ex:eightnodenet}
{\em A particular eight-node power network}.
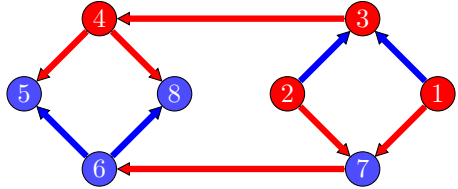
\begin{figure}[h]
\centering
\begin{tikzpicture}[tbcircle/.style={circle,minimum size=13pt,inner sep=0pt},
forward tbcircle/.style={tbcircle,fill=blue!70},
backward tbcircle/.style={tbcircle,fill=red!100},
tbsquare/.style={rectangle,rounded corners,minimum width=25pt,minimum height=25pt,fill=orange!50},
tbcarrow/.style={/utils/exec=\tikzset{tbcarrow pars/.cd,#1},
line width=0.1*\pgfkeysvalueof{/tikz/tbcarrow pars/width},
draw=\pgfkeysvalueof{/tikz/tbcarrow pars/draw},
-{Triangle[line width=\pgfkeysvalueof{/tikz/tbcarrow pars/line width},
length=1.25*\pgfkeysvalueof{/tikz/tbcarrow pars/width},
width=1.3*\pgfkeysvalueof{/tikz/tbcarrow pars/width},
fill=\pgfkeysvalueof{/tikz/tbcarrow pars/fill}]},
postaction={draw=\pgfkeysvalueof{/tikz/tbcarrow pars/fill},-,
line width=\pgfkeysvalueof{/tikz/tbcarrow pars/width}-4*\pgfkeysvalueof{/tikz/tbcarrow pars/line width},
shorten <=\pgfkeysvalueof{/tikz/tbcarrow pars/line width},
shorten >=1.25*\pgfkeysvalueof{/tikz/tbcarrow pars/width}-1.5*\pgfkeysvalueof{/tikz/tbcarrow pars/line width}}
},
tbcarrow pars/.cd,
fill/.initial=white,draw/.initial=black,width/.initial=4pt,
line width/.initial=0.4pt,]\label{kkkk}

\node[backward tbcircle,draw]  at (1,0) {$\textcolor{white}{2}$};
\node[backward tbcircle,draw]  at (3,0) {$\textcolor{white}{1}$};
\node[backward tbcircle,draw]  at (2,1) {$\textcolor{white}{3}$};
\node[forward tbcircle,draw]  at (2,-1) {$\textcolor{white}{7}$};
\node[forward tbcircle,draw]  at (-0.5,0) {$\textcolor{white}{8}$};
\node[forward tbcircle,draw]  at (-2.5,0) {$\textcolor{white}{5}$};
\node[backward tbcircle,draw]  at (-1.5,1) {$\textcolor{white}{4}$};
\node[forward tbcircle,draw]  at (-1.5,-1) {$\textcolor{white}{6}$};

\draw[tbcarrow={fill=blue}] (1+0.16,0+0.16) to (2-0.16,1-0.16);
 
\draw[tbcarrow={fill=blue}] (3-0.16,0+0.16) to (2+0.16,1-0.16);
 
\draw[tbcarrow={fill=red}] (1+0.16,0-0.16) to (2-0.16,-1+0.16) ;

\draw[tbcarrow={fill=red}] (3-0.16,0-0.16) to (2+0.16,-1+0.16) ;

\draw[tbcarrow={fill=red}] (2-0.23,1) to  (-1.5+0.23,1);

\draw[tbcarrow={fill=red}] (2-0.23,-1) to (-1.5+0.23,-1);

\draw[tbcarrow={fill=red}] (-1.5-0.16,1-0.16) to (-2.5+0.16,0+0.16);

\draw[tbcarrow={fill=red}](-1.5+0.16,1-0.16) to (-0.5-0.16,0+0.16);

\draw[tbcarrow={fill=blue}] (-1.5-0.16,-1+0.16) to (-2.5+0.16,0-0.16);

\draw[tbcarrow={fill=blue}] (-1.5+0.16,-1+0.16) to (-0.5-0.16,0-0.16);

 
 \end{tikzpicture}

\caption{A particular eight-node academic example}
\label{fig:figureeight}
\end{figure}

\begin{figure*}
  \centering
  \begin{tikzpicture}
   \draw[->, line width=1pt, blue!50!black, >=stealth] (0,0) -- (11,0);
    \draw[->, line width=1pt, blue!50!black, >=stealth] (0,0) -- (0,4);
        
    \fill[blue] (0.5,0) rectangle (0.7,2*0.5274); 
    \fill[red] (0.7,0) rectangle (0.9,2*0.2681); 
    \fill[green] (0.9,0) rectangle (1.1,2*1.3530); 
      \node at (0.9,-0.3) {{\scriptsize 6-7}};
      \fill[blue] (0.5+1,0) rectangle (0.7+1,2*0.5217); 
    \fill[red] (0.7+1,0) rectangle (0.9+1,2*0.2699); 
    \fill[green] (0.9+1,0) rectangle (1.1+1,2*1.3529); 
     \node at (0.9+1,-0.3) {{\scriptsize 4-8}};
    \fill[blue] (0.5+2,0) rectangle (0.7+2,2*0.4988); 
    \fill[red] (0.7+2,0) rectangle (0.9+2,2*0.2694); 
    \fill[green] (0.9+2,0) rectangle (1.1+2,2*1.3284); 
     \node at (0.9+2,-0.3) {{\scriptsize 4-5}};
\fill[blue] (0.5+3,0) rectangle (0.7+3,2*0.5605); 
    \fill[red] (0.7+3,0) rectangle (0.9+3,2*0.2239); 
    \fill[green] (0.9+3,0) rectangle (1.1+3,2*1.2500); 
    \node at (0.9+3,-0.3) {{\scriptsize 2-7}};
 \fill[blue] (0.5+4,0) rectangle (0.7+4,2*0.5140); 
    \fill[red] (0.7+4,0) rectangle (0.9+4,2*0.2230); 
    \fill[green] (0.9+4,0) rectangle (1.1+4,2*1.2007); 
     \node at (0.9+4,-0.3) {{\scriptsize 1-7}};
    \fill[blue] (0.5+5,0) rectangle (0.7+5,2*0.5695); 
    \fill[red] (0.7+5,0) rectangle (0.9+5,2*0.1648); 
    \fill[green] (0.9+5,0) rectangle (1.1+5,2*1.0770); 
    \node at (0.9+5,-0.3) {{\scriptsize 3-4}};
\fill[blue] (0.5+6,0) rectangle (0.7+6,2*0.0216); 
    \fill[red] (0.7+6,0) rectangle (0.9+6,2*0.2658); 
    \fill[green] (0.9+6,0) rectangle (1.1+6,2*0.8402); 
    \node at (0.9+6,-0.3) {{\scriptsize 6-8}};

\fill[blue] (0.5+7,0) rectangle (0.7+7,2*0.0016); 
    \fill[red] (0.7+7,0) rectangle (0.9+7,2*0.2659); 
    \fill[green] (0.9+7,0) rectangle (1.1+7,2*0.8205); 
    \node at (0.9+7,-0.3) {{\scriptsize 5-6}};
 \fill[blue] (0.5+8,0) rectangle (0.7+8,2*0.0189); 
    \fill[red] (0.7+8,0) rectangle (0.9+8,2*0.1267); 
    \fill[green] (0.9+8,0) rectangle (1.1+8,2*0.4091); 
    \node at (0.9+8,-0.3) {{\scriptsize 5-6}};
     \fill[blue] (0.5+9,0) rectangle (0.7+9,2*0.0210); 
    \fill[red] (0.7+9,0) rectangle (0.9+9,2*0.1247); 
    \fill[green] (0.9+9,0) rectangle (1.1+9,2*0.4051); 
     \node at (0.9+9,-0.3) {{\scriptsize 2-3}};
%

   \fill[blue] (10.5-2.9,3.5) rectangle (10.7-2.9,3.9);   
    \node at (11.7-3.0,3.7) {{\scriptsize $|\theta_i-\theta_j|$}};
  \fill[red] (10.5-2.9,3.5-0.5) rectangle (10.7-2.9,3.9-0.5);   
    \node at (11.4-3.0,3.7-0.5) {{\scriptsize $\sigma_{ij}$}};
    \fill[green] (10.5-2.9,3.5-1) rectangle (10.7-2.9,3.9-1);   
    \node at (12.5-3.2,3.7-1) {{\scriptsize $|\theta_i-\theta_j|+3.08\:\sigma_{ij}$}};

    \node at (10.5,-0.7) {{\scriptsize Line}};
     \node at (0.1,2) {{\scriptsize -1}};
             \node at (0.2,1.57*2) {{\scriptsize-$\frac{\pi}{2}$}};
              \node at (0.15,0.5*2) {{\scriptsize-$\frac{1}{2}$}};

  \end{tikzpicture}
  \caption{The outputs of the particular eight-node academic network}
  \label{fig:eightnodenetcomputations}
\end{figure*}
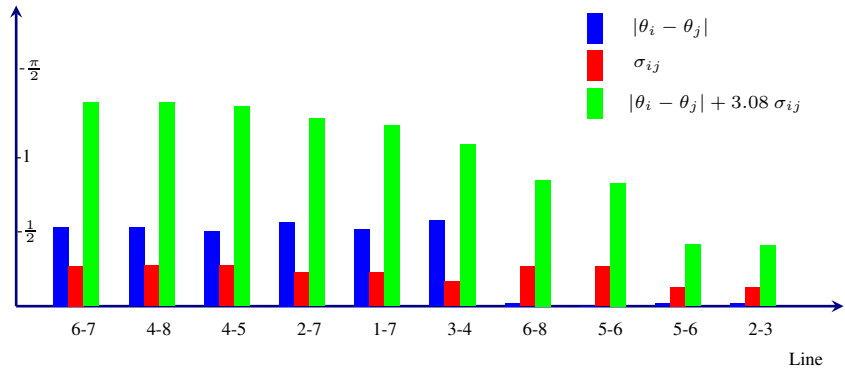
This academic power network is borrowed from the paper \cite{Witthaut2012}.
It is often used to illustrate the Braess paradox.
The network is displayed in Fig.~\ref{fig:figureeight}, where
the nodes which are colored red, $1,2,3,4$, provide power supply
while nodes which are colored blue, $5,6,7,8$, have only power loads.
\par
The parameters of the power system are 
the inertias, the damping coefficients, 
the standard deviations of the disturbances, 
the maximal power supplies and the power demands,
and the line capacities.
In a second part of the example for other cases, 
the parameter values are changed to different values. 
One computation of the first part 
is displayed in Fig.~\ref{fig:eightnodenetcomputations}.
The columns of that column chart 
are ordered according to the values of 
$|\theta_i-\theta_j|+3.08\:\sigma_{ij}$ 
in the decreasing order of magnitude. 
\par
The conclusions for this example are:
\begin{enumerate}
\item
There are six lines with relatively high power flows which are colored red and
four lines with relatively low power flows 
which are colored blue in
Fig.~ \ref{fig:figureeight}.
\item
The computation has been additionally evaluated by discretizing the domain of the power supply vectors for each dimension into a grid with 150 steps. The minimum value computed by this grid method is 1.3529 which is smaller than the value in Table \ref{table:OriginalTheminimumvalueandoptimalpowervector1} while the time for computation  is longer than that of our proposed method.
\item
From different initial states, the local minimizers of the control objective function might be different, and that is due to the non-convexity property of
the control objective function, see the Tables \ref{table:OriginalTheminimumvalueandoptimalpowervector1} and \ref{table:OriginalTheminimumvalueandoptimalpowervector2}.
\item
It is very amazing that different minimizers have 
almost the same minimum values, see the minimum values in
Tables~\ref{table:OriginalTheminimumvalueandoptimalpowervector1} and
\ref{table:OriginalTheminimumvalueandoptimalpowervector2}. Due to this phenomenon,
 other economic considerations can decide the "best" power injection among these safe ones.
\end{enumerate}
\begin{remark}
A {\em Braess paradox of Example 4.1.}
If one adds to the power network the line $(2, 4)$
between node 2 and node 4
or if one doubles only the capacity of line $(3, 4)$
then one expects that the value of the minimum will decrease.
Yet, the value of the minimum increases, 
see the
Tables~\ref{table:AddingLine24theminimumvalueandoptimalpowervector} and
\ref{table:DoublingLine34theminimumvalueandoptimalpowervector} 
in contrast with  
Table~\ref{table:OriginalTheminimumvalueandoptimalpowervector1}.
This is a paradoxical phenomenon,
see the Braess paradox described in
\cite{frank1981braess} and
\cite{steinberg1983prevalence}.
\end{remark}

\begin{remark}\label{re:contingency}
Dealing with a contingency which often happens in power systems. When large power consumers like steel factories or data centers start to work, the power demand increases immediately. Consider the case that the maximal available power supply cannot meet the power demand. By a summation of equation (\ref{eq:synchronousstateequationcomponent}) over all nodes, we can get $\omega_s=\frac{\sum_{i=1}^{n_\mathcal{V}}{p_{sp,i}}}{\sum_{i=1}^{n_\mathcal{V}}{D_{(i, i)}}}$, thus $\omega_s<0$ and there is a need of load shedding.  Consider the case that the maximal available power supply can still meet the power demand, but the amount of the power demand approaches the power supply, the computation result of our proposed procedure is shown in Table \ref{increasepowerdemandextremeoutput}, while the comparison of with and without control is shown in Table \ref{increasepowerdemandextremecomparison}. Furthermore, the tail probability computed according to Table \ref{increasepowerdemandextremecomparison} is shown in Table \ref{TailProbabilityEightNodeNetworkExtreme}. 
\par
Note that  $f_{a,k}=P(\omega\in \Omega~|~\theta_{i_k,j_k}(\omega,t)<-\pi/2),f_{b,k}=P(\omega\in \Omega~|~\theta_{i_k,j_k}(\omega,t)>\pi/2)$, and $p_{out,k}$ is the exit probability computed by our proposed criterion, and it equals two times an upper bound of $f_{a,k}$ and $f_{b,k}$. 
From these comparison, one can see indeed the power system is more stable after control in a practical sense.
\end{remark}
\centering
\vspace{-2mm}
\begin{table}[H]
\centering
  \caption{Example~\ref{ex:eightnodenet}.
  The minimum value and the optimal power vector from the initial power supply vector $[12,12,12]$.}
\begin{tabular}{|c|c|c|c|c|} 
\hline
\text{Minimum} & $p_1$   & $p_2$   & $p_3$   & $p_4$ \\
\hline
1.3530         & 11.7679 & 13.7632 & 13.5247& 10.9441 \\
\hline
               & $p_5$   & $p_6$   & $p_7$   & $p_8$\\
\hline
               & -12     & -12     & -13     & -13 \\ 
\hline
\end{tabular}
  \label{table:OriginalTheminimumvalueandoptimalpowervector1}
\end{table}
\vspace{-0.5cm}

\begin{table}[H]
\centering
  \caption{Example~\ref{ex:eightnodenet}.
  The minimum value and the optimal power vector from the initial power supply vector $[11,11,14]$.}
\begin{tabular}{|c|c|c|c|c|} 
\hline
\text{Minimum} & $p_1$   & $p_2$   & $p_3$   & $p_4$ \\
\hline
1.3532         & 11.7085 & 12.6649 & 15.2572 & 10.3694 \\
\hline
               & $p_5$   & $p_6$   & $p_7$   & $p_8$\\
\hline
               & -12     & -12     & -13     & -13 \\ 
\hline
\end{tabular}
  \label{table:OriginalTheminimumvalueandoptimalpowervector2}
\end{table}

\begin{table}[H]
\centering
  \caption{Example~\ref{ex:eightnodenet}.
  Added is line $2-4$. 
  Results of the minimum value and the optimal power vector from the initial power supply vector $[12,12,12]$.}
\begin{tabular}{|c|c|c|c|c|} 
\hline
\text{Minimum} & $p_1$   & $p_2$   & $p_3$   & $p_4$ \\
\hline
1.3953         & 12.0000 & 14.0000 & 16.0000 & 8.0000 \\
\hline
               & $p_5$   & $p_6$   & $p_7$   & $p_8$\\
\hline
               & -12     & -12     & -13     & -13 \\ 
\hline
\end{tabular}
\label{table:AddingLine24theminimumvalueandoptimalpowervector}
\end{table}

\begin{table}[H]
\centering
  \caption{Example~\ref{ex:eightnodenet}.
      Change doubling the capacity of Line $3-4$. 
      Results of the minimum and the optimal power vector from the initial power supply vector $[12,12,12]$.}
\begin{tabular}{|c|c|c|c|c|} 
\hline
\text{Minimum} & $p_1$   & $p_2$   & $p_3$   & $p_4$   \\
\hline
1.3712         & 12.0000 & 14.0000 & 16.0000 &8.0000\\
\hline
               & $p_5$   & $p_6$   & $p_7$   & $p_8$ \\
\hline
               & -12     & -12     & -13     & -13 \\
\hline
\end{tabular}
\label{table:DoublingLine34theminimumvalueandoptimalpowervector}
\end{table}

 \begin{table}[H]
 \centering
  \caption{Insrease the power demand to 59, where the maximal power demand is 60, the minimum value and optimal power vector start from the power supply vector is $[5,5,5]$}
  \begin{tabular}{|c|c|c|c|c|} 
\hline
\text{Minimum} & $p_1$   & $p_2$   & $p_3$   & $p_4$   \\
\hline
1.5306& 12.0001 & 14.001 & 16.0000 & 16.9998\\
\hline
               & $p_5$   & $p_6$   & $p_7$   & $p_8$ \\
\hline
               & -16 & -13 & -14 & -16\\
\hline
\end{tabular}
  \label{increasepowerdemandextremeoutput}
\end{table}

\begin{table}[h]
\centering
  \caption{Example~\ref{ex:eightnodenet} The particular eight-node network. Increasing the power demand to almost the maximal power supply,
      comparison of performance without and with control.}
\vspace{1mm}
\begin{tabular} {|r|r|l|r|r|c|} 
\hline
\multicolumn{2}{|l|}{Nodes} 
          & Control & $|\theta_{i_k} - \theta_{j_k}|$ 
                              & $\sigma_k$ & $|\theta_{i_k} - \theta_{j_k}| + r_{\epsilon}~  \sigma_k$ \\
\hline
$i_k$ 
    & $j_k$ 
          &         &         &            & \\
\hline
 4 & 5 &without & 0.6746 & 0.2801 & 1.5373 \\ 
4 & 5&with & 0.6687 & 0.2798 & 1.5306 \\\hline 
4 & 8 &without & 0.6746 & 0.2796 & 1.5358 \\
4 & 8&with & 0.6687 & 0.2793 & 1.5291 \\\hline
6 & 7 &without & 0.5834 & 0.2727 & 1.4234 \\  
6 & 7 &with & 0.5944 & 0.2734 & 1.4366 \\\hline
2 & 7 &without & 0.6127 & 0.2272 & 1.3126 \\ 
2 & 7 &with & 0.6187 & 0.2276 & 1.3198 \\\hline
\end{tabular}
\label{increasepowerdemandextremecomparison}
\end{table}

\begin{table}
\begin{center}
\caption{The exit probabilities per component of the vector of phase-angle differences
for the two sides of the safe subset of the particular eight-node network according to Table \ref{increasepowerdemandextremecomparison},
both without control and with control. }
\begin{tabular}{|r|r|r|r|r|r|r|}
\hline
Power line & \multicolumn{2}{|l}{Without control} & \multicolumn{3}{|l|}{With control} \\
           & \multicolumn{2}{|l}{probabilities}   & \multicolumn{3}{|l|}{probabilities} \\
\hline
$k$        & $f_{a,k}$ & $ f_{b,k}$  & $f_{a,k}$ & $f_{b,k}$ & $p_{out,k}$ \\
\hline
4-5         & 5.44e-16 & 6.88e-04&6.03e-16 & 6.32e-04 & 2× 6.33e-04     \\
4-8        &  4.84e-16 & 6.75e-04&             5.36e-16 & 6.19e-04 & 2× 6.32e-04     \\
6-7         &1.47e-04 & 1.40e-15&  1.78e-04 & 1.19e-15 &2× 6.25e-04      \\
2-7         &3.61e-22 & 1.24e-05&3.29e-22 & 1.44e-05 &2× 5.64e-04     \\
\hline
\end{tabular}
\label{TailProbabilityEightNodeNetworkExtreme}
\end{center}
\end{table}

\end{example}
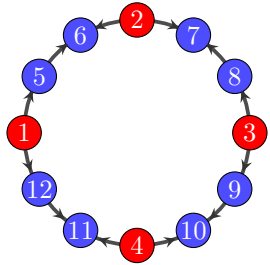
\begin{figure}[h]
\begin{center} 
\xdef\Rad{1.5}
\newcommand{\ARW}[2][]{%
    \foreach \ang in {#2}{%
        \draw[#1] (\ang:\Rad)--(\ang+1:\Rad) ;
    }
}

\begin{tikzpicture}[tbcircle/.style={circle,minimum size=13pt,inner sep=0pt},
forward tbcircle/.style={tbcircle,fill=blue!70},
backward tbcircle/.style={tbcircle,fill=red!100},>=stealth]

\draw[line width=0.5mm, color=black!70] (0,0) circle (\Rad) ;

\node[backward tbcircle,draw] at (0:\Rad) {$\textcolor{white}{3}$} ;
\node[forward tbcircle,draw] at (30:\Rad) {$\textcolor{white}{8}$} ;
\node[forward tbcircle,draw] at (60:\Rad) {$\textcolor{white}{7}$} ;
\node[backward tbcircle,draw] at (90:\Rad) {$\textcolor{white}{2}$} ;
\node[forward tbcircle,draw] at (120:\Rad) {$\textcolor{white}{6}$} ;
\node[forward tbcircle,draw] at (150:\Rad) {$\textcolor{white}{5}$} ;
\node[backward tbcircle,draw] at (180:\Rad){$\textcolor{white}{1}$} ;
\node[forward tbcircle,draw] at (210:\Rad) {$\textcolor{white}{12}$} ;
\node[forward tbcircle,draw] at (240:\Rad) {$\textcolor{white}{11}$} ;
\node[backward tbcircle,draw] at (270:\Rad) {$\textcolor{white}{4}$} ;
\node[forward tbcircle,draw] at (300:\Rad) {$\textcolor{white}{10}$} ;
\node[forward tbcircle,draw] at (330:\Rad) {$\textcolor{white}{9}$} ;


\ARW[line width=0.3mm,black!80,->]{21,51,111,199,229,289}
\ARW[line width=0.3mm,black!80,<-]{68,340,310,128,158,250}
\end{tikzpicture}
\caption{A ring network}
\label{fig:ringnetwork}
\end{center}
\end{figure}
\par\vspace{1\baselineskip}
\begin{example}\label{ex:ringnettwelvenodes}
{\em A ring network with twelve nodes}.
\begin{figure*}[h]
  \centering
  \begin{tikzpicture}
   \draw[->, line width=1pt, blue!50!black, >=stealth] (0,0) -- (13,0);
    \draw[->, line width=1pt, blue!50!black, >=stealth] (0,0) -- (0,4);
        
    \fill[blue] (0.5,0) rectangle (0.7,2* 0.6747); 
    \fill[red] (0.7,0) rectangle (0.9,2*0.2503); 
    \fill[green] (0.9,0) rectangle (1.1,2*1.4455); 
      \node at (0.9,-0.3) {{\scriptsize 4-10}};
      \fill[blue] (0.5+1,0) rectangle (0.7+1,2*0.6755); 
    \fill[red] (0.7+1,0) rectangle (0.9+1,2* 0.2500); 
    \fill[green] (0.9+1,0) rectangle (1.1+1,2*1.4455); 
     \node at (0.9+1,-0.3) {{\scriptsize 1-12}};
    \fill[blue] (0.5+2,0) rectangle (0.7+2,2*0.5742); 
    \fill[red] (0.7+2,0) rectangle (0.9+2,2*0.2647); 
    \fill[green] (0.9+2,0) rectangle (1.1+2,2*1.3896); 
     \node at (0.9+2,-0.3) {{\scriptsize 3-9}};
\fill[blue] (0.5+3,0) rectangle (0.7+3,2*0.5236); 
    \fill[red] (0.7+3,0) rectangle (0.9+3,2* 0.2496); 
    \fill[green] (0.9+3,0) rectangle (1.1+3,2*1.2923); 
    \node at (0.9+3,-0.3) {{\scriptsize 2-7}};
 \fill[blue] (0.5+4,0) rectangle (0.7+4,2*0.3398); 
    \fill[red] (0.7+4,0) rectangle (0.9+4,2*0.2433); 
    \fill[green] (0.9+4,0) rectangle (1.1+4,2*1.0891); 
     \node at (0.9+4,-0.3) {{\scriptsize 1-5}};
    \fill[blue] (0.5+5,0) rectangle (0.7+5,2*0.3707); 
    \fill[red] (0.7+5,0) rectangle (0.9+5,2*0.2229); 
    \fill[green] (0.9+5,0) rectangle (1.1+5,2*1.0573); 
    \node at (0.9+5,-0.3) {{\scriptsize 4-11}};
\fill[blue] (0.5+6,0) rectangle (0.7+6,2*0.3093); 
    \fill[red] (0.7+6,0) rectangle (0.9+6,2*0.2214); 
    \fill[green] (0.9+6,0) rectangle (1.1+6,2*0.9913); 
    \node at (0.9+6,-0.3) {{\scriptsize 2-6}};

\fill[blue] (0.5+7,0) rectangle (0.7+7,2*0.2083 ); 
    \fill[red] (0.7+7,0) rectangle (0.9+7,2*0.2502); 
    \fill[green] (0.9+7,0) rectangle (1.1+7,2*0.9790); 
    \node at (0.9+7,-0.3) {{\scriptsize 3-8}};
 \fill[blue] (0.5+8,0) rectangle (0.7+8,2*0.1675); 
    \fill[red] (0.7+8,0) rectangle (0.9+8,2*0.2398); 
    \fill[green] (0.9+8,0) rectangle (1.1+8,2*0.9060); 
    \node at (0.9+8,-0.3) {{\scriptsize 7-8}};
     \fill[blue] (0.5+9,0) rectangle (0.7+9,2*0.0849); 
    \fill[red] (0.7+9,0) rectangle (0.9+9,2*0.2490); 
    \fill[green] (0.9+9,0) rectangle (1.1+9,2*0.8519); 
     \node at (0.9+9,-0.3) {{\scriptsize 11-12}};
 \fill[blue] (0.5+10,0) rectangle (0.7+10,2*0.1125); 
    \fill[red] (0.7+10,0) rectangle (0.9+10,2*0.2172); 
    \fill[green] (0.9+10,0) rectangle (1.1+10,2* 0.7814); 
    \node at (0.9+10,-0.3) {{\scriptsize 5-6}};
     \fill[blue] (0.5+11,0) rectangle (0.7+11,2*0.0831); 
    \fill[red] (0.7+11,0) rectangle (0.9+11,2*0.2265); 
    \fill[green] (0.9+11,0) rectangle (1.1+11,2*0.7807); 
     \node at (0.9+11,-0.3) {{\scriptsize 9-10}};


   \fill[blue] (10.5-1.3,5-1.5) rectangle (10.7-1.3,5.4-1.5);   
    \node at (11.7-1.3,5.2-1.5) {{\scriptsize $|\theta_i-\theta_j|$}};
  \fill[red] (10.5-1.3,5-0.5-1.5) rectangle (10.7-1.3,5.4-0.5-1.5);   
    \node at (11.4-1.3,5.2-0.5-1.5) {{\scriptsize $\sigma_{ij}$}};
    \fill[green] (10.5-1.3,5-1-1.5) rectangle (10.7-1.3,5.4-1-1.5);   
    \node at (12.5-1.5,5.2-1-1.5) {{\scriptsize $|\theta_i-\theta_j|+3.08\sigma_{ij}$}};   
    
    \node at (12.5,-0.7) {{\scriptsize Line}};
     \node at (0.1,2) {{\scriptsize -1}};
     \node at (0.15,1) {{\scriptsize -$\frac{1}{2}$}};
             \node at (0.2,1.57*2) {{\scriptsize -$\frac{\pi}{2}$}};
  \end{tikzpicture}
  \caption{The outputs of the ring network starting from $[20,18,25]$}
  \label{fig:ringoutputcomputation}
\end{figure*}
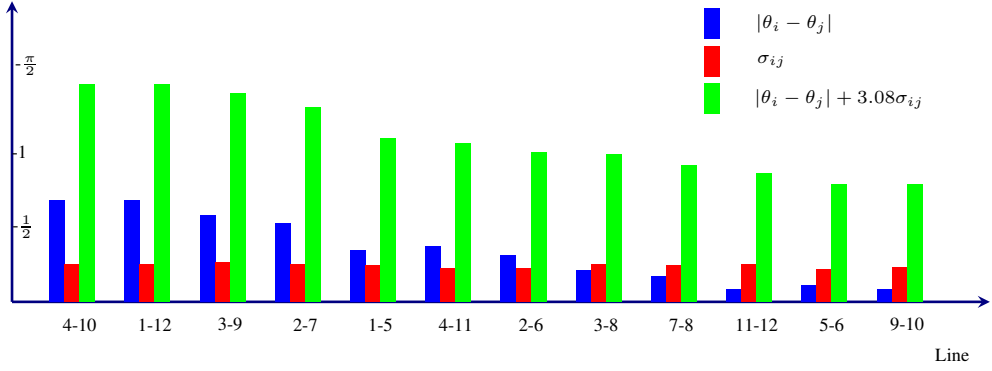
The power network shown in 
Fig.~\ref{fig:ringnetwork}
consists of a ring with
four red-colored nodes with power supply, $1,2,3$ and $4$,
and eight blue-colored nodes with power demand. 
As for the  Manhattan-grid network, 
the inertias and damping coefficients for nodes with power supply 
are relatively larger than those of power demand nodes.
Due to the ring network structure,
every neighborhood is connected to only two other neighborhoods, 
for example, 
the neighborhood $(2, 6, 7)$ only directly connects 
to the neighborhoods $(1, 5, 12)$ and $(3, 8, 9)$.
The power network is inspired by \cite{Xi2016}, and 
the variance of a ring network structure 
is less than that of a tree-like structure for a rate $O(1/N)$ 
when they are subject to the same disturbances \cite{WANG2023110884}, 
where $N$ is the number of nodes of these two network structures. 
Also, the authors in \cite{menck2014dead} introduced 
a method of curing dead ends in power network 
by formulating a small ring structure.
The outcomes for a group of asymmetric network parameters 
starting from the power supply vector $[20,18,25]$ 
are plotted in Fig.~\ref{fig:ringoutputcomputation}
and exhibited in Table~\ref{table:RingNetworkNewMinimum1}
and 
the outcomes from another initial vector  $[23,19,24]$ are presented in
Table~\ref{table:RingNetworkNewMinimum2}.
\par
Table~\ref{table:ringnetcomparisoncontrolwithoutwith}
and Table~\ref{TailProbabilityRingNetwork}
show the comparison of 
the performance for several lines of the power system {\em without and with control}
as described in the introduction of Section~\ref{sec:examples}.
The tables show that,
for the power line with the highest values of $f_k(p_s)$,
there is a significant reduction
of the mean value $|\theta_{i_k} - \theta_{j_k}|$
and a small change of the standard deviation $\sigma_k$, 
while the probability exiting the safe set of that line decreases a lot.
Table~\ref{TailProbabilityRingNetwork} displays the tail probabilities computed according to Table~\ref{table:ringnetcomparisoncontrolwithoutwith}.
Fig.~\ref{fig:ringnetcontrolwithoutwith} 
shows the effect of control 
on the probability density function of a power flow. 
%
\par
The conclusions of this example are:
\begin{enumerate}
\item 
The nonconvexity property of the control objective function has been additionally illustrated in this example,
See Tables~\ref{table:RingNetworkNewMinimum1} and
\ref{table:RingNetworkNewMinimum2} for two power supply vectors starting from different initial state.
\item 
There is a subset of two power lines 
of which both members have approximately the same values 
for the variable $|\theta_{i_k}-\theta_{j_k}| + r_\epsilon~ \sigma_{i_k,j_k}$. See Fig. \ref{fig:ringoutputcomputation}.
\item
In case the parameter values correspond to a symmetric network, 
then the power lines which connect neighborhoods,
$l_{5-6},\:l_{7-8},\:l_{9-10},\:l_{11-12}$ in 
Fig.~\ref{fig:ringnetwork},
have little power flow.
The fluctuations of the power flows on those lines are not so high. 
For parameter values of an asymmetric network, 
there are relatively small power flows in those power lines. See the length of the blue bars of these lines in Fig.~\ref{fig:ringoutputcomputation}, the sum of the absolute value of the phase-angle differences plus a multiplier of their standard deviation can not be neglected at all. Therefore, it is always necessary to consider the mean and the standard deviation of the power flow together. 
\end{enumerate}
\begin{table}[H]
\centering
  \caption{Example~\ref{ex:ringnettwelvenodes}.
      The minimum value and optimal power vector starting 
      from the power supply vector [20,18,25].}
\vspace{1mm}
\begin{footnotesize}
\begin{tabular} {|c|c|c|c|c|c|c|} 
\hline
\text{Minimum} & $p_1$   & $p_2$   & $p_3$   & $p_4$   & $p_5$    & $p_6$ \\
\hline
1.4455 & 21.7314 & 19.3050 & 23.0071 & 19.9564 & -6 & -10   \\
\hline
               & $p_7$   & $p_8$   & $p_9$   & $p_{10}$ &$p_{11}$&$p_{12}$\\
\hline
               & -8      & -12     & -17     & -13     & -7       & -11 \\
\hline
\end{tabular}
\label{table:RingNetworkNewMinimum1}
\end{footnotesize}
\end{table}

\vspace{-5mm}

\begin{table}[H]
\centering
\caption{Example~\ref{ex:ringnettwelvenodes}.
  The minimum and optimal power vector starting 
  from the power supply vector [23,19,24].}
\vspace{1mm}
\begin{footnotesize}
\begin{tabular} {|c|c|c|c|c|c|c|} 
\hline
\text{Minimum} & $p_1$   & $p_2$   & $p_3$   & $p_4$   & $p_5$    & $p_6$ \\
\hline
1.4455 & 21.6905 & 19.2546 & 23.0549 & 20.0000 & -6 & -10     \\
\hline
               & $p_7$   & $p_8$   & $p_9$   & $p_{10}$& $p_{11}$ & $p_{12}$\\
\hline
               & -8      & -12     & -17     & -13     & -7       & -11  \\
\hline
\end{tabular}
\label{table:RingNetworkNewMinimum2}
\end{footnotesize}
\end{table}
%
%
%
\begin{table}[h]
\centering
  \caption{Example~\ref{ex:ringnettwelvenodes} The ring network.
      Comparison of performance without and with control.}
\label{table:ringnetcomparisoncontrolwithoutwith}
\vspace{1mm}
\begin{tabular} {|r|r|l|r|r|c|} 
\hline
\multicolumn{2}{|l|}{Nodes} 
          & Control & $|\theta_{i_k} - \theta_{j_k}|$ 
                              & $\sigma_k$ & $|\theta_{i_k} - \theta_{j_k}| + r_{\epsilon} ~ \sigma_k$ \\
\hline
$i_k$ 
    & $j_k$ 
          &         &         &            & \\
\hline
 1 & 12 &without& 0.7074 & 0.2753 & 1.5552\\ 
1 & 12&with & 0.5720 & 0.2646 & 1.3870 \\ \hline 
4 & 10 &without& 0.6902 & 0.2517 & 1.4656\\ 
4 & 10 &with& 0.6747 & 0.2502 & 1.4455\\ \hline 
3 & 9 &without& 0.6602 & 0.2492 & 1.4278 \\ 
3 & 9 &with& 0.6755 & 0.2500 & 1.4455 \\ \hline 
2 & 7 &without& 0.5534 & 0.2512 & 1.3272\\ 
2 & 7 &with& 0.5213 & 0.2495 & 1.2896 \\ \hline 
11 & 12 &without& 0.1927 & 0.2510 & 0.9659 \\ 
11 & 12 &with& 0.0831 & 0.2490 & 0.8500 \\ \hline 
\end{tabular}
\end{table}

\begin{table}
\begin{center}
\caption{The exit probabilities 
of the ring network according to Table \ref{table:ringnetcomparisoncontrolwithoutwith},
both without control and with control. See the notations in Remark \ref{re:contingency}.}
\label{TailProbabilityRingNetwork}
\begin{tabular}{|r|r|r|r|r|r|r|}
\hline
Power line & \multicolumn{2}{|l}{Without control} & \multicolumn{3}{|l|}{With control} \\
           & \multicolumn{2}{|l}{probabilities}   & \multicolumn{3}{|l|}{probabilities} \\
\hline
$k$        & $f_{a,k}$ & $ f_{b,k}$  & $f_{a,k}$ & $f_{b,k}$ & $p_{out,k}$ \\
\hline
1-12         & 6.41e-17    & 8.56e-04            &2.79e-16    & 8.01e-05    & 2× 1.90e-04      \\
4-10         & 1.32e-19    & 2.34e-04             & 11.42e-19   & 1.71e-04   & 2× 1.71e-04     \\
3-9         & 1.74e-19    & 1.29e-04             & 1.29e-19    & 1.71e-04    & 2× 1.71e-04      \\
2-7         &1.38e-17    & 2.56e-05   & 2.53e-17   & 1.30e-05    & 2× 1.70e-04     \\
11-12&2.00e-08&1.06e-12&1.15e-09&1.55e-11&2×1.70e-04\\
\hline
\end{tabular}
\end{center}
\end{table}

\par
\par
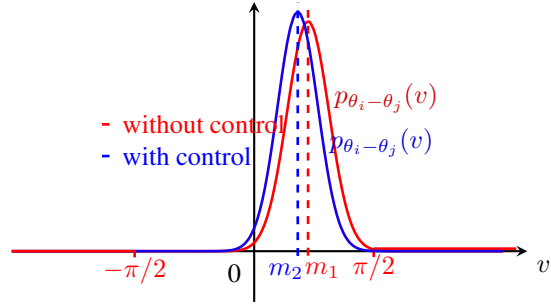
\begin{figure}
\begin{center}
\begin{tikzpicture}
 \message{Expected sensitivty & p-value^^J}
  \def\N{100}
  \def\q{0.7074}
   \def\qn{0.5720}
  \def\Bs{2}
  \def\S{\q}
   \def\Sn{\qn}
   \def\Ss{0.2753}
  \def\Ssn{ 0.2646}   
     \def\xmin{-4.5*\q}
  \def\xmax{3.5*\q}
  \def\ymin{{-0.16/\Bs}}
  \def\ymax{{0.70/\Bs}}
   \begin{axis}[every axis plot post/.append style={
               mark=none,domain=\xmin:{1.32*\xmax},samples=\N,smooth},
               xmin=\xmin, xmax=1.20*\xmax,
               ymin=\ymin, ymax=\ymax,
               axis lines=middle,
               axis line style=thick,
               enlargelimits=upper, 
               ticks=none,
               xlabel=$v$,
               every axis x label/.style={at={(current axis.right of origin)},anchor=north west},
               y=240pt,
              ]
   \addplot[name path=S,line width=0.35mm,red!90!red ] {gauss(x,\S,\Ss)};
         \addplot[red,dashed,line width=0.35mm]
      coordinates {(\q,{-0.02*exp(-\q/\Bs)/\Bs}) (\q, {1.08*gauss(\q,\S,\Ss)})}
       node[below=-4pt,right=-5pt,pos=-0.06] {$m_1$};
    \addplot[red,line width=0.35mm]
      coordinates {(1.57,{-0.04*exp(-1.57/\Bs)/\Bs}) (1.57, {1.38*gauss(1.57,\S,\Ss)})}
       node[below=-5pt,pos=0] {$\pi/2$};
          \addplot[red,line width=0.35mm]
      coordinates {(1.57, {1.5*gauss(1.57,\S,\Ss)}) (1.39*\xmax, {1.5*gauss(1.57,\S,\Ss)})};
  \addplot[name path=S2,line width=0.35mm,blue!90!red ] {gauss(x,\Sn,\Ssn)};
   \addplot[blue,dashed,line width=0.35mm]
      coordinates {(\qn,{-0.02*exp(-\qn/\Bs)/\Bs}) (\qn, {1.08*gauss(\qn,\Sn,\Ssn)})}
       node[below=1pt,left=-6pt,pos=-0.06] {$m_2$};

       \addplot[red,line width=0.35mm]
      coordinates {(-1.57,{-0.009*exp(1.57/\Bs)/\Bs}) (-1.57, {10.5*gauss(-1.57,\S,\Ss)})}
       node[below=-4pt,pos=0] {$-\pi/2$};    
              \addplot[red,line width=0.35mm]
      coordinates {(\xmin,{2.5*gauss(-1.57,\S,\Ss)}) (-1.57, {2.5*gauss(-1.57,\S,\Ss)})};

    \path[name path=xaxis1]
     (0,0) -- (1.08*\xmax,0);
    \addplot[red!60!blue] fill between[of=xaxis1 and S, soft clip={domain=1.57:1.50*\xmax}];
       \path[name path=xaxis2]
     (-1.08*\xmax,0) -- (0,0);
    \addplot[red!60!blue] fill between[of=xaxis2 and S, soft clip={domain=-1.08*\xmax:-1.57}];
    \node[below left=2pt] at (0,0) {0};

    \node[above right=1.3pt,black!20!blue] at ( 0.42*\Bs,0.13) {$p_{\theta_i - \theta_j}(v)$};
     \node[above right=1.3pt,black!20!red] at ( 0.45*\Bs,0.20) {$p_{\theta_i - \theta_j}(v)$};
     \addplot[red,line width=0.35mm]
      coordinates {(-2.0,0.2) (-1.9, 0.2)}
       node[right=4pt,pos=0] {without control};    
       \addplot[blue,line width=0.35mm]
      coordinates {(-2.0,0.15) (-1.9, 0.15)}
       node[right=4pt,pos=0] {with control};

  \end{axis}
  
  \end{tikzpicture}
\end{center}
\caption{Example. Ring network, line 1-12, without and with control.}
\label{fig:ringnetcontrolwithoutwith}
\end{figure}
\end{example}
\par\vspace{1\baselineskip}
\begin{example}\label{ex:manhattengrid}
{\em A Manhattan-grid network}.
\begin{figure}[h]
\centering
\begin{tikzpicture}[tbcircle/.style={circle,minimum size=13pt,inner sep=0pt},
forward tbcircle/.style={tbcircle,fill=blue!70},
backward tbcircle/.style={tbcircle,fill=red!100},
tbsquare/.style={rectangle,rounded corners,minimum width=15pt,minimum height=15pt,fill=orange!50},
tbcarrow/.style={/utils/exec=\tikzset{tbcarrow pars/.cd,#1},
line width=0.1*\pgfkeysvalueof{/tikz/tbcarrow pars/width},
draw=\pgfkeysvalueof{/tikz/tbcarrow pars/draw},
-{Triangle[line width=\pgfkeysvalueof{/tikz/tbcarrow pars/line width},
length=1.1*\pgfkeysvalueof{/tikz/tbcarrow pars/width},
width=1.1*\pgfkeysvalueof{/tikz/tbcarrow pars/width},
fill=\pgfkeysvalueof{/tikz/tbcarrow pars/fill}]},
postaction={draw=\pgfkeysvalueof{/tikz/tbcarrow pars/fill},-,
line width=\pgfkeysvalueof{/tikz/tbcarrow pars/width}-5*\pgfkeysvalueof{/tikz/tbcarrow pars/line width},
shorten <=\pgfkeysvalueof{/tikz/tbcarrow pars/line width},
shorten >=1.25*\pgfkeysvalueof{/tikz/tbcarrow pars/width}-1.5*\pgfkeysvalueof{/tikz/tbcarrow pars/line width}}
},
tbcarrow pars/.cd,
fill/.initial=white,draw/.initial=black,width/.initial=4pt,
line width/.initial=0.4pt,]

\node[forward tbcircle,draw]  at (0,0) {$\textcolor{white}{21}$};
\node[forward tbcircle,draw]  at (1,0) {$\textcolor{white}{22}$};
\node[forward tbcircle,draw]  at (2,0) {$\textcolor{white}{23}$};
\node[forward tbcircle,draw]  at (3,0) {$\textcolor{white}{24}$};
\node[forward tbcircle,draw]  at (4,0) {$\textcolor{white}{25}$};

\node[forward tbcircle,draw]  at (0,1) {$\textcolor{white}{17}$};
\node[forward tbcircle,draw]  at (1,1) {$\textcolor{white}{18}$};
\node[backward tbcircle,draw]  at (2,1) {$\textcolor{white}{4}$};
\node[forward tbcircle,draw]  at (3,1) {$\textcolor{white}{19}$};
\node[forward tbcircle,draw]  at (4,1) {$\textcolor{white}{20}$};

\node[forward tbcircle,draw]  at (0,2) {$\textcolor{white}{14}$};
\node[backward tbcircle,draw]  at (1,2) {$\textcolor{white}{2}$};
\node[forward tbcircle,draw]  at (2,2) {$\textcolor{white}{15}$};
\node[backward tbcircle,draw]  at (3,2) {$\textcolor{white}{3}$};
\node[forward tbcircle,draw]  at (4,2) {$\textcolor{white}{16}$};

\node[forward tbcircle,draw]  at (0,3) {$\textcolor{white}{10}$};
\node[forward tbcircle,draw]  at (1,3) {$\textcolor{white}{11}$};
\node[backward tbcircle,draw]  at (2,3) {$\textcolor{white}{1}$};
\node[forward tbcircle,draw]  at (3,3) {$\textcolor{white}{12}$};
\node[forward tbcircle,draw]  at (4,3) {$\textcolor{white}{13}$};

\node[forward tbcircle,draw]  at (0,4) {$\textcolor{white}{5}$};
\node[forward tbcircle,draw]  at (1,4) {$\textcolor{white}{6}$};
\node[forward tbcircle,draw]  at (2,4) {$\textcolor{white}{7}$};
\node[forward tbcircle,draw]  at (3,4) {$\textcolor{white}{8}$};
\node[forward tbcircle,draw]  at (4,4) {$\textcolor{white}{9}$};

\draw[tbcarrow={fill=black!70}] (0+0.2,0) to (1-0.2,0);
\draw[tbcarrow={fill=black!70}] (1+0.2,0) to (2-0.2,0);
\draw[tbcarrow={fill=black!70}] (2+0.2,0) to (3-0.2,0);
\draw[tbcarrow={fill=black!70}] (3+0.2,0) to (4-0.2,0);

\draw[tbcarrow={fill=black!70}] (0+0.2,1) to (1-0.2,1);
\draw[tbcarrow={fill=black!70}] (2-0.2,1)  to (1+0.2,1);
\draw[tbcarrow={fill=black!70}] (2+0.2,1) to (3-0.2,1);
\draw[tbcarrow={fill=black!70}] (3+0.2,1) to (4-0.2,1);

\draw[tbcarrow={fill=black!70}] (1-0.2,2) to (0+0.2,2);
\draw[tbcarrow={fill=black!70}] (1+0.2,2) to (2-0.2,2);
\draw[tbcarrow={fill=black!70}]  (3-0.2,2) to (2+0.2,2);
\draw[tbcarrow={fill=black!70}] (3+0.2,2) to (4-0.2,2);

\draw[tbcarrow={fill=black!70}] (0+0.2,3) to (1-0.2,3);
\draw[tbcarrow={fill=black!70}] (2-0.2,3) to (1+0.2,3);
\draw[tbcarrow={fill=black!70}] (2+0.2,3) to (3-0.2,3);
\draw[tbcarrow={fill=black!70}] (3+0.2,3) to (4-0.2,3);

\draw[tbcarrow={fill=black!70}] (0+0.2,4) to (1-0.2,4);
\draw[tbcarrow={fill=black!70}] (1+0.2,4) to (2-0.2,4);
\draw[tbcarrow={fill=black!70}] (2+0.2,4) to (3-0.2,4);
\draw[tbcarrow={fill=black!70}] (3+0.2,4) to (4-0.2,4);
\draw[tbcarrow={fill=black!70}] (0,0+0.2) to (0,1-0.2);
\draw[tbcarrow={fill=black!70}] (0,1+0.2) to (0,2-0.2);
\draw[tbcarrow={fill=black!70}] (0,2+0.2) to (0,3-0.2);
\draw[tbcarrow={fill=black!70}] (0,3+0.2) to (0,4-0.2);

\draw[tbcarrow={fill=black!70}] (1,0+0.2) to (1,1-0.2);
\draw[tbcarrow={fill=black!70}] (1,2-0.2) to (1,1+0.2);
\draw[tbcarrow={fill=black!70}] (1,2+0.2) to (1,3-0.2);
\draw[tbcarrow={fill=black!70}] (1,3+0.2) to (1,4-0.2);

\draw[tbcarrow={fill=black!70}] (2,1-0.2) to (2,0+0.2);
\draw[tbcarrow={fill=black!70}] (2,1+0.2) to (2,2-0.2);
\draw[tbcarrow={fill=black!70}] (2,2+0.2) to (2,3-0.2);
\draw[tbcarrow={fill=black!70}] (2,3+0.2) to (2,4-0.2);

\draw[tbcarrow={fill=black!70}] (3,0+0.2) to (3,1-0.2);
\draw[tbcarrow={fill=black!70}] (3,2-0.2) to (3,1+0.2);
\draw[tbcarrow={fill=black!70}] (3,2+0.2) to (3,3-0.2);
\draw[tbcarrow={fill=black!70}] (3,3+0.2) to (3,4-0.2);

\draw[tbcarrow={fill=black!70}] (4,0+0.2) to (4,1-0.2);
\draw[tbcarrow={fill=black!70}] (4,1+0.2) to (4,2-0.2);
\draw[tbcarrow={fill=black!70}] (4,2+0.2) to (4,3-0.2);
\draw[tbcarrow={fill=black!70}] (4,3+0.2) to (4,4-0.2);

\end{tikzpicture}
\caption{A Manhattan-grid like network}
\label{fig:manhattengrid}
\end{figure}
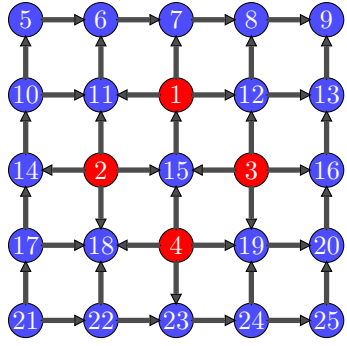

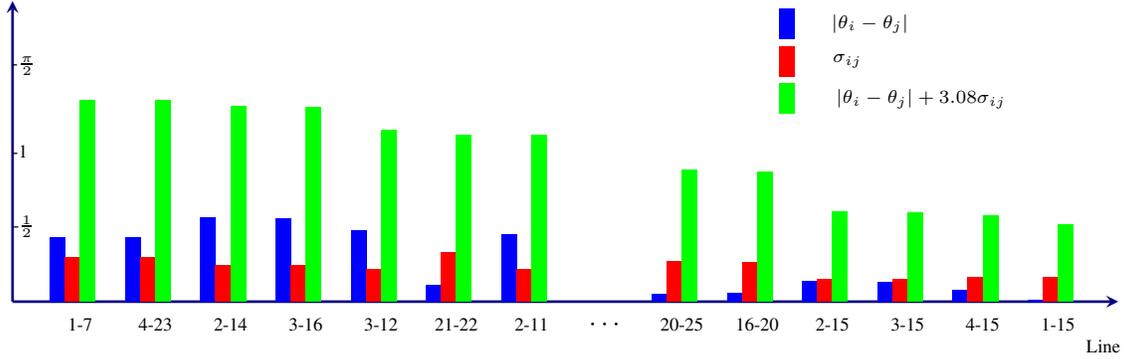
\begin{figure*}[h]
  \centering
  \begin{tikzpicture}
   \draw[->, line width=1pt, blue!50!black, >=stealth] (0,0) -- (14.7,0);
    \draw[->, line width=1pt, blue!50!black, >=stealth] (0,0) -- (0,4);
        
    \fill[blue] (0.5,0) rectangle (0.7,2*0.4285); 
    \fill[red] (0.7,0) rectangle (0.9,2*0.2964); 
    \fill[green] (0.9,0) rectangle (1.1,2*1.3413); 
      \node at (0.9,-0.3) {{\scriptsize 1-7}};
      \fill[blue] (0.5+1,0) rectangle (0.7+1,2*0.4285); 
    \fill[red] (0.7+1,0) rectangle (0.9+1,2*0.2964); 
    \fill[green] (0.9+1,0) rectangle (1.1+1,2*1.3413); 
     \node at (0.9+1,-0.3) {{\scriptsize 4-23}};
    \fill[blue] (0.5+2,0) rectangle (0.7+2,2*0.5614); 
    \fill[red] (0.7+2,0) rectangle (0.9+2,2*0.2400); 
    \fill[green] (0.9+2,0) rectangle (1.1+2,2*1.3004); 
     \node at (0.9+2,-0.3) {{\scriptsize 2-14}};
\fill[blue] (0.5+3,0) rectangle (0.7+3,2*0.5547); 
    \fill[red] (0.7+3,0) rectangle (0.9+3,2*0.2397); 
    \fill[green] (0.9+3,0) rectangle (1.1+3,2*1.2929); 
    \node at (0.9+3,-0.3) {{\scriptsize 3-16}};
 \fill[blue] (0.5+4,0) rectangle (0.7+4,2*0.4742); 
    \fill[red] (0.7+4,0) rectangle (0.9+4,2*0.2169); 
    \fill[green] (0.9+4,0) rectangle (1.1+4,2*1.1423); 
     \node at (0.9+4,-0.3) {{\scriptsize 3-12}};
    \fill[blue] (0.5+5,0) rectangle (0.7+5,2*0.1100); 
    \fill[red] (0.7+5,0) rectangle (0.9+5,2*0.3248); 
    \fill[green] (0.9+5,0) rectangle (1.1+5,2*1.1103); 
    \node at (0.9+5,-0.3) {{\scriptsize 21-22}};
\fill[blue] (0.5+6,0) rectangle (0.7+6,2*0.4453); 
    \fill[red] (0.7+6,0) rectangle (0.9+6,2*0.2159); 
    \fill[green] (0.9+6,0) rectangle (1.1+6,2*1.1102); 
    \node at (0.9+6,-0.3) {{\scriptsize 2-11}};

    \node at (0.9+7,-0.3) {$\cdots$};
%
 \fill[blue] (0.5+8,0) rectangle (0.7+8,2*0.0529); 
    \fill[red] (0.7+8,0) rectangle (0.9+8,2*0.2683); 
    \fill[green] (0.9+8,0) rectangle (1.1+8,2*0.8792); 
    \node at (0.9+8,-0.3) {{\scriptsize 20-25}};
     \fill[blue] (0.5+9,0) rectangle (0.7+9,2*0.0571); 
    \fill[red] (0.7+9,0) rectangle (0.9+9,2*0.2609); 
    \fill[green] (0.9+9,0) rectangle (1.1+9,2*0.8605); 
     \node at (0.9+9,-0.3) {{\scriptsize 16-20}};
     \fill[blue] (0.5+10,0) rectangle (0.7+10,2*0.1337); 
    \fill[red] (0.7+10,0) rectangle (0.9+10,2*0.1507); 
    \fill[green] (0.9+10,0) rectangle (1.1+10,2*0.5978); 
     \node at (0.9+10,-0.3) {{\scriptsize 2-15}};
      \fill[blue] (0.5+1+10,0) rectangle (0.7+1+10,2*0.1291); 
    \fill[red] (0.7+1+10,0) rectangle (0.9+1+10,2*0.1507); 
    \fill[green] (0.9+1+10,0) rectangle (1.1+1+10,2*0.5934); 
      \node at (0.9+1+10,-0.3) {{\scriptsize 3-15}};
    \fill[blue] (0.5+2+10,0) rectangle (0.7+2+10,2*0.0753); 
    \fill[red] (0.7+2+10,0) rectangle (0.9+2+10,2*0.1621); 
    \fill[green] (0.9+2+10,0) rectangle (1.1+2+10,2*0.5744); 
     \node at (0.9+2+10,-0.3) {{\scriptsize 4-15}};
\fill[blue] (0.5+3+10,0) rectangle (0.7+3+10,2*0.0128); 
    \fill[red] (0.7+3+10,0) rectangle (0.9+3+10,2*0.1624); 
    \fill[green] (0.9+3+10,0) rectangle (1.1+3+10,2*0.5131); 
     \node at (0.9+3+10,-0.3) {{\scriptsize 1-15}};

   \fill[blue] (10.5-0.3,5-1.5) rectangle (10.7-0.3,5.4-1.5);   
    \node at (11.7-0.3,5.2-1.5) {{\scriptsize $|\theta_i-\theta_j|$}};
  \fill[red] (10.5-0.3,5-0.5-1.5) rectangle (10.7-0.3,5.4-0.5-1.5);   
    \node at (11.4-0.3,5.2-0.5-1.5) {{\scriptsize $\sigma_{ij}$}};
    \fill[green] (10.5-0.3,5-1-1.5) rectangle (10.7-0.3,5.4-1-1.5);   
    \node at (12.5-0.42,5.2-1-1.5) {{\scriptsize $|\theta_i-\theta_j|+3.08\sigma_{ij}$}};

    \node at (14.5,-0.6) {{\scriptsize Line}};
     \node at (0.1,2) {{\scriptsize -1}};
       \node at (0.15,1) {{\scriptsize -$\frac{1}{2}$}};
          \node at (0.15,1.57*2) {{\scriptsize -$\frac{\pi}{2}$}};

  \end{tikzpicture}
  \caption{The outputs of the Manhattan-grid like network}
  \label{fig:manhattengridcomputations}
\end{figure*}
\par 
The power network is displayed in Fig.~\ref{fig:manhattengrid}.
There are four red-colored nodes 1, 2, 3, and 4 with power supply 
with high inertias and damping coefficients
which provide power to 21 blue-colored nodes 
with power demand with low virtual inertias and low damping coefficients.
One may define for each node with only a power source,
an imaginary neighborhood of nodes with only power demand
which are largely supplied power by that particular power source.
Then the power network is partitioned into several such neighborhoods
and between such neighborhoods there is little power exchange.
The Manhattan-grid power network has such an interpretation.
The computation results for the case of parameter values 
of an asymmetric power network, 
are shown in 
Fig.~\ref{fig:manhattengridcomputations}.
\par
The reader finds in 
Table~\ref{table:manhattangridcomparisoncontrolwithoutwith}
the results for the comparison without and with control and the corresponding exit probability from the safe set in Table \ref{TailProbabilityManhattanNetwork}.
Here we can see the exit probability 
from the safe set of the most vulnerable line $l_{1-7}$ 
does not decrease as much as line $l_{1-12}$ of the ring network. 
That's because due to the network structure, the power grid is already very stable.
%
%
%
\begin{table}[h]
\centering
  \caption{Example~\ref{ex:manhattengrid}. The Manhattan-grid network. 
  Comparison of performance without and with control.}
\label{table:manhattangridcomparisoncontrolwithoutwith}
\vspace{1mm}
\begin{tabular} {|r|r|l|r|r|c|} 
\hline
\multicolumn{2}{|l|}{Nodes} 
          & Control & $|\theta_{i_k} - \theta_{j_k}|$ 
                              & $\sigma_k$ & $|\theta_{i_k} - \theta_{j_k}| + r_\epsilon~ \sigma_k$ \\
\hline
$i_k$ 
    & $j_k$ 
          &         &         &            & \\
\hline
1   & 7   & without & 0.5363  & 0.3149     & 1.5062 \\
1   & 7   & with    & 0.4990  & 0.3128     & 1.4625 \\
\hline
4   & 23  & without & 0.5363  & 0.3128     & 1.4997 \\
4   & 23  & with    & 0.5046  & 0.3110     & 1.4625 \\
\hline
3   & 16  & without & 0.5574  & 0.2589     & 1.3550 \\
3   & 16  & with    & 0.5932  & 0.2601     & 1.3943 \\
\hline
2   & 14  & without & 0.5574  & 0.2590     & 1.3550 \\
2   & 14  & with    & 0.5932  & 0.2600     & 1.3941 \\
\hline
\end{tabular}
\end{table}
\begin{table}
\begin{center}
\caption{The exit probabilities of the Manhattan-grid network according to Table \ref{table:manhattangridcomparisoncontrolwithoutwith},
both without control and with control. See the notations in Remark \ref{re:contingency}.}
\label{TailProbabilityManhattanNetwork}
\begin{tabular}{|r|r|r|r|r|r|r|}
\hline
Power line & \multicolumn{2}{|l}{Without control} & \multicolumn{3}{|l|}{With control} \\
           & \multicolumn{2}{|l}{probabilities}   & \multicolumn{3}{|l|}{probabilities} \\
\hline
$k$        & $f_{a,k}$ & $ f_{b,k}$  & $f_{a,k}$ & $f_{b,k}$ & $p_{out,k}$ \\
\hline
1-7         & 1.11e-11    & 5.10e-04            &1.83e-11    & 3.06e-04    & 2×3.06e-04      \\
4-23         & 8.13e-12    & 4.71e-04             & 1.25e-11  & 3.04e-04   & 2×3.04e-04     \\
3-16         & 1.02-16    & 4.53e-05             & 4.40e-17    & 8.55e-05   & 2×2.36e-04      \\
2-14         &1.04e-16    & 4.56e-05   &4.29e-17  & 8.50e-05    & 2×2.36e-04     \\
\hline
\end{tabular}
\end{center}
\end{table}

\par
The conclusions of this example are:
\begin{enumerate}
\item
For this particular example the values of 
the standard deviations $\sigma_{i,j}$
are relatively high compared with those of $|\theta_i - \theta_j|$.
\item
The reader may observe that for this particular power system,
the power flowing from a node with power supply
is larger than the power flow of any power line 
connected to that node,
and, similarly, for the nodes with power demand.
But no power outage is likely to occur,
due to the tight interconnections of the Manhattan power grid, 
see Fig.~\ref{fig:manhattengridcomputations}.
Thus, this particular power network has a very stable dynamic behavior 
due to the grid structure and the large number of power lines.
\item 
The optimization algorithm used 
is better than a method based 
on a grid of the feasible set of power supply vectors.
This conclusion has been verified by a computation based 
on a grid of power supply vectors and by computation of the control
objective function at each grid point. 
The minimal values of the second method are in general larger 
than those of our proposed algorithm
while the computation time is also larger.
\item
A starting point for a vector of power supplies 
for our proposed algorithm
may be taken as the sum of all nodes with power demands 
in a neighborhood of each node with power supply. 
The computational efficiency can be improved by such an initial choice.
\end{enumerate}
\end{example}
\subsection{Conclusions of All Three Examples}
\begin{enumerate}
\item 
The descending order of the magnitude 
of the values of $f_k(p_s)$ of power lines 
of a stochastic power system,
differ significantly from those of a deterministic power system,
Figs.~\ref{fig:eightnodenetcomputations},
\ref{fig:ringoutputcomputation}, 
\ref{fig:manhattengridcomputations},
where the columns are ordered in the descending order for the variable 
$f_{k}(p_s) = |\theta_{i_k}-\theta_{j_k}|+r_\epsilon~\sigma_{(i_k, j_k)}$. 
The orders are different 
in case of the values of the absolute mean 
$|\theta_{i_k} - \theta_{j_k}|$ and 
of the standard deviation $\sigma_{(i_k, j_k)}$. 
\item 
A byproduct of the computations is that it is directly clear 
which power lines have to be monitored by the power system operators
during the short horizon for an exit from the safe set.
For example, 
the first four power lines with the highest values 
in Fig.~\ref{fig:manhattengridcomputations},
are best monitored.
\item 
The reader sees from the length of the bars 
$f_{k}(p_s) = |\theta_i-\theta_j|+r_\epsilon~\sigma_{ij}$,
Figs.~\ref{fig:eightnodenetcomputations},
\ref{fig:ringoutputcomputation}, 
\ref{fig:manhattengridcomputations},
that there is a subset of power lines 
of which all members have approximately the same values. 
This phenomenon is a property of the optimal solution
for the considered examples. 
\item 
The Manhatten grid network is much stable than the ring network,
due to its network structure. 
\item 
In general, the proposed procedure
is better than a computation method based on 
gridding the feasible set of power supply vectors into equal lengths in terms of efficiency, while their accuracies are comparable.
\item
The Braess paradox also occurs in a stochastic power system.
\item
The power line with the highest exit probability from the safe subset
without control is reduced in probability.
Because the power supply has to equal power demand,
the control input has the effect
that several power lines with exit probabilities
below the value of $\epsilon$,
will with control have a higher probability of exiting
than without control,
though all values are still below the threshold of $\epsilon$.
See line $l_{3- 9}$ in Table \ref{TailProbabilityRingNetwork}, and 
line $l_{3- 16}$, line $l_{2- 14}$ 
in Table \ref{TailProbabilityManhattanNetwork}.
This issue is due to the choice of the control objective function.
\item
The control objective function values of the local minimizers computed by the proposed procedure approach each other, and therefore, 
one can determine the economic power injection among this set of safe options.
\end{enumerate}
\section{Conclusions and Further Research}\label{conclusionfurtherinvestigation}
The emphasis of this paper on 
control of the phase-angle differences of power lines
of a power system,
is regarded as a useful focus for control of power systems.
The power flows through all power lines of the power network
are the main characteristics of a power system as a network.
The concept of a safe set 
and the probability that the phase-angle difference of any power line
will exit the safe set,
are useful concepts for the control of a stochastic power system.
The results for the three examples show clearly that 
the performance as measured by 
the sum of the mean value and of a multiple of the standard deviation,
differs from that of only the mean value or of only the standard deviation.
\par
Needed for control theory of power systems and of stochastic system
is further research 
on how the performance of the power system depends on:
the graph of the power network and
the distribution over the nodes 
of the power supplies
and of the standard deviations of the disturbances.
In addition, further research is needed 
on the performance of a power system with very low inertia in all nodes.
Moreover, as mentioned in the literature review, this research may be extended to the framework of the Security-Constrained Optimal Power Flow methodology.
Especially in a probabilistic way, where the economic cost on power dispatch can also be investigated.
%
\section*{References}
%
\bibliographystyle{abbrv}
\bibliography{conpowerflows}
%
%
{\bf Zhen Wang} was born in China, September 1995. 
He received 
his bachelor diploma in computational mathematics 
from Ningxia University, Yinchuan, China, in 2017.
During 2015--2016, 
he attended the School of Mathematics of Jilin University.
He has received in June 2024 a Ph.D. degree 
from the School of Mathematics of Shandong University, 
in Jinan, Shandong Province, China.
From November 2021 till February 2024, 
he has been a visting Ph.D. student 
at the Department of Applied Mathematics of Delft University of Technology
in Delft, The Netherlands.
\par
His research interests include
computational mathematics,
control theory, optimization, and stability analysis of power systems.
%
%
\par
{\bf Kaihua Xi }
was awarded a Ph.D. degree 
by the Delft Institute of Applied Mathematics,
Delft University of Technology, Delft, The Netherlands, in 2018.
Currently,
he is an associate professor at 
the School of Mathematics of Shandong University, 
Jinan, Shandong University, China.
\par
His research interests include 
control, optimization and stability analysis of power systems,
numerical algorithms of differential equations, and 
data assimilation of oil reservoir simulation.
%
\par
{\bf Aijie Cheng }is Full Professor 
at the School of Mathematics, Shandong University,
in Jinan, Shandong Province, China.
\par
His research interests include
numerical approximation of partial differential equations,
and computational problems of science and engineering.
%
\par
{\bf Hai Xiang Lin} received a Ph.D. degree 
from Delft University of Technology. 
He is an associate professor 
at the Department of Mathematical Physics, 
Delft Institute of Applied Mathematics, 
Delft University of Technology. 
He is also a professor in Data Analytics for Environmental Modelling 
at the Institute of Environmental Sciences of Leiden University, 
on behalf of the R. Timman Foundation. 
\par
His research interests include high-performance computing,
atmospheric and environmental modelling, 
data assimilation, big data, machine learning, simulation, and 
control of power systems. 
He has published more than 150 peer-reviewed Journal and 
international conference papers. 
\par
Dr. Lin is an Associate Editor of 
the journal Algorithms and Computational Technology, and 
an editorial board member of Big Data Mining and Analytics.
\par
%
{\bf J. H. van Schuppen} 
(Life Member, IEEE).
Jan H. van Schuppen 
was born in Veenendaal, The Netherlands, October 1947.
He was awarded an engineering diploma by
Delft University of Technology in 1970 and
a Ph.D. diploma in Electrical Engineering and Computer Science
by the University of California at Berkeley, CA, USA in 1973.
\par
As Professor Emeritus he is affiliated 
with the Department of Applied Mathematics 
of Delft University of Technology
in Delft, The Netherlands 
since his retirement in October 2012 
from the Centre of Mathematics and Computer Science (CWI).
Since then he is active as a researcher with his consulting company
Van Schuppen Control Research in Amsterdam, The Netherlands.
His research interests include
stochastic control and control of power systems.
\par
Van Schuppen is a member of the
IEEE Societies of Control Systems, Computers, and Information Theory,
and of the Society for Industrial and Applied Mathematics (SIAM).
He was an Associate-Editor-at-large of the
{\em IEEE Transactions on Automatic Control}, 
a co-editor-in-Chief of the journal
{\em Mathematics of Control, Signals, and Systems}, and
was a Department Editor of {\em J. of Discrete-Event Dynamic Systems}.
%
%
\appendix
The following appendices are included in the paper
only for the review process.
If the paper is accepted for publication
then these appendices will be removed from the paper
unless the Editor-in-Chief or the Associate Editor
requests differently.
\subsection{Theory and Proofs of Section~\ref{sec:intropowersystem} }
\label{appendix:laplacianmatrixcompletenetwork}
\subsubsection{The Laplacian Matrix for a Complete Power Network}\label{Jacobian}
A complete network is defined as a network 
where every pair of distinct vertices is connected by a unique edge. 
In this case, the product $BWF(\theta_s)$ in equation (\ref{eq:jacobianmatrix}) satisfies,
\begin{align*}
    (BWF(\theta_s))_{i,j}
    & = - L_{i,j} \cos (\theta_{s,i} - \theta_{s,j}), \\
    &   \forall ~ (i, ~ j) \in \mathcal{E} ~ 
        \mbox{with} ~ i \neq j; \\
    (BWF(\theta_s))_{k,k}
    & = - \sum_{j=1, ~ j \neq k}^{n_{\mathcal{V}}} ~ 
        (BWF(\theta_s))_{k,j}, ~
        \forall ~ k \in \mathcal{E}.
\end{align*}
\subsubsection{Standard Deviation Strictly Positive}%
\label{apsubsec:stdevstrictlypositive}
\begin{proposition}\label{prop:standarddeviationpositive}
\begin{itemize}
\item[(a)]
Consider the deterministic linear system
from an input $u$ to an out\-put $y$,
\begin{align*}
        dx(t)/dt
    & = J(\theta_s, ~ 0) x(t) + K u(t), ~ x(0) = x_0, \\
        y(t)
    & = C x(t).
\end{align*}
Recall Assumption~\ref{assumption:existencestablesynchronousstate}.
Assume that the diagonal matrix $K$
has strictly positive diagonal elements, 
hence, for all $ k \in \mathcal{\mathcal{V}}$, 
$K_{k,k} > 0$.
A weaker condition is that $(J_r,~K_r)$ is a controllable  pair.
Then this system is a controllable system.
\item[(b)]
	For any power line $k \in \mathcal{\mathcal{E}}$ and any
fixed power supply vector $p_s \in P^+$,
the standard deviation $\sigma_k(p_s)$ is strictly positive,
$\sigma_k(p_s) = (Q_{y,(k,k))})^{1/2} > 0$.
\end{itemize}
\end{proposition}
\par\vspace{1\baselineskip}\par\noindent
{\bf Proof}
(a)
(a.1) Consider the following Lyapunov equation
for the variance of the phase-angle differences over the power lines,
\begin{align*}
    0
    & = J(\theta_s, ~0) Q_x + Q_x J(\theta_s, ~ 0)^{\top} + K K^{\top}, \\
        Q_y
    & = C Q_x C^{\top}, \\
        J(\theta_s, ~ 0 ) 
    & = \begin{bmatrix}
            0                      & I_{n_\mathcal{V}} \\
            M^{-1} B W F(\theta_s) & - M^{-1} D
          \end{bmatrix}.
\end{align*}
Because the Laplacian matrix $B W F(\theta_s)$
has one zero eigenvalue,
it is required to carry out a transformation 
and to truncate the system matrix $J(\theta_s, ~ 0)$
to eliminate the zero eigenvalue,
as described in \cite{WANG2023110884}.
Denote the combined transformation and truncation matrix
as follows and then note the transformation 
of the Lyapunov equation,
\begin{align*}
        n_{x_r}
    & = n_x - 1, ~~
          L \in \mathbb{R}^{n_{x_r} \times n_x}, \\
        J_r
    & = L J(\theta_s, ~ 0) L^{\top}, ~
          K_r =  L K,  ~
          C_r = C L^{\top}, \\
        Q_{x_r} 
    & = L Q_x L^{\top}; \\
        0
    & = J_r Q_{x_r} + Q_{x_r} J_r^{\top} + K_r K_r^{\top}; ~~
          \spec(J_r) \subset \mathbb{C}^-; \\
        Q_y
    & = C Q_x C^{\top} 
          = C L^{\top} L Q_{x_r} L^{\top} L C^{\top}
          = C_r Q_{x_r} C_r^{\top}.
\end{align*}
\par
(a.2) 
It will be argued that the tuple
$(J_r, ~ K_r)$ is a controllable pair.
\par
Define the undirected state graph 
$G_X = (V_X, ~ E_X)$ 
with node set $V_X = \mathbb{Z}_{n_x}$ and edge set 
$E_X$
 by
$(i, ~ j) \in E_X$ 
if $J(\theta_s, ~ 0)_{i,j} \neq 0$.
Recall that $n_x = 2 n_\mathcal{V}$.
Define the input-to-state graph as the undirected graph
$G_{UX} = (V_{UX}, ~ E_{UX})$
with $V_{UX} = \mathbb{Z}_{n_\mathcal{V}} \times \mathbb{Z}_{n_\mathcal{V}}$ and
$(k, ~ i) \in E_{UX}$ if
$K_{i,k} \neq 0$.
\par
Recall \ref{sec:powersystem},
that the graph of the power network is a connected set.
It then follows from this assumption 
that, for any tuple $(i, ~ j) \in E_X$,
there exists a path from $i$ to $j$.
\par
Recall from the assumption of part (a)
that for all $k \in \mathbb{Z}_{n_\mathcal{V}}$, $K_{k,k} > 0$.
It then follows that,
for any $i+n_\mathcal{V} \in \mathbb{Z}_{n_x}$,
there exists a $k = i \in \mathbb{Z}_{n_\mathcal{V}}$ 
and a path from $u_k$ to $x_i$.
Because the differential equation of the power system
includes the equation that, for all $i \in \mathbb{Z}_{n_\mathcal{V}}$,
$d\theta_i(t)/dt = \omega_i(t)$,
there exists for each such $i$ an integer $k=i \in \mathbb{Z}_{n_\mathcal{V}}$
and a path from $u_k$ to $x_i$ via $x_{i+n_\mathcal{V}} = \omega_i$.
\par
(a.3)
From Theorem~\cite[Thm. 6.4.2]{murota:2000}
follows that the time-invariant linear system is controllable
if and only if
(1) there exists a set of mutually disjoint cycles and stems
such that all nodes of $V_X$ are covered; and
(2) for every node of a state there exists a path
from an input to that state.
This is also proven in \cite{glover:silverman:1976}.
That condition (1) holds follows from the
assumption that the power network is a connected set.
It was proven above that condition (2) holds.
Thus the tuple $(J(\theta_s, ~ 0), ~ K)$ is a controllable pair.
The same conclusion holds for the tuple $(J_r, ~ K_r)$
because the deletion of a component with a zero eigenvalue
for all concerned matrices
does not affect the controlability property,
except that it now applies to the truncated linear system.
\par
(b)
It follows from 
$\spec(J_r) \subset \mathbb{C}^-$,
$(J_r, ~ K_r)$ a controllable pair, and
\cite[Thm. 3.28]{trentelman:stoorvogel:hautus:2001}
for the direction that (1) and (2) imply (3),
that $Q_{x_r} \succ 0$,
hence is strictly positive-definite and symmetric.
Recall the notation, $C(k)$ represent the $k$-th row of the matrix $C$,
the first part of which equals the $k$-th column of the incidence matrix $B$.
It then follows from the definition of the matrix $B$
that there are two nonzero elements of $C(k)$.
This result and $Q_{x_r} \succ 0$ imply that,
   $ \sigma_k 
       = (Q_{y,(k,k))})^{1/2}
        = ( C(k) Q_{x_r} C(k)^{\top})^{1/2} > 0
        , ~
        \forall ~ k \in \mathbb{Z}_{n_\mathcal{E}}.$
%
\subsubsection{Formulas of $A_1, b_1,b_2$ of Def.~\ref{def:domain}}\label{ExpressionOfMatricesA1b1b2}
\begin{align*}
  i,j&\in \mathbb{Z}_{n^+-1},
      b_1,\:b_2\in \mathbb{R}^{n^+-1},\\
                                      b_1\left(i\right)
    & = p_{sum}^--p_{n^+}^{+,max}-\cdots-p_{i+1}^{+,max}, ~
                                                                 b_2 \left(i\right) 
     = p_{i}^{+,max},~\\
             A_1\left(i,j\right)
    & =
      \begin{cases}
        1 \quad \text{if} ~ i \leq j\\
        0 \quad \text{if} ~ i> j
      \end{cases}, ~
      A_1\in \mathbb{R}^{(n^+-1)\times (n^+-1)}.
         \end{align*}
\subsubsection{Proof of Proposition~\ref{prop:probabilitybound}}
\label{proofofpropprobabilitybound}
%
\vspace{-1mm}
{\bf Proof}
(a)
Denote respectively the probability density function
and the probability distribution function of $G(0, ~ 1)$
by $p_G: \mathbb{R} \rightarrow \mathbb{R}_+$ and 
$f_G: \mathbb{R} \rightarrow \mathbb{R}_+$.
Because $p_G$ is symmetric when mirrored at $0$,
thus for all $v \in \mathbb{R}$, $p_G(v) = p_G(-v)$,
it follows that, for all $r \in \mathbb{R}$, $1 - f_G(r) = f_G(-r)$. 
\par
For all $k \in \mathbb{Z}_{n_\mathcal{E}}$, and for all $p_s \in P^+$,
the probability that the power flow
of the $k$-th power line goes into the unstable region 
according to the invariant probability distribution
of the power line flow,
is equal to,
\begin{align*}
         \lefteqn{
           p_{out,k}(p_s)
         } \\
     & = P
           \left(
           \left\{
             \omega \in \Omega |~
               \theta_{i_k}(\omega, t) - \theta_{j_k}(\omega, t) 
               < - \pi/2
           \right\}
           \right)  + \\
     &   + P
           \left(
           \left\{
             \omega \in \Omega |~
               \theta_{i_k}(\omega, t) - \theta_{j_k}(\omega, t)
               > + \pi/2
           \right\}
           \right) \\
     & = P
           \left(
           \left\{
           \begin{array}{l}
             \omega \in \Omega |~
               \frac{ 
                      ( \theta_{i_k}(\omega, t) - \theta_{j_k}(\omega, t) )
                      - m_k(p_s)
                    }{
                      \sigma_k(p_s)
                    } 
           \end{array}
           \right.
           \right. \\
     &   \left.
           \left.
           \begin{array}{l}
             ~~~~~~ < \frac{ 
                      - \pi/2
                      - m_k(p_s) |}{
                      \sigma_k(p_s)
                    }
           \end{array}
           \right\}
           \right) + P \ldots  \\
             & = f_{G(0, 1)}(r_{a,k}(p_s)) 
           + [ 1 - f_{G(0,1)}(r_{b,k}(p_s)) ]  \\
     & = f_{G(0, 1)}(r_{a,k}(p_s)) + f_{G(0,1)}(- r_{b,k}(p_s)), \\
                     r_{a,k}(p_s)
     & = \frac{- \pi/2 - m_k(p_s)}{\sigma_k(p_s)}, \\
         \end{align*}
\begin{align*}
        - r_{b,k}(p_s) 
     & = \frac{- \pi/2 + m_k(p_s)}{\sigma_k(p_s)},  \\
               r_{a,b,k}(p_s) 
     & = \frac{ - \pi/2 + | m_k(p_s) |}{\sigma_k(p_s)},  \\
           &   m_k(p_s) \in (-\pi/2, ~ +\pi/2) ~
          \Rightarrow \\
     &   r_{a,k}(p_s) < 0, ~ - r_{b,k}(p_s) < 0.
\end{align*}
Because the probability distribution function
$f_G$ is strictly monotone,
the definition of $r_{a,b,k}$ and the monotonicity of $f_G$
{imply the inequality} of $f_G(r_{a,b,k}(p_s))$.
\par
(b)
By definition $f_{G(0,1)}(r_{\epsilon/2}) \leq \epsilon/2$,
for example, 
if $\epsilon/2 = 10^{-3}$ then $r_{\epsilon/2} = - 3.08$.
Assume that $r_{a,b,k}(p_s) \leq r_{\epsilon/2}$.
Then,
\begin{align*}
          r_{a,k}(p_s)
    & \leq \max \{ r_{a,k}(p_s), ~ - r_{b,k}(p_s) \}  \\
    & = r_{a,b,k}(p_s) \leq r_{\epsilon/2}, ~
             \mbox{and similarly,} \\
          - r_{b,k}(p_s)
    & \leq \max \{ r_{a,k}(p_s), ~ - r_{b,k}(p_s) \} \\
    & = r_{a,b,k}(p_s) \leq r_{\epsilon/2}, ~
          \Rightarrow \\
          p_{out,k}(p_s)
    & = f_{G(0,1)}(r_{a,k}(p_s)) + f_{G(0,1)}(- r_{b,k}(p_s))\\
    & \leq \epsilon.
\end{align*}
\par
The existence of $ p_s^*\in P^+$ follows from 
a result of \cite{zhenwang:reportthree:2023}
and the lower bound from the implications,
\begin{align}
             f_d^* 
    & < \pi/2 ~
             \Rightarrow ~
             \forall ~ k \in \mathbb{Z}_{n_\mathcal{E}}, ~ \nonumber\\
           r_{a,b,k}(p_s^*)
    & = \frac{- \pi/2 + |m_k(p_s^*)|}{\sigma_k(p_s^*)} 
        \leq 
          \frac{- \pi/2 + f_d^*}{\sigma_k(p_s^*)}  + r_{\epsilon}
          \leq r_{\epsilon} \label{importantinequality}\\
        p_{out,k}(p_s^*)   \nonumber
    & \leq 2 ~ \epsilon.
\end{align}
Note that if $f_d^* < \pi/2$, let $ \frac{- \pi/2 + f_d^*}{\sigma_k(p_s^*)}  + r_{\epsilon}=r_{\epsilon_{new}}$, then 
it satisfies that $r_{\epsilon_{new}}<r_{\epsilon}$. We can further get that $p_{out,k}(p_s^*)  \leq 2 ~ \epsilon_{new}$. 
Furthernmore, in the first inquality of (\ref{importantinequality}), the `=' is satisfied if and only only if line $k$ is the most vulnerable line according to our criterion.
%
\subsection{Theory of Section \ref{sec:problem}}
\label{appendix:theoryofsectionthree}
\subsubsection{Computation of the matrices $A$ and $b$
of $f_{as,k}(p_s)$ in Def.~\ref{def:controlobjectivefunction}}
\label{computingAb}
By Theorem \ref{criterion} and \cite[Prop. 1]{fazlyab2017optimal}, 
the sin values of phase-angle differences 
satisfy $sin\:(B^{\top}\theta^*)=B^{\top}(BWB^\top)^{\dag}\:p_{sp}$ and 
then we transform the matrix $(BWB^\top)^{\dag}$ 
into a diagonal matrix: 
$\hat{U}^{\top}(BWB^\top)\hat{U}
=\Lambda\Rightarrow (BWB^\top)^{\dagger}
=\hat{U}\Lambda^{\dagger}\hat{U}^{\top}$.\\
Define $U=\hat{U}^{\top}$, 
which makes $(BWB^\top)^{\dagger}=U^\top \Lambda^{\dagger}U$.
Recall the notation that $U_i$ represents the i-th column of the matrix $U$.
%
%
 \text{Define}
        ~$p_s\in \mathbb{R}^{(n^+-1)\times 1},p_d\in \mathbb{R}^{(n_\mathcal{V}-n^+)\times 1}$, $  A\in \mathcal{R}^{n_\mathcal{E}\times (n^+-1)},  b\in \mathcal{R}^{n_\mathcal{E}\times 1}$ by the following formulas:
\begin{equation*}
\begin{aligned}
    p_s 
    & = \begin{bmatrix}
            p_1, & p_2, & \cdots & p_{n^+-1}
          \end{bmatrix}^\top, \\
    -p_d
    & = \begin{bmatrix}
              -p_{n^++1}^{-}, & -p_{n^++2}^{-}, & \cdots &
              -p_{n}^{-}
            \end{bmatrix}^\top;\\
    &  = B^{\top} U^{\top} \Lambda^{\dagger}
       \begin{bmatrix}
         U_1 - U_{n^+}, & \cdots & U_{n^+-1}-U_{n^+}
       \end{bmatrix}
       p_s+\\
    &  \quad 
       + B^{\top}U^{\top}\Lambda^{\dagger}
       \begin{bmatrix}
         U_{n^++1}-U_{n^+}, &\cdots & U_{n}-U_{n^+}
       \end{bmatrix}p_d\\
     A
    &  = B^{\top} U^{\top} \Lambda^{\dagger}
         \begin{bmatrix}
           U_{1}-U_{n^+}, & \cdots & U_{n^+-1}-U_{n^+}
         \end{bmatrix}, \\
    b
    &  = B^{\top} U^{\top}\Lambda^{\dagger}
       \begin{bmatrix}
         U_{n^++1}-U_{n^+}, & \cdots & U_{n}-U_{n^+}
       \end{bmatrix}p_d;\\
       \end{aligned}
       \end{equation*}
       then, it follows that,
       \begin{equation}\label{table:formulasab}
       \begin{aligned}
    \lefteqn{
      B^{\top}(BWB^\top)^{\dagger}p_{sp}
       = B^{\top} U^\top \Lambda^{\dagger} U p_{sp}
    } \\
    &  = B^{\top} U^{\top} \Lambda^{\dagger} U
         \left[
           p_{1},  \cdots  ,p_{n^+-1}, p_{sum}^- 
           -p_1-\cdots-p_{n^+-1},\right.\\
           &\qquad\qquad\qquad\qquad\left.
             -p_{n^++1}^{-,max},  \cdots , -p_n^{-,max}
         \right]^\top\\
    &  = B^{\top} U^{\top} \Lambda^{\dagger}
         \left(
           p_1
           \left(
             U_1-U_{n^+}
           \right) + \cdots + p_{n^+-1}
           \left(
             U_{n^+-1}-U_{n^+}
           \right)
         \right.\\
    &  \quad
       \left.  
         + p_{sum}^-U_{n^+}-p_{n^++1}^{-,max} U_{n^++1}
           - \cdots - p_{n}^{-,max} U_n
       \right)\\
    &  = B^{\top}U^{\top} \Lambda^{\dagger}
         \left(
           p_1(U_1 - U_{n^+}) + \cdots
           + p_{n^+-1} (U_{n^+-1} - U_{n^+})
         \right.\\
    &  \qquad
       \left. 
         + p_{n^++1}^{-,max}
         \left(
           U_{n^+}-U_{n^++1}
         \right) + \cdots+p_{n}^{-,max}
         \left(
           U_{n^+}-U_{n}
         \right)
       \right)\\
     &  = A ~ p_s + b.
\end{aligned}
\end{equation}
%
%

%
%
\end{document}


\title{{\LARGE\sf 
Supplement of\\
`Control of the Power Flows\\
\vspace{2mm}
of a Stochastic Power System'}}
\author{Zhen Wang
\thanks{
Zhen Wang is with the School of Mathematics, Shandong University, 
Jinan, 250100, Shandong Province, China, and visited the Dept. of Applied Mathematics, Delft University of Technology, 
Delft, The Netherlands from November 2021 to February 2024 (email: wangzhen17@mail.sdu.edu.cn, wangzhen17.sdu@vip.163.com). Zhen Wang is grateful to the China Scholarship Council (No. 202106220104)
for financial support for his stay at Delft University of Technology
in Delft, The Netherlands.
}
}
\vspace{-3mm}
%
%
%
%

\maketitle
\section{Introduction}
The purpose for this supplement is to provide the direct outcomes of the proposed procedure for the three examples investigated in the main body of the paper. The parameters of specific power systems can be used for computations and are listed so as to allow reproduction. From these results , the readers may learn the proposed procedure. Furthermore, the figures in the manuscript are plotted based on the particular tables listed in this supplement. Fig. 4 is based on Table~\ref{OriginalPhaseangledifferencesandotheroutput}; 
 Fig. 6 is based on Table~\ref{RingNetworkNewPhaseAngleDifference1}; and Fig. 9 is based on Table~\ref{ManhattanGridPhaseAngleDifferences3}.
\paragraph{Paper organization} Section~\ref{sec:table1} deals with the particular eight-node academic example. Its original parameters and original outcomes are listed. Afterward, some of the parameters of this particular power network or the network structure are changed in order to investigate their influence on the outcomes.  
\par 
Section~\ref{sec:table2} deals with a twelve-node ring network, its network structure is symmetric, we first make every kind of parameter symmetric to investigate the outcomes. Secondly, every kind of parameter is changed in order to see its influence on the outcomes.
\par
 Section~\ref{sec:table3} deals with a Manhattan-grid network. Since it is a symmetric network structure, we first make every kind of parameter symmetric to see what will the outcomes look like. Next, we make the power demand vector asymmetric and similarly the disturbance vector, to compare the results.  Since this network is large, it will cost more time to compute the minimizer than small networks, hence we do not investigate the influence on the outcomes by changing the network parameters of this power system. Additionally, Sections~\ref{sec:table2} and \ref{sec:table3} also compare the computation results of a power system with and without applying the proposed procedure.

\newpage
\section{Tables of the Particular Eight-Node Academic Example}\label{sec:table1}
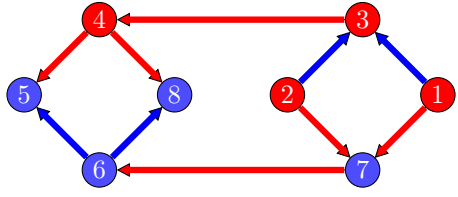
\begin{figure}[h]
\centering
\begin{tikzpicture}[tbcircle/.style={circle,minimum size=13pt,inner sep=0pt},
forward tbcircle/.style={tbcircle,fill=blue!70},
backward tbcircle/.style={tbcircle,fill=red!100},
tbsquare/.style={rectangle,rounded corners,minimum width=25pt,minimum height=25pt,fill=orange!50},
tbcarrow/.style={/utils/exec=\tikzset{tbcarrow pars/.cd,#1},
line width=0.1*\pgfkeysvalueof{/tikz/tbcarrow pars/width},
draw=\pgfkeysvalueof{/tikz/tbcarrow pars/draw},
-{Triangle[line width=\pgfkeysvalueof{/tikz/tbcarrow pars/line width},
length=1.25*\pgfkeysvalueof{/tikz/tbcarrow pars/width},
width=1.3*\pgfkeysvalueof{/tikz/tbcarrow pars/width},
fill=\pgfkeysvalueof{/tikz/tbcarrow pars/fill}]},
postaction={draw=\pgfkeysvalueof{/tikz/tbcarrow pars/fill},-,
line width=\pgfkeysvalueof{/tikz/tbcarrow pars/width}-4*\pgfkeysvalueof{/tikz/tbcarrow pars/line width},
shorten <=\pgfkeysvalueof{/tikz/tbcarrow pars/line width},
shorten >=1.25*\pgfkeysvalueof{/tikz/tbcarrow pars/width}-1.5*\pgfkeysvalueof{/tikz/tbcarrow pars/line width}}
},
tbcarrow pars/.cd,
fill/.initial=white,draw/.initial=black,width/.initial=4pt,
line width/.initial=0.4pt,]\label{kkkk}

\node[backward tbcircle,draw]  at (1,0) {$\textcolor{white}{2}$};
\node[backward tbcircle,draw]  at (3,0) {$\textcolor{white}{1}$};
\node[backward tbcircle,draw]  at (2,1) {$\textcolor{white}{3}$};
\node[forward tbcircle,draw]  at (2,-1) {$\textcolor{white}{7}$};
\node[forward tbcircle,draw]  at (-0.5,0) {$\textcolor{white}{8}$};
\node[forward tbcircle,draw]  at (-2.5,0) {$\textcolor{white}{5}$};
\node[backward tbcircle,draw]  at (-1.5,1) {$\textcolor{white}{4}$};
\node[forward tbcircle,draw]  at (-1.5,-1) {$\textcolor{white}{6}$};

\draw[tbcarrow={fill=blue}] (1+0.16,0+0.16) to (2-0.16,1-0.16);
 
\draw[tbcarrow={fill=blue}] (3-0.16,0+0.16) to (2+0.16,1-0.16);
 
\draw[tbcarrow={fill=red}] (1+0.16,0-0.16) to (2-0.16,-1+0.16) ;

\draw[tbcarrow={fill=red}] (3-0.16,0-0.16) to (2+0.16,-1+0.16) ;

\draw[tbcarrow={fill=red}] (2-0.23,1) to  (-1.5+0.23,1);

\draw[tbcarrow={fill=red}] (2-0.23,-1) to (-1.5+0.23,-1);

\draw[tbcarrow={fill=red}] (-1.5-0.16,1-0.16) to (-2.5+0.16,0+0.16);

\draw[tbcarrow={fill=red}](-1.5+0.16,1-0.16) to (-0.5-0.16,0+0.16);

\draw[tbcarrow={fill=blue}] (-1.5-0.16,-1+0.16) to (-2.5+0.16,0-0.16);

\draw[tbcarrow={fill=blue}] (-1.5+0.16,-1+0.16) to (-0.5-0.16,0-0.16);

 
 \end{tikzpicture}

\caption{The particular eight-node academic example}
\label{fig:figureeight}
\end{figure}

The picture of the particular eight-node academic network is displayed in Fig. ~\ref{fig:figureeight} where the nodes $1,2,3,4$ have only power supply and the other nodes have only power demand. Tables~\ref{Thenetworkparameters} and \ref{Informationoflines} show the network parameters of this particular eight-node academic power system, to make these parameters realistic, the inertias and the damping coefficients of the power supply nodes are set to be larger than those of the power demand nodes. Table \ref{Powersupplyandpowerdemand} displays the maximum power supply and the predicted power demand of each node, one can check that these specified values indeed satisfy our Definition 2.3 in the main body of the paper. 
\par Table~\ref{OriginalTheminimumvalueandoptimalpowervector} shows the minimum of the control objective function and the optimal power supply vector computed by our proposed procedure. Table~\ref{OriginalPhaseangledifferencesandotheroutput} shows the absolute value of mean of the phase-angle difference of each power line: $|\theta_i-\theta_j|$, the standard deviation of the phase-angle difference $\sigma_{ij}$, the component of our object function $|\theta_i-\theta_j|+3.08\:\sigma_{ij}$ and the power flow $l_{ij}\:sin(|\theta_i-\theta_j|+3.08\:\sigma_{ij})$ of each power line and the rows of this table are sorted by $|\theta_i-\theta_j|+3.08\:\sigma_{ij}$ in a descending order. What's more, this computation result is by applying the initial point $[12, 14, 13]$.  In this example, the parameter $r_\epsilon$ is chosen as $3.08$, which leads to the probability that the phase-angle differences of all power lines exceed the safety subset is less than $0.2\%$.  As comparison groups, the other two results are computed by our procedure through the initial point $[12,12,12],~[11, 11, 14]$, separately and the results are displayed in Tables~ \ref{OriginalTheminimumvalueandoptimalpowervectornewstart1} and \ref{OriginalPhaseangledifferencesandotheroutputnewstart1} and Tables~ \ref{OriginalTheminimumvalueandoptimalpowervectornewstart2} and \ref{OriginalPhaseangledifferencesandotheroutputnewstart2}. With the comparisons, the nonconvexity of the control objective function has been verified, which means that starting from different initial points, one can find different local minimizers. The computation has been additionally evaluated by discretizing the domain of the power supply vectors for each dimension into a grid with 150 steps. The minimum value computed by this grid method is $1.3529$ which is smaller or equal than the values computed by our proposed procedure from these three initial points. The results of this gridding method are shown in Tables \ref{OriginalTheminimumvalueandoptimalpowervectorgrid} and \ref{OriginalPhaseangledifferencesandotheroutputgrid}. 
\par
By increasing the strength or intensity of the disturbances to 1.2 times before, the computational results are shown in Tables \ref{IncreaseTheStrengthofRheDisturbancetobe1.2timesbeforetheminimumvalueandoptimalpowervector1}, \ref{IncreaseTheStrengthofTheDisturbancetobe1.2timesbeforephaseangledifferencesandotheroutput1}, 
Tables \ref{IncreaseTheStrengthofRheDisturbancetobe1.2timesbeforetheminimumvalueandoptimalpowervector2}, \ref{IncreaseTheStrengthofTheDisturbancetobe1.2timesbeforephaseangledifferencesandotheroutput2}, and
Tables \ref{IncreaseTheStrengthofRheDisturbancetobe1.2timesbeforetheminimumvalueandoptimalpowervector3}, \ref{IncreaseTheStrengthofTheDisturbancetobe1.2timesbeforephaseangledifferencesandotheroutput3}, starting from $[12,14,13],[12,12,12],[11,11,14]$ seperately. 
From these tables, one can see the values of the minima have increased and are closer to $\pi/2$ now. Additionally, one can see the minimizers in Table \ref{IncreaseTheStrengthofRheDisturbancetobe1.2timesbeforetheminimumvalueandoptimalpowervector1} and Table \ref{IncreaseTheStrengthofTheDisturbancetobe1.2timesbeforephaseangledifferencesandotheroutput2} differ a little while the minimizer in Table \ref{IncreaseTheStrengthofTheDisturbancetobe1.2timesbeforephaseangledifferencesandotheroutput3} differs a lot with these two vectors of minimizers.
\par
Tables~\ref{AddingLine24theminimumvalueandoptimalpowervector} and \ref{AfteraddingLine24thephaseangledifferencesandotheroutput} show the outcomes after adding line $2-4$ while Tables \ref{DoublingLine34theminimumvalueandoptimalpowervector} and \ref{DoublingLine34thephaseangledifferencesandotheroutput}  present the consequences after doubling the capacity of line $3-4$. The results of the comparison groups of these two cases are shown in Tables \ref{AddingLine24theminimumvalueandoptimalpowervectornewstart}, \ref{AfteraddingLine24thephaseangledifferencesandotheroutputnewstart}, and Tables \ref{Doublingline34theminimumvalueandoptimalpowervectornewstart}, \ref{Doublingline34phaseangledifferencesandotheroutputnewstart}, separately. After adding a new line or doubling the capacity of one line, one expects the power system to become more stable than before, but in fact it becomes worse. See the comparisons of Table~\ref{AddingLine24theminimumvalueandoptimalpowervector} with Table~\ref{OriginalPhaseangledifferencesandotheroutput} and Table \ref{DoublingLine34theminimumvalueandoptimalpowervector} with Table~\ref{OriginalPhaseangledifferencesandotheroutput}. 
\par
\par 
Furthermore, readers can also conclude that the minimum values are very close to each other even if the minimizers differ obviously from these tables. Moreover, the largest value and second largest value of vector $|\theta_i-\theta_j|+3.08~\sigma$ differ very little and this may be due to the optimal solution.  Last but not least, the control objective is to maintain the phase-angle differences over power lines inside the safety region, and a comparison should be made of the stochastic power system without and with using the proposed procedure. Here, we use the {\em proportional control law} as the form of without using the proposed procedure, and the comparison results are shown in Table \ref{OriginalPhaseangledifferencesandotheroutputcompare}, followed by the tail probability computed by this table. Here, with the tail probability we mean the probability of the power flow going outside the safety region. Note that the tail probability of line $i-j$ computed by our control objective function is two times an upper bound of the largest value of $P(\omega\in \Omega~|~\theta_{ij}(\omega,t)<-\pi/2)$ and $P(\omega\in \Omega~|~\theta_{ij}(\omega,t)>\pi/2)$, and these two probabilities are computed according to the real expectations of the phase-angle differences. Some of them are positive while the rest of them are negative. From Table \ref{OriginalPhaseangledifferencesandotheroutputcompare}, we can see the probabilities of the two vulnerable lines going out of the safety region drop down tremendously, while this tail probability of the other lines may increase. This is because of the choice of our control objective function, what we want to adjust is the most vulnerable line while the other lines may be less stable than before, see line 6-7, line 2-7, line 1-7 etc.
\par
Finally, we provide evidence that the power system is more stable after using our proposed procedure than without using the procedure in a practical sense, by increasing the power demand to approach the maximal available power supply. This is a contingency that happens a lot in practice when large consumer centers like steel factories or data centers change the value of their power demand due to a change in activities, and the same for steel factories. We still assume the maximal power supply is larger than the sum of power demand, otherwise, there is a need for load shedding. We set the sum of power demand is 59 which almost reaches the maximal power supply 60. The computation results are shown in Tables \ref{increasepowerdemandextremeoutput}, \ref{increasepowerdemandextremecomparison}, \ref{TailProbabilityEightNodeNetworkExtreme}.

\vspace{-3mm}

 \begin{table}[H]
 \begin{footnotesize}
 \caption{The network parameters:}
 \begin{tabular}{|l|l|l|l|l|l|l|l|l|}
 \hline
$i,$\qquad~~~~~Node&1&2&3&4&5&6&7&8\\
 \hline
$m_i,$\qquad~Inertia&10&10&10&10&1&1&1&1\\
 \hline
$d_i,$\qquad~ ~Damping&$4$&4&4&4&1&1&1&1\\
 \hline
$K_2(i,i),$~~Disturbance&1.7668&1.9482&1.5724&1.4638&2.2385&2.4133&2.1685&2.2157\\
 \hline
 \end{tabular}
 \label{Thenetworkparameters}
  \end{footnotesize}
 \end{table}

 \begin{table}[H]
  \caption{Line capacities}
  \begin{tabular}{|l|l|l|l|l|l|l|l|l|l|l|l|l|}
 \hline
 Line&1-3&1-7&2-3&2-7&3-4&4-5&4-8&5-6&6-7&6-8&2-4\\
  \hline
  Capacities&25&25&25&25&25 or 50&25&25&25&25&25&0 or 25\\
  \hline
  \end{tabular}
  \label{Informationoflines}
  \end{table}

\begin{table}[H]
  \caption{Power supply and power demand}
  \begin{tabular}{|l|l|l|l|l|l|l|l|l|}
 \hline
 Power&$p_1^{+,max}$&$p_2^{+,max}$&$p_3^{+,max}$&$p_4^{+,max}$&$p_5^{-,max}$&$p_6^{-,max}$&$p_7^{-,max}$&$p_8^{-,max}$\\
 \hline
 &12&14&16&18&12&12&13&13\\
 \hline
 \end{tabular}
  \label{Powersupplyandpowerdemand}
 \end{table}

   \begin{table}[H]
  \caption{The minimum and optimal power vector start from the power supply vector $[12,12,12]$}
\begin{tabular}{|l|l|l|l|l|l|l|l|l|} 
\hline
\text{Minimum}&$p_1$&$p_2$&$p_3$&$p_4$&$p_5$&$p_6$&$p_7$&$p_8$\\
\hline
1.3530 & 11.7679 & 13.7632 & 13.5247 & 10.9441 & -12 & -12 & -13 & -13\\
\hline
\end{tabular}
  \label{OriginalTheminimumvalueandoptimalpowervector}
\end{table}

 \begin{table}[H]
  \caption{Phase-angle differences and other outputs start from the power supply vector $[12,12,12]$}
 \begin{tabular}{|ll|l|l|l|r|} \hline  $i$&$j$&$|\theta_i -\theta_j| $&$\sigma_{ij}$&$|\theta_i -\theta_j|+ 3.08\sigma_{ij}$&$ l_{ij}sin(|\theta_i -\theta_j|+ 3.08\sigma_{ij})$\\\hline           6 & 7 & 0.5274 & 0.2681 & 1.3530 & 24.4096\\ 4 & 8 & 0.5217 & 0.2699 & 1.3529 & 24.4089\\ 4 & 5 & 0.4988 & 0.2694 & 1.3284 & 24.2692\\ 2 & 7 & 0.5605 & 0.2239 & 1.2500 & 23.7248\\ 1 & 7 & 0.5140 & 0.2230 & 1.2007 & 23.3076\\ 3 & 4 & 0.5695 & 0.1648 & 1.0770 & 22.0132\\ 6 & 8 & 0.0216 & 0.2658 & 0.8402 & 18.6193\\ 5 & 6 & 0.0016 & 0.2659 & 0.8205 & 18.2871\\ 2 & 3 & 0.0189 & 0.1267 & 0.4091 & 9.9452\\ 1 & 3 & 0.0210 & 0.1247 & 0.4051 & 9.8535 \\\hline &  &  & & $1.57 \approx \pi/2$&\\
\hline
\end{tabular}
\label{OriginalPhaseangledifferencesandotheroutput}
\end{table}

 \begin{table}[H]
  \caption{The minimum and optimal power vector start from the power supply vector $[12,14,13]$}
\begin{tabular}{|l|l|l|l|l|l|l|l|l|} 
\hline
\text{Minimum}&$p_1$&$p_2$&$p_3$&$p_4$&$p_5$&$p_6$&$p_7$&$p_8$\\
\hline
1.3529 & 11.8411 & 13.8306 & 13.3145 & 11.0138 & -12 & -12 & -13 & -13\\
\hline
\end{tabular}
  \label{OriginalTheminimumvalueandoptimalpowervectornewstart1}
\end{table}

 \begin{table}[H]
  \caption{Phase-angle differences and other outputs start from the power supply vector $[12,14,13]$}
 \begin{tabular}{|ll|l|l|l|r|} \hline  $i$&$j$&$|\theta_i -\theta_j| $&$\sigma_{ij}$&$|\theta_i -\theta_j|+ 3.08\sigma_{ij}$&$ l_{ij}sin(|\theta_i -\theta_j|+ 3.08\sigma_{ij})$\\\hline          
6 & 7 & 0.5273 & 0.2681 & 1.3529 & 24.4090\\ 4 & 8 & 0.5217 & 0.2699 & 1.3529 & 24.4089\\ 4 & 5 & 0.4988 & 0.2693 & 1.3284 & 24.2692\\ 2 & 7 & 0.5604 & 0.2239 & 1.2499 & 23.7239\\ 1 & 7 & 0.5141 & 0.2229 & 1.2007 & 23.3075\\ 3 & 4 & 0.5660 & 0.1646 & 1.0730 & 21.9663\\ 6 & 8 & 0.0216 & 0.2658 & 0.8402 & 18.6186\\ 5 & 6 & 0.0016 & 0.2659 & 0.8205 & 18.2864\\ 2 & 3 & 0.0217 & 0.1267 & 0.4121 & 10.0126\\ 1 & 3 & 0.0181 & 0.1248 & 0.4023 & 9.7891 \\
\hline
\end{tabular}
\label{OriginalPhaseangledifferencesandotheroutputnewstart1}
\end{table}
 \begin{table}[H]
  \caption{The minimum and optimal power vector start from the power supply vector $[11,11,14]$}
\begin{tabular}{|l|l|l|l|l|l|l|l|l|} 
\hline
\text{Minimum}&$p_1$&$p_2$&$p_3$&$p_4$&$p_5$&$p_6$&$p_7$&$p_8$\\
\hline
1.3532 & 11.7085 & 12.6649 & 15.2572 & 10.3694 & -12 & -12 & -13 & -13 \\
\hline
\end{tabular}
  \label{OriginalTheminimumvalueandoptimalpowervectornewstart2}
\end{table}

 \begin{table}[H]
  \caption{Phase-angle differences and other outputs start from the power supply vector $[11,11,14]$}
 \begin{tabular}{|ll|l|l|l|r|} \hline  $i$&$j$&$|\theta_i -\theta_j| $&$\sigma_{ij}$&$|\theta_i -\theta_j|+ 3.08\sigma_{ij}$&$ l_{ij}sin(|\theta_i -\theta_j|+ 3.08\sigma_{ij})$\\\hline          
4 & 8 & 0.5218 & 0.2700 & 1.3532 & 24.4107\\ 6 & 7 & 0.5273 & 0.2682 & 1.3532 & 24.4106\\ 4 & 5 & 0.4988 & 0.2694 & 1.3287 & 24.2711\\ 2 & 7 & 0.5482 & 0.2238 & 1.2375 & 23.6240\\ 1 & 7 & 0.5259 & 0.2231 & 1.2130 & 23.4168\\ 3 & 4 & 0.5969 & 0.1657 & 1.1071 & 22.3606\\ 6 & 8 & 0.0216 & 0.2658 & 0.8402 & 18.6191\\ 5 & 6 & 0.0016 & 0.2659 & 0.8205 & 18.2869\\ 1 & 3 & 0.0337 & 0.1245 & 0.4172 & 10.1289\\ 2 & 3 & 0.0146 & 0.1263 & 0.4036 & 9.8180 \\
\hline
\end{tabular}
\label{OriginalPhaseangledifferencesandotheroutputnewstart2}
\end{table}

 \begin{table}[H]
  \caption{The minimum and optimal power vector by gridding method}
\begin{tabular}{|l|l|l|l|l|l|l|l|l|} 
\hline
\text{Minimum}&$p_1$&$p_2$&$p_3$&$p_4$&$p_5$&$p_6$&$p_7$&$p_8$\\
\hline
1.3529 & 11.9329& 13.9333& 13.0203 & 11.1135 & -12 & -12 & -13 & -13\\
\hline
\end{tabular}
  \label{OriginalTheminimumvalueandoptimalpowervectorgrid}
\end{table}

 \begin{table}[H]
  \caption{Phase-angle differences and other outputs by gridding method}
 \begin{tabular}{|ll|l|l|l|r|} \hline  $i$&$j$&$|\theta_i -\theta_j| $&$\sigma_{ij}$&$|\theta_i -\theta_j|+ 3.08\sigma_{ij}$&$ l_{ij}sin(|\theta_i -\theta_j|+ 3.08\sigma_{ij})$\\\hline          
4 & 8 & 0.5218 & 0.2698 & 1.3529 & 24.4088\\ 6 & 7 & 0.5273 & 0.2680 & 1.3528 & 24.4085\\ 4 & 5 & 0.4988 & 0.2693 & 1.3284 & 24.2691\\ 2 & 7 & 0.5605 & 0.2239 & 1.2500 & 23.7246\\ 1 & 7 & 0.5139 & 0.2229 & 1.2005 & 23.3059\\ 3 & 4 & 0.5613 & 0.1645 & 1.0679 & 21.9044\\ 6 & 8 & 0.0216 & 0.2658 & 0.8401 & 18.6183\\ 5 & 6 & 0.0016 & 0.2659 & 0.8204 & 18.2861\\ 2 & 3 & 0.0257 & 0.1268 & 0.4163 & 10.1085\\ 1 & 3 & 0.0143 & 0.1248 & 0.3987 & 9.7051 \\
\hline
\end{tabular}
\label{OriginalPhaseangledifferencesandotheroutputgrid}
\end{table}

 \begin{table}[H]
  \caption{Increase the strength of the disturbance to 1.2 times before, the minimum value and optimal power vector start from the power supply vector $[12,12,12]$}
\begin{tabular}{|l|l|l|l|l|l|l|l|l|} \hline
\text{Minimum}&$p_1$&$p_2$&$p_3$&$p_4$&$p_5$&$p_6$&$p_7$&$p_8$\\
\hline
1.5188 & 11.8169 & 13.7957 & 13.4565 & 10.9308 & -12 & -12 & -13 & -13 \\
\hline
\end{tabular}
\label{IncreaseTheStrengthofRheDisturbancetobe1.2timesbeforetheminimumvalueandoptimalpowervector1}
\end{table}

\begin{table}[H]
 \caption{Increase the strength of the disturbance to be 1.2 times before, phase-angle differences and other output start from the power supply vector $[12,12,12]$}
 \begin{tabular}{|ll|l|l|l|r|} \hline  $i$&$j$&$|\theta_i -\theta_j| $&$\sigma_{ij}$&$|\theta_i -\theta_j|+ 3.08\sigma_{ij}$&$ l_{ij}sin(|\theta_i -\theta_j|+ 3.08\sigma_{ij})$\\\hline       
4 & 8 & 0.5214 & 0.3238 & 1.5188 & 24.9663\\ 6 & 7 & 0.5279 & 0.3217 & 1.5188 & 24.9662\\ 4 & 5 & 0.4985 & 0.3232 & 1.4940 & 24.9264\\ 2 & 7 & 0.5606 & 0.2687 & 1.3881 & 24.5838\\ 1 & 7 & 0.5145 & 0.2676 & 1.3386 & 24.3289\\ 3 & 4 & 0.5693 & 0.1977 & 1.1782 & 23.0978\\ 6 & 8 & 0.0219 & 0.3189 & 1.0041 & 21.0925\\ 5 & 6 & 0.0019 & 0.3190 & 0.9845 & 20.8249\\ 2 & 3 & 0.0202 & 0.1520 & 0.4884 & 11.7313\\ 1 & 3 & 0.0194 & 0.1497 & 0.4804 & 11.5535 \\
  \hline
\end{tabular}
\label{IncreaseTheStrengthofTheDisturbancetobe1.2timesbeforephaseangledifferencesandotheroutput1}
\end{table}

\begin{table}[H]
  \caption{Increase the strength of the disturbance to be 1.2 times before, the minimum value and optimal power vector start from the power supply vector $[12,14,13]$}
\begin{tabular}{|l|l|l|l|l|l|l|l|l|} \hline
\text{Minimum}&$p_1$&$p_2$&$p_3$&$p_4$&$p_5$&$p_6$&$p_7$&$p_8$\\
\hline
1.5188 & 11.8003 & 13.7608 & 13.5348 & 10.9041 & -12 & -12 & -13 & -13 \\
\hline
\end{tabular}
\label{IncreaseTheStrengthofRheDisturbancetobe1.2timesbeforetheminimumvalueandoptimalpowervector2}
\end{table}

\begin{table}[H]
 \caption{Increase the strength of the disturbance to be 1.2 times before, phase-angle differences and other output start from the power supply vector $[12,14,13]$}
 \begin{tabular}{|ll|l|l|l|r|} \hline  $i$&$j$&$|\theta_i -\theta_j| $&$\sigma_{ij}$&$|\theta_i -\theta_j|+ 3.08\sigma_{ij}$&$ l_{ij}sin(|\theta_i -\theta_j|+ 3.08\sigma_{ij})$\\\hline       
6 & 7 & 0.5280 & 0.3217 & 1.5188 & 24.9663\\ 4 & 8 & 0.5214 & 0.3238 & 1.5188 & 24.9663\\ 4 & 5 & 0.4985 & 0.3232 & 1.4940 & 24.9264\\ 2 & 7 & 0.5604 & 0.2687 & 1.3879 & 24.5829\\ 1 & 7 & 0.5147 & 0.2676 & 1.3388 & 24.3302\\ 3 & 4 & 0.5705 & 0.1978 & 1.1796 & 23.1113\\ 6 & 8 & 0.0219 & 0.3189 & 1.0041 & 21.0926\\ 5 & 6 & 0.0019 & 0.3190 & 0.9845 & 20.8250\\ 2 & 3 & 0.0189 & 0.1520 & 0.4872 & 11.7030\\ 1 & 3 & 0.0203 & 0.1497 & 0.4812 & 11.5715 \\
  \hline
\end{tabular}
\label{IncreaseTheStrengthofTheDisturbancetobe1.2timesbeforephaseangledifferencesandotheroutput2}
\end{table}

\begin{table}[H]
  \caption{Increase the strength of the disturbance to be 1.2 times before, the minimum value and optimal power vector start from the power supply vector $[11,11,14]$}
\begin{tabular}{|l|l|l|l|l|l|l|l|l|} \hline
\text{Minimum}&$p_1$&$p_2$&$p_3$&$p_4$&$p_5$&$p_6$&$p_7$&$p_8$\\
\hline
 1.5192 & 11.7376 & 12.7167 & 15.1915 & 10.3543 & -12 & -12 & -13 & -13 \\
\hline
\end{tabular}
\label{IncreaseTheStrengthofRheDisturbancetobe1.2timesbeforetheminimumvalueandoptimalpowervector3}
\end{table}

\begin{table}[H]
 \caption{Increase the strength of the disturbance to be 1.2 times before, phase-angle differences and other output start from the power supply vector $[11,11,14]$}
 \begin{tabular}{|ll|l|l|l|r|} \hline  $i$&$j$&$|\theta_i -\theta_j| $&$\sigma_{ij}$&$|\theta_i -\theta_j|+ 3.08\sigma_{ij}$&$ l_{ij}sin(|\theta_i -\theta_j|+ 3.08\sigma_{ij})$\\\hline       
6 & 7 & 0.5279 & 0.3218 & 1.5192 & 24.9667\\ 4 & 8 & 0.5214 & 0.3239 & 1.5192 & 24.9667\\ 4 & 5 & 0.4985 & 0.3233 & 1.4943 & 24.9270\\ 2 & 7 & 0.5488 & 0.2686 & 1.3760 & 24.5272\\ 1 & 7 & 0.5260 & 0.2677 & 1.3505 & 24.3960\\ 3 & 4 & 0.5969 & 0.1988 & 1.2093 & 23.3842\\ 6 & 8 & 0.0219 & 0.3189 & 1.0042 & 21.0931\\ 5 & 6 & 0.0019 & 0.3191 & 0.9845 & 20.8255\\ 1 & 3 & 0.0326 & 0.1494 & 0.4927 & 11.8257\\ 2 & 3 & 0.0130 & 0.1516 & 0.4798 & 11.5408 \\
  \hline
\end{tabular}
\label{IncreaseTheStrengthofTheDisturbancetobe1.2timesbeforephaseangledifferencesandotheroutput3}
\end{table}

\begin{table}[H]
  \caption{Doubling capacity of line $3-4$, the minimum and optimal power vector start from the power supply vector $[12,12,12]$}
\begin{tabular}{|l|l|l|l|l|l|l|l|l|} \hline
\text{Minimum}&$p_1$&$p_2$&$p_3$&$p_4$&$p_5$&$p_6$&$p_7$&$p_8$\\
\hline
1.3712 & 12.0001 & 14.0000 & 16.0000 & 7.9999 & -12 & -12 & -13 & -13 \\
 \hline
 \end{tabular}
 \label{DoublingLine34theminimumvalueandoptimalpowervector}
\end{table}

\begin{table}[H]
  \caption{Doubling capacity of line $3-4$, the phase-angle differences and other outputs start from the power supply vector $[12,12,12]$}
 \begin{tabular}{|ll|l|l|l|r|} \hline  $i$&$j$&$|\theta_i -\theta_j| $&$\sigma_{ij}$&$|\theta_i -\theta_j|+ 3.08\sigma_{ij}$&$ l_{ij}sin(|\theta_i -\theta_j|+3.08\sigma_{ij})$\\\hline   
 4 & 8 & 0.5552 & 0.2649 & 1.3712 & 24.5037\\ 4 & 5 & 0.5319 & 0.2644 & 1.3463 & 24.3728\\ 6 & 7 & 0.4620 & 0.2600 & 1.2628 & 23.8239\\ 2 & 7 & 0.5269 & 0.2198 & 1.2040 & 23.3372\\ 1 & 7 & 0.4812 & 0.2186 & 1.1544 & 22.8638\\ 5 & 6 & 0.0271 & 0.2651 & 0.8435 & 18.6750\\ 6 & 8 & 0.0071 & 0.2650 & 0.8233 & 18.3348\\ 3 & 4 & 0.3652 & 0.1371 & 0.7875 & 35.4282\\ 2 & 3 & 0.0572 & 0.1418 & 0.4939 & 11.8509\\ 1 & 3 & 0.0171 & 0.1377 & 0.4412 & 10.6757\\          
  \hline
\end{tabular}
\label{DoublingLine34thephaseangledifferencesandotheroutput}
\end{table}

\begin{table}[H]
  \caption{Doubling the capacity of line $3-4$, the minimum value and optimal power vector start from the power supply vector $[12,14,13]$}
\begin{tabular}{|l|l|l|l|l|l|l|l|l|} \hline
\text{Minimum}&$p_1$&$p_2$&$p_3$&$p_4$&$p_5$&$p_6$&$p_7$&$p_8$\\
\hline
1.3712 & 12.0001 & 14.0000 & 16.0000 & 7.9999 & -12 & -12 & -13 & -13 \\
\hline
\end{tabular}
\label{Doublingline34theminimumvalueandoptimalpowervectornewstart}
\end{table}

\begin{table}[H]
  \caption{Doubling the capacity of line $3-4$, the phase-angle differences and other outputs start from the power supply vector $[12,14,13]$}
 \begin{tabular}{|ll|l|l|l|r|} \hline  $i$&$j$&$|\theta_i -\theta_j| $&$\sigma_{ij}$&$|\theta_i -\theta_j|+ 3.08\sigma_{ij}$&$ l_{ij}sin(|\theta_i -\theta_j|+ 3.08\sigma_{ij})$\\\hline             
 4 & 8 & 0.5552 & 0.2649 & 1.3712 & 24.5037\\ 4 & 5 & 0.5319 & 0.2644 & 1.3463 & 24.3728\\ 6 & 7 & 0.4620 & 0.2600 & 1.2628 & 23.8239\\ 2 & 7 & 0.5269 & 0.2198 & 1.2040 & 23.3372\\ 1 & 7 & 0.4812 & 0.2186 & 1.1544 & 22.8638\\ 5 & 6 & 0.0271 & 0.2651 & 0.8435 & 18.6750\\ 6 & 8 & 0.0071 & 0.2650 & 0.8233 & 18.3348\\ 3 & 4 & 0.3652 & 0.1371 & 0.7875 & 35.4283\\ 2 & 3 & 0.0572 & 0.1418 & 0.4939 & 11.8510\\ 1 & 3 & 0.0171 & 0.1377 & 0.4412 & 10.6758 \\
 \hline
\end{tabular}
\label{Doublingline34phaseangledifferencesandotheroutputnewstart}
\end{table}

\begin{table}[H]
  \caption{Adding line $2-4$, the minimum value and optimal power vector start from the power supply vector $[12,12,12]$}
\begin{tabular}{|l|l|l|l|l|l|l|l|l|} \hline
\text{Minimum}&$p_1$&$p_2$&$p_3$&$p_4$&$p_5$&$p_6$&$p_7$&$p_8$\\
\hline
1.3953 & 12.0000 & 14.0000 & 16.0000 & 7.9999 & -12 & -12 & -13 & -13 \\
\hline
\end{tabular}
\label{AddingLine24theminimumvalueandoptimalpowervector}
\end{table}

\begin{table}[H]
  \caption{Adding line $2-4$, the phase-angle differences and other outputs start from the power supply vector $[12,12,12]$}
 \begin{tabular}{|ll|l|l|l|r|} \hline  $i$&$j$&$|\theta_i -\theta_j| $&$\sigma_{ij}$&$|\theta_i -\theta_j|+ 3.08\sigma_{ij}$&$ l_{ij}sin(|\theta_i -\theta_j|+ 3.08\sigma_{ij})$\\\hline     
 4 & 8 & 0.5670 & 0.2689 & 1.3953 & 24.6161\\ 4 & 5 & 0.5435 & 0.2683 & 1.3700 & 24.4975\\ 1 & 7 & 0.5536 & 0.2236 & 1.2424 & 23.6636\\ 6 & 7 & 0.4398 & 0.2584 & 1.2356 & 23.6089\\ 2 & 7 & 0.4334 & 0.2197 & 1.1101 & 22.3934\\ 3 & 4 & 0.4588 & 0.1444 & 0.9036 & 19.6386\\ 5 & 6 & 0.0372 & 0.2649 & 0.8529 & 18.8299\\ 6 & 8 & 0.0171 & 0.2648 & 0.8326 & 18.4926\\ 2 & 4 & 0.2957 & 0.1356 & 0.7135 & 16.3612\\ 2 & 3 & 0.1520 & 0.1055 & 0.4770 & 11.4789\\ 1 & 3 & 0.0457 & 0.1293 & 0.4439 & 10.7365  \\     
  \hline
\end{tabular}
\label{AfteraddingLine24thephaseangledifferencesandotheroutput}
\end{table}

\begin{table}[H]
  \caption{Adding line $2-4$, the minimum value and optimal power vector start from the power supply vector $[12,14,13]$}
\begin{tabular}{|l|l|l|l|l|l|l|l|l|} \hline
\text{Minimum}&$p_1$&$p_2$&$p_3$&$p_4$&$p_5$&$p_6$&$p_7$&$p_8$\\
\hline
1.3953 & 12.0001 & 14.0000 & 16.0000 & 7.9999 & -12 & -12 & -13 & -13 \\
\hline
\end{tabular}
\label{AddingLine24theminimumvalueandoptimalpowervectornewstart}
\end{table}

\begin{table}[H]
  \caption{Adding line $2-4$, the phase-angle differences and other outputs start from the power supply vector $[12,14,13]$}
 \begin{tabular}{|ll|l|l|l|r|} \hline  $i$&$j$&$|\theta_i -\theta_j| $&$\sigma_{ij}$&$|\theta_i -\theta_j|+ 3.08\sigma_{ij}$&$ l_{ij}sin(|\theta_i -\theta_j|+ 3.08\sigma_{ij})$\\\hline     
4 & 8 & 0.5670 & 0.2689 & 1.3953 & 24.6161\\ 4 & 5 & 0.5435 & 0.2683 & 1.3700 & 24.4975\\ 1 & 7 & 0.5536 & 0.2236 & 1.2424 & 23.6636\\ 6 & 7 & 0.4398 & 0.2584 & 1.2356 & 23.6089\\ 2 & 7 & 0.4334 & 0.2197 & 1.1101 & 22.3934\\ 3 & 4 & 0.4588 & 0.1444 & 0.9036 & 19.6386\\ 5 & 6 & 0.0372 & 0.2649 & 0.8529 & 18.8299\\ 6 & 8 & 0.0171 & 0.2648 & 0.8326 & 18.4925\\ 2 & 4 & 0.2957 & 0.1356 & 0.7135 & 16.3612\\ 2 & 3 & 0.1520 & 0.1055 & 0.4770 & 11.4789\\ 1 & 3 & 0.0457 & 0.1293 & 0.4439 & 10.7365 \\
  \hline
\end{tabular}
\label{AfteraddingLine24thephaseangledifferencesandotheroutputnewstart}
\end{table}

\newpage
\begin{table}[H]
  \caption{Phase-angle differences and other outputs for particular eight-node network without and with applying the procedure, the initial power supply vector is $[12,12,12]$}
 \begin{tabular}{|l|ll|l|l|l|c|} \hline Procedure&$i$&$j$&$|\theta_i -\theta_j| $&$\sigma_{ij}$&\begin{tabular}{l}$|\theta_i-\theta_j|$+ \\~$3.08\:\sigma_{ij}$\end{tabular}&\begin{tabular}{l}$ l_{ij}\:sin\:(|\theta_i -\theta_j|$\\~$+ 3.08~\sigma_{ij})$\end{tabular}\\\hline
 without&4 & 8 & 0.5566 & 0.2711 & 1.3917 & 24.5999\\ 
with&4 & 8 & 0.5217 & 0.2699 & 1.3529 & 24.4089\\ \hline
 without&4 & 5 & 0.5332 & 0.2705 & 1.3664 & 24.4796\\ 
 with&4 & 5 & 0.4988 & 0.2694 & 1.3284 & 24.2692\\  \hline
without& 6 & 7 & 0.4593 & 0.2647 & 1.2747 & 23.9121\\ 
with&6 & 7 & 0.5274 & 0.2681 & 1.3530 & 24.4096\\ \hline
 without&2 & 7 & 0.5217 & 0.2219 & 1.2051 & 23.3472\\
with&2 & 7 & 0.5605 & 0.2239 & 1.2500 & 23.7248\\ \hline 
without& 1 & 7 & 0.4836 & 0.2210 & 1.1644 & 22.9642\\ 
 with&1 & 7 & 0.5140 & 0.2230 & 1.2007 & 23.3076\\ \hline
 without&3 & 4 & 0.4519 & 0.1605 & 0.9463 & 20.2809\\ 
  with&3 & 4 & 0.5695 & 0.1648 & 1.0770 & 22.0132\\  \hline
 without&5 & 6 & 0.0283 & 0.2658 & 0.8470 & 18.7324\\ 
   with& 5 & 6 & 0.0016 & 0.2659 & 0.8205 & 18.2871\\\hline
without& 6 & 8 & 0.0083 & 0.2657 & 0.8267 & 18.3919\\ 
with&6 & 8 & 0.0216 & 0.2658 & 0.8402 & 18.6193\\   \hline
 without&1 & 3 & 0.0650 & 0.1258 & 0.4525 & 10.9301\\ 
   with& 1 & 3 & 0.0210 & 0.1247 & 0.4051 & 9.8535 \\   \hline
 without&2 & 3 & 0.0317 & 0.1278 & 0.4253 & 10.3137 \\
    with&2 & 3 & 0.0189 & 0.1267 & 0.4091 & 9.9452\\  \hline
\end{tabular}
\label{OriginalPhaseangledifferencesandotheroutputcompare}
\end{table}

  \begin{table}[h]
\caption{Tail Probability of the particular Eight-node network according to Table \ref{OriginalPhaseangledifferencesandotheroutputcompare}}
  \begin{small}
  \begin{tabular}{|l|ll|l|l|l|}
  \hline
Procdure&i&j&  $P(\omega\in \Omega~|~\theta_{ij}(\omega,t)<-\pi/2)$&$P(\omega\in \Omega~|~\theta_{ij}(\omega,t)>\pi/2)$&\begin{tabular}{l}\mbox{by our criterion}\\ \mbox{(Upper bound)}\end{tabular}\\\hline
 without&4&8& 2.1259e-15 & 9.1630e-05&\\ 
 with& 4&8&4.4924e-15 & 5.0749e-05 & $2\times 5.0755e-05$\\\hline
 without& 4&5& 3.6789e-15 & 6.2566e-05&\\
  with&4&5&7.8163e-15 & 3.4574e-05 & $2\times 5.0443e-05$\\\hline
 without& 6&7&1.3400e-05 & 8.6386e-15&\\ 
 with&6&7&4.9748e-05 & 2.5148e-15 & $2\times 4.9635e-05$\\\hline
  without&2&7& 2.0523e-21 & 1.1349e-06&\\
with&2&7& 8.7456e-22 & 3.2069e-06 & $2\times 2.5311e-05$\\\hline
without& 1&7&  7.2975e-21 & 4.3396e-07&\\
  with&1&7&4.4298e-21 & 1.0739e-06 & $2\times 2.4889e-05$\\\hline
  without& 3&4&1.0232e-36 & 1.5699e-12&\\
  with&3&4& 7.2278e-39 & 6.1675e-10 & $2\times 5.3733e-06$\\\hline
   without&5&6&8.9299e-10 & 3.2526e-09&\\
 with&  5&6&1.8016e-09 & 1.6747e-09 &$2\times 4.8277e-05$\\\hline
 without&  6&8&2.0429e-09 & 1.3979e-09&\\
  with&  6&8& 1.0431e-09 & 2.7976e-09 & $2\times 4.8216e-05$\\\hline
 without& 1&3&   2.5587e-33 & 5.8703e-39&\\ 
  with&1&3&  9.1864e-36 & 1.2856e-37 &$2\times 6.9455e-07$\\\hline
 without&   2&3&   1.0556e-33 & 2.2809e-36& \\
  with& 2&3&   2.0661e-36 & 8.5471e-35 & $2\times 7.9733e-07$\\\hline 
  \end{tabular}
  \label{TailProbabilityEightNodeNetwork}
  \end{small}
  \end{table}
\newpage

 \begin{table}[H]
  \caption{Increase the power demand to 59, the minimum value and optimal power vector start from the power supply vector is $[5,5,5]$}
\begin{tabular}{|l|l|l|l|l|l|l|l|l|} 
\hline
\text{Minimum}&$p_1$&$p_2$&$p_3$&$p_4$&$p_5$&$p_6$&$p_7$&$p_8$\\
\hline
1.5306& 12.0001 & 14.001 & 16.0000 & 16.9998 & -16 & -13 & -14 & -16\\
\hline
\end{tabular}
  \label{increasepowerdemandextremeoutput}
\end{table}

\begin{table}[H]
  \caption{Increase the power demand to 59, without and with applying the procedure, the initial power supply vector is $[5,5,5]$} 
 \begin{tabular}{|l|ll|l|l|l|c|} \hline Procedure&$i$&$j$&$|\theta_i -\theta_j| $&$\sigma_{ij}$&\begin{tabular}{l}$|\theta_i-\theta_j|$+ \\~$3.08\:\sigma_{ij}$\end{tabular}&\begin{tabular}{l}$ l_{ij}\:sin\:(|\theta_i -\theta_j|$\\~$+ 3.08~\sigma_{ij})$\end{tabular}\\\hline
 without& 4 & 5 & 0.6746 & 0.2801 & 1.5373 & 24.9860\\ 
with &4 & 5& 0.6687 & 0.2798 & 1.5306 & 24.9798\\\hline
without&4 & 8 & 0.6746 & 0.2796 & 1.5358 & 24.9847\\
with &4 & 8& 0.6687 & 0.2793 & 1.5291 & 24.9783\\\hline
without & 6 & 7& 0.5834 & 0.2727 & 1.4234 & 24.7290\\ 
 with & 6 & 7& 0.5944 & 0.2734 & 1.4366 & 24.7751\\\hline
without& 2 & 7 & 0.6127 & 0.2272 & 1.3126 & 24.1712\\ 
with &  2 & 7& 0.6187 & 0.2276 & 1.3198 & 24.2166\\\hline
without& 1 & 7 & 0.5654 & 0.2263 & 1.2623 & 23.8196\\
 with &1 & 7& 0.5704 & 0.2266 & 1.2685 & 23.8663\\\hline
  without &3 & 4& 0.5718 & 0.1662 & 1.0838 & 22.0939\\ 
 with & 3 & 4& 0.5944 & 0.1672 & 1.1093 & 22.3848\\ \hline
  without &5 & 6& 0.0154 & 0.2694 & 0.8450 & 18.6998\\ 
  with & 5 & 6& 0.0200 & 0.2694 & 0.8496 & 18.7759\\\hline
  without &6 & 8& 0.0154 & 0.2690 & 0.8441 & 18.6841\\ 
 with & 6 & 8& 0.0200 & 0.2691 & 0.8487 & 18.7603\\\hline
  without&1 & 3 & 0.0638 & 0.1269 & 0.4548 & 10.9811\\ 
  with &1 & 3& 0.0600 & 0.1267 & 0.4502 & 10.8785\\\hline
  without &2 & 3& 0.0244 & 0.1289 & 0.4214 & 10.2253 \\
  with&2 & 3 & 0.0200 & 0.1286 & 0.4161 & 10.1048
\\\hline
\end{tabular}
\label{increasepowerdemandextremecomparison}
\end{table}

\begin{table}[h]
\caption{Tail Probability of the particular Eight-node network according to Table \ref{increasepowerdemandextremecomparison}}
  \begin{small}
  \begin{tabular}{|l|ll|l|l|l|}
  \hline
Procdure&i&j&  $P(\omega\in \Omega~|~\theta_{ij}(\omega,t)<-\pi/2)$&$P(\omega\in \Omega~|~\theta_{ij}(\omega,t)>\pi/2)$&\begin{tabular}{l}\mbox{by our criterion}\\ \mbox{(Upper bound)}\end{tabular}\\\hline
  without& 4 & 5&  5.4441e-16 & 6.8819e-04&\\ 
  with& 4 & 5&6.0261e-16 & 6.3190e-04 & $2\times 6.3282e-04$\\\hline
  without&4 & 8 & 4.8441e-16 & 6.7466e-04&\\ 
  with&4 & 8 &5.3633e-16 & 6.1928e-04 & $2\times 6.3225e-04$\\ \hline
  without & 6 & 7&1.4684e-04 & 1.4000e-15&\\ 
  with & 6 & 7&1.7760e-04 & 1.1922e-15 &$2\times 6.2543e-04$\\ \hline
  without& 2 & 7 &3.6113e-22 & 1.2380e-05&\\ 
  with& 2 & 7 &3.2938e-22 & 1.4372e-05 &$2\times 5.6376e-04$\\ \hline
  without& 1 & 7 &1.8699e-21 & 4.4406e-06&\\ 
  with& 1 & 7 & 1.7066e-21 & 5.0549e-06 &$2\times 5.6221e-04$\\ \hline
 without &3 & 4& 2.5071e-38 & 9.2300e-10&\\
 with&3 & 4&1.1788e-38 & 2.6149e-09 &$2\times 4.4943e-04$\\ \hline
 without &5 & 6&   3.8808e-09 & 1.9558e-09&\\ 
 with&5 & 6& 4.2942e-09 & 1.7637e-09 &$2\times 6.2067e-04$\\ \hline
  without &6 & 8& 1.8549e-09 & 3.6878e-09&\\ 
  with &6 & 8& 1.6946e-09 & 4.1340e-09 &$2\times 6.2031e-04$\\ \hline
   without&1 & 3 &7.9376e-33 & 2.8800e-38&\\ 
   with&1 & 3 &4.4264e-33 & 3.2638e-38 &$2\times 3.4033e-04$\\\hline
    without &2 & 3&1.8450e-33 & 1.7744e-35&\\
   with&2 & 3&  8.6867e-34 & 1.8965e-35 &$2\times 3.4620e-04 $\\\hline 
  \end{tabular}
  \label{TailProbabilityEightNodeNetworkExtreme}
  \end{small}
  \end{table}

\newpage
 \clearpage
\section{Tables of a Ring Network Example}\label{sec:table2}
The picture of the twelve-node ring network is displayed in Fig.~\ref{Aringnetwork}. First, a group of symmetric parameters are shown in \ref{RingNetworkParameter}, which leads to the symmetric outputs in Tables~\ref{RingNetworkMinimum} and \ref{RingNetworkPhaseAngleDifferences}. From the three columns of $|\theta_i -\theta_j| ,\sigma_{ij},|\theta_i -\theta_j| +3.08\sigma_{ij}$ in Table~\ref{RingNetworkPhaseAngleDifferences}, one can see that several transmission lines have very small almost zero power flows but at the same time their fluctuations are serious. Thus, we have justified our choice of the control objective function, i.e. to consider both the mean value and the standard deviation of the phase-angle difference together.
\par Secondly, several of the parameters are changed to be asymmetric as shown in Tables~\ref{RingNetworkNewParameter}. The computation results are exhibited in Tables~\ref{RingNetworkNewMinimum1} and \ref{RingNetworkNewPhaseAngleDifference1} starting from the power supply vector $[20,28,15]$ and in Tables~\ref{RingNetworkNewMinimum2} and \ref{RingNetworkNewPhaseAngleDifference2} starting from the power supply vector $[23,19,24]$. From these four tables, one can see that the lines which connect two neighbourhoods have now small power flows. What's more, the two different initial points lead to slightly different minimizers. 
\par Moreover, as for the particular eight-node academic network, the numerical values for without and with applying our proposed procedure are shown in Table~\ref{RingNetworkNewPhaseAngleDifference3}.  In Table~\ref{RingNetworkNewPhaseAngleDifference3}, it is observable that the two highest values of $|\theta_i -\theta_j|+ 3.08~\sigma_{ij}$ have decreased as well. We generate a figure displaying the probability density function of line $1-12$ for both the `without applying' and `with applying' the proposed procedure, as shown in Fig.~\ref{fig:Ringnetcontrolwithoutwith}, where the values are borrowed from Table~\ref{RingNetworkNewPhaseAngleDifference3}. In this depiction, the reduction in both standard deviation and mean value from the `without applying the proposed procedure to the `with applying the proposed procedure' is evident. Same as the particular eight-node example, we show the comparison results of applying the proposed procedure and not applying the proposed procedure in Table \ref{RingNetworkNewPhaseAngleDifference3}, while the corresponding tail probabilities are shown in Table \ref{TailProbabilityoftheringnetwork}. From Table \ref{TailProbabilityoftheringnetwork}, we can see the tail probability of the most vulnerable line $1-12$ decreases quite a lot and that of the second vulnerable line decreases a little. What's more, some of the lines may be less stable than before, see line $3-9$ and line $3-8$, and the reason of that scenario is due to the choice of the control objective function.
\vspace{-4mm}
\begin{figure}[h]
\centering
\xdef\Rad{1.5}
\newcommand{\ARW}[2][]{%
    \foreach \ang in {#2}{%
        \draw[#1] (\ang:\Rad)--(\ang+1:\Rad) ;
    }
}

\begin{tikzpicture}[tbcircle/.style={circle,minimum size=13pt,inner sep=0pt},
forward tbcircle/.style={tbcircle,fill=blue!70},
backward tbcircle/.style={tbcircle,fill=red!100},>=stealth]

\draw[line width=0.5mm, color=black!70] (0,0) circle (\Rad) ;

\node[backward tbcircle,draw] at (0:\Rad) {$\textcolor{white}{3}$} ;
\node[forward tbcircle,draw] at (30:\Rad) {$\textcolor{white}{8}$} ;
\node[forward tbcircle,draw] at (60:\Rad) {$\textcolor{white}{7}$} ;
\node[backward tbcircle,draw] at (90:\Rad) {$\textcolor{white}{2}$} ;
\node[forward tbcircle,draw] at (120:\Rad) {$\textcolor{white}{6}$} ;
\node[forward tbcircle,draw] at (150:\Rad) {$\textcolor{white}{5}$} ;
\node[backward tbcircle,draw] at (180:\Rad){$\textcolor{white}{1}$} ;
\node[forward tbcircle,draw] at (210:\Rad) {$\textcolor{white}{12}$} ;
\node[forward tbcircle,draw] at (240:\Rad) {$\textcolor{white}{11}$} ;
\node[backward tbcircle,draw] at (270:\Rad) {$\textcolor{white}{4}$} ;
\node[forward tbcircle,draw] at (300:\Rad) {$\textcolor{white}{10}$} ;
\node[forward tbcircle,draw] at (330:\Rad) {$\textcolor{white}{9}$} ;


\ARW[line width=0.3mm,black!80,->]{21,51,111,199,229,289}
\ARW[line width=0.3mm,black!80,<-]{68,340,310,128,158,250}

\end{tikzpicture}
\caption{A ring network}
\label{Aringnetwork}
\end{figure}
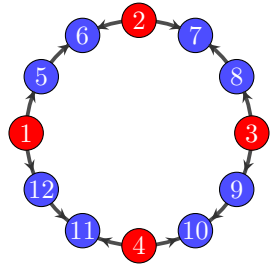

\begin{figure}
\begin{center}
\begin{tikzpicture}
 \message{Expected sensitivty & p-value^^J}
  \def\N{100}
  \def\q{0.7074}
   \def\qn{0.5720}
  \def\Bs{2}
  \def\S{\q}
   \def\Sn{\qn}
   \def\Ss{0.2753}
  \def\Ssn{0.2646}   
     \def\xmin{-4*\q}
  \def\xmax{3.2*\q}
  \def\ymin{{-0.16/\Bs}}
  \def\ymax{{0.78/\Bs}}
   \begin{axis}[every axis plot post/.append style={
               mark=none,domain=\xmin:{1.32*\xmax},samples=\N,smooth},
               xmin=\xmin, xmax=1.20*\xmax,
               ymin=\ymin, ymax=\ymax,
               axis lines=middle,
               axis line style=thick,
               enlargelimits=upper, 
               ticks=none,
               xlabel=$v$,
               every axis x label/.style={at={(current axis.right of origin)},anchor=north west},
               y=240pt,
              ]
   \addplot[name path=S,line width=0.35mm,red!90!red ] {gauss(x,\S,\Ss)};
         \addplot[red,dashed,line width=0.35mm]
      coordinates {(\q,{-0.02*exp(-\q/\Bs)/\Bs}) (\q, {1.08*gauss(\q,\S,\Ss)})}
       node[below=-4pt,right=-5pt,pos=-0.06] {$m_1$};
    \addplot[red,line width=0.35mm]
      coordinates {(1.57,{-0.04*exp(-1.57/\Bs)/\Bs}) (1.57, {1.38*gauss(1.57,\S,\Ss)})}
       node[below=-5pt,pos=0] {$\pi/2$};
          \addplot[red,line width=0.35mm]
      coordinates {(1.57, {1.5*gauss(1.57,\S,\Ss)}) (1.39*\xmax, {1.5*gauss(1.57,\S,\Ss)})};
  \addplot[name path=S2,line width=0.35mm,blue!90!red ] {gauss(x,\Sn,\Ssn)};
   \addplot[blue,dashed,line width=0.35mm]
      coordinates {(\qn,{-0.02*exp(-\qn/\Bs)/\Bs}) (\qn, {1.08*gauss(\qn,\Sn,\Ssn)})}
       node[below=1pt,left=-6pt,pos=-0.06] {$m_2$};

       \addplot[red,line width=0.35mm]
      coordinates {(-1.57,{-0.009*exp(1.57/\Bs)/\Bs}) (-1.57, {10.5*gauss(-1.57,\S,\Ss)})}
       node[below=-4pt,pos=0] {$-\pi/2$};    
              \addplot[red,line width=0.35mm]
      coordinates {(\xmin,{2.5*gauss(-1.57,\S,\Ss)}) (-1.57, {2.5*gauss(-1.57,\S,\Ss)})};

    \path[name path=xaxis1]
     (0,0) -- (1.08*\xmax,0);
    \addplot[red!60!blue] fill between[of=xaxis1 and S, soft clip={domain=1.57:1.50*\xmax}];
       \path[name path=xaxis2]
     (-1.08*\xmax,0) -- (0,0);
    \addplot[red!60!blue] fill between[of=xaxis2 and S, soft clip={domain=-1.08*\xmax:-1.57}];
    \node[below left=2pt] at (0,0) {0};

    \node[above right=1.3pt,black!20!blue] at ( 0.48*\Bs,0.13) {$p_{\theta_i - \theta_j}(v)$};
     \node[above right=1.3pt,black!20!red] at ( 0.50*\Bs,0.20) {$p_{\theta_i - \theta_j}(v)$};
     \addplot[red,line width=0.35mm]
      coordinates {(-2.5,0.2) (-2.4, 0.2)}
       node[right=4pt,pos=0] {without applying};    
       \addplot[blue,line width=0.35mm]
      coordinates {(-2.5,0.15) (-2.4, 0.15)}
       node[right=4pt,pos=0] {with applying};

  \end{axis}
  
  \end{tikzpicture}
\end{center}
\caption{Example. ring network, line 1-12, without and with control.}
 \vspace{2cm}
 \label{fig:Ringnetcontrolwithoutwith}
\end{figure}

\begin{table}[H]
\begin{small}
  \caption{Network parameters, where the line capacities are not shown and they are all equal 20.}
\begin{tabular} {|l|l|l|l|l|}
\hline Node&$p^{+,max},$&Inertia&Damping&Disturbance\\
 i&$-p^{-,max}$&$m_i$&$d_i$&$K_2(i,i)$
\\\hline1&30&10&4&2
\\\hline2&20&10&4&2
\\\hline3&25&10&4&2
\\\hline4&20&10&4&2
\\\hline5&-10&1&1&2
\\\hline6&-10&1&1&2
\\\hline7&-10&1&1&2
\\\hline8&-10&1&1&2
\\\hline9&-10&1&1&2
\\\hline10&-10&1&1&2
\\\hline11&-10&1&1&2
\\\hline12&-10&1&1&2
\\
\hline
\end{tabular}
\label{RingNetworkParameter}
\end{small}
\end{table}

 \begin{table}[H]
  \caption{The minimum and optimal power vector starting from the power supply vector  [20,18,25]}
   \begin{small}
\begin{tabular} {|l|l|l|l|l|l|l|l|l|l|l|l|l|} \hline
\text{Minimum}&$p_1$&$p_2$&$p_3$&$p_4$&$p_5$&$p_6$&$p_7$&$p_8$&$p_9$&$p_{10}$&$p_{11}$&$p_{12}$\\
\hline
1.4221 & 20.0011 & 20.0000 & 19.9989 & 20.0000 & -10 & -10 & -10 & -10 & -10 & -10 & -10 & -10 
\\ \hline
\end{tabular}
\label{RingNetworkMinimum}
\end{small}
\end{table}

 \begin{table}[H]
  \caption{Phase-angle differences and other outputs}
 \begin{tabular}{|ll|l|l|l|l|}\hline  $i$&$j$&$|\theta_i -\theta_j| $&$\sigma_{ij}$&$|\theta_i -\theta_j|+ 3.08~\sigma_{ij}$&$ l_{ij}~sin(|\theta_i -\theta_j|+ 3.08~\sigma_{ij})$\\\hline 
4 & 10 & 0.5236 & 0.2917 & 1.4221 & 19.7792\\ 2 & 7 & 0.5236 & 0.2917 & 1.4221 & 19.7792\\ 1 & 5 & 0.5236 & 0.2917 & 1.4221 & 19.7792\\ 1 & 12 & 0.5236 & 0.2917 & 1.4221 & 19.7792\\ 3 & 9 & 0.5236 & 0.2917 & 1.4220 & 19.7789\\ 3 & 8 & 0.5236 & 0.2917 & 1.4220 & 19.7789\\ 2 & 6 & 0.5236 & 0.2917 & 1.4220 & 19.7789\\ 4 & 11 & 0.5236 & 0.2917 & 1.4220 & 19.7789\\ 7 & 8 & 2.8213e-05 & 0.2782 & 0.8570 & 15.1180\\ 9 & 10 & 2.8205e-05 & 0.2782 & 0.8570 & 15.1180\\ 11 & 12 & 2.8217e-05 & 0.2782 & 0.8570 & 15.1180\\ 5 & 6 & 2.8208e-05 & 0.2782 & 0.8570 & 15.1180\\\hline &  & & &$1.57 \approx \pi/2$ & 
 \\\hline
 \end{tabular}
 \label{RingNetworkPhaseAngleDifferences}
 \end{table}

\paragraph{Changing the parameters:} Several parameters are changed to be asymmetric.
\vspace{-0.3cm}
\begin{table}[H]
  \caption{Network parameters, what not listed are that all line capacities are increased to 24}
\begin{tabular} {|l|l|l|l|l|}
\hline Node&$p^{+,max},$&Inertia&Damping&Disturbance\\
 i&$-p^{-,max}$&$m_i$&$d_i$&$K_2(i,i)$
\\\hline1&30&10&4&2.00
\\\hline2&20&10&4&2.30
\\\hline3&25&10&4&2.50
\\\hline4&20&10&4&2.70
\\\hline5&-6&1&1&1.60
\\\hline6&-10&1&1&1.70
\\\hline7&-8&1&1&1.80
\\\hline8&-12&1&1&1.90
\\\hline9&-17&1&1&1.65
\\\hline10&-13&1&1&1.75
\\\hline11&-7&1&1&1.85
\\\hline12&-11&1&1&2.05
\\
\hline
\end{tabular}
\label{RingNetworkNewParameter}
\end{table}
\vspace{-6mm}

 \begin{table}[H]
 \begin{small}
  \caption{The minimum value and optimal power vector starting from the power supply vector [20,18,25]}
\begin{tabular}{|l|l|l|l|l|l|l|l|l|l|l|l|l|} \hline
\text{Minimum}&$p_1$&$p_2$&$p_3$&$p_4$&$p_5$&$p_6$&$p_7$&$p_8$&$p_9$&$p_{10}$&$p_{11}$&$p_{12}$\\
\hline
1.4455 & 21.7314 & 19.3050 & 23.0071 & 19.9564 & -6 & -10 & -8 & -12 & -17 & -13 & -7 & -11
\\ \hline
\end{tabular}
 \label{RingNetworkNewMinimum1}
 \end{small}
\end{table}

 \begin{table}[H]
 \begin{small}
  \caption{Phase-angle differences and other outputs}
 \begin{tabular}{|ll|l|l|l|l|}\hline  $i$&$j$&$|\theta_i -\theta_j| $&$\sigma_{ij}$&$|\theta_i -\theta_j|+ 3.08~\sigma_{ij}$&$ l_{ij}~sin~(|\theta_i -\theta_j|+ 3.08~\sigma_{ij})$\\\hline 
4 & 10 & 0.6747 & 0.2503 & 1.4455 & 23.8119\\ 
3 & 9 & 0.6755 & 0.2500 & 1.4455 & 23.8118\\
 1 & 12 & 0.5742 & 0.2647 & 1.3896 & 23.6071\\ 
 2 & 7 & 0.5236 & 0.2496 & 1.2923 & 23.0753\\ 
 3 & 8 & 0.3398 & 0.2433 & 1.0891 & 21.2693\\ 
 1 & 5 & 0.3707 & 0.2229 & 1.0573 & 20.9044\\ 
 2 & 6 & 0.3093 & 0.2214 & 0.9913 & 20.0813\\ 
 4 & 11 & 0.2083 & 0.2502 & 0.9790 & 19.9190\\
  7 & 8 & 0.1675 & 0.2398 & 0.9060 & 18.8894\\ 
  11 & 12 & 0.0849 & 0.2490 & 0.8519 & 18.0606\\ 
  5 & 6 & 0.1125 & 0.2172 & 0.7814 & 16.9022\\ 
  9 & 10 & 0.0831 & 0.2265 & 0.7807 & 16.8903
 \\\hline
 \end{tabular}
  \label{RingNetworkNewPhaseAngleDifference1}
  \end{small}
 \end{table}

  \begin{table}[H]
  \begin{small}
  \caption{The minimum value and optimal power vector starting from the power supply vector [23,19,24]}
\begin{tabular} {|l|l|l|l|l|l|l|l|l|l|l|l|l|} \hline
\text{Minimum}&$p_1$&$p_2$&$p_3$&$p_4$&$p_5$&$p_6$&$p_7$&$p_8$&$p_9$&$p_{10}$&$p_{11}$&$p_{12}$\\
\hline
1.4455 & 21.6905 & 19.2546 & 23.0549 & 20.0000 & -6 & -10 & -8 & -12 & -17 & -13 & -7 & -11 
\\ \hline
\end{tabular}
 \label{RingNetworkNewMinimum2}
 \end{small}
\end{table}

 \begin{table}[H]
  \caption{Phase-angle differences and other outputs}
 \begin{tabular}{|ll|l|l|l|l|}\hline  $i$&$j$&$|\theta_i -\theta_j| $&$\sigma_{ij}$&$|\theta_i -\theta_j|+ 3.08~\sigma_{ij}$&$ l_{ij}~sin(|\theta_i -\theta_j|+ 3.08~\sigma_{ij})$\\\hline 
3 & 9 & 0.6755 & 0.2500 & 1.4455 & 23.8118\\ 4 & 10 & 0.6747 & 0.2502 & 1.4455 & 23.8118\\ 1 & 12 & 0.5720 & 0.2646 & 1.3870 & 23.5957\\ 2 & 7 & 0.5213 & 0.2495 & 1.2896 & 23.0577\\ 3 & 8 & 0.3419 & 0.2433 & 1.0914 & 21.2945\\ 1 & 5 & 0.3709 & 0.2229 & 1.0574 & 20.9063\\ 2 & 6 & 0.3091 & 0.2214 & 0.9911 & 20.0796\\ 4 & 11 & 0.2102 & 0.2502 & 0.9810 & 19.9447\\ 7 & 8 & 0.1655 & 0.2398 & 0.9039 & 18.8577\\ 11 & 12 & 0.0831 & 0.2490 & 0.8500 & 18.0306\\ 5 & 6 & 0.1127 & 0.2172 & 0.7815 & 16.9045\\ 9 & 10 & 0.0831 & 0.2265 & 0.7807 & 16.8899 
 \\\hline
 \end{tabular}
 \label{RingNetworkNewPhaseAngleDifference2}
 \end{table}

\newpage
\begin{table}[h]
  \caption{Phase-angle differences and other outputs for ring network without and with applying the procedure, the initial power supply vector is  [23,19,24]}
  \begin{footnotesize}
 \begin{tabular}{|l|ll|l|l|l|l|}\hline The procedure& $i$&$j$&$|\theta_i -\theta_j| $&$\sigma_{ij}$&$|\theta_i-\theta_j|$+ ~$3.08\:\sigma_{ij}$&$ l_{ij}\:sin\:(|\theta_i -\theta_j|$~$+ 3.08~\sigma_{ij})$\\\hline 
 without&1 & 12 & 0.7074 & 0.2753 & 1.5552 & 23.9971\\ 
with&1 & 12 & 0.5720 & 0.2646 & 1.3870 & 23.5957\\ \hline 
without&4 & 10 & 0.6902 & 0.2517 & 1.4656 & 23.8672\\ 
with&4 & 10 & 0.6747 & 0.2502 & 1.4455 & 23.8118\\ \hline 
without&3 & 9 & 0.6602 & 0.2492 & 1.4278 & 23.7549\\ 
with&3 & 9 & 0.6755 & 0.2500 & 1.4455 & 23.8118\\ \hline 
without&2 & 7 & 0.5534 & 0.2512 & 1.3272 & 23.2917\\ 
with&2 & 7 & 0.5213 & 0.2495 & 1.2896 & 23.0577\\ \hline 
without&1 & 5 & 0.4728 & 0.2268 & 1.1715 & 22.1121\\
with&1 & 5 & 0.3709 & 0.2229 & 1.0574 & 20.9063\\ \hline 
without&3 & 8 & 0.3128 & 0.2427 & 1.0602 & 20.9389\\ 
with&3 & 8 & 0.3419 & 0.2433 & 1.0914 & 21.2945\\ \hline 
without&11 & 12 & 0.1927 & 0.2510 & 0.9659 & 19.7410\\ 
with&11 & 12 & 0.0831 & 0.2490 & 0.8500 & 18.0306\\ \hline 
without&7 & 8 & 0.1935 & 0.2405 & 0.9341 & 19.2974\\ 
with&7 & 8 & 0.1655 & 0.2398 & 0.9039 & 18.8577\\ \hline 
without&2 & 6 & 0.2129 & 0.2193 & 0.8882 & 18.6222\\ 
with&2 & 6 & 0.3091 & 0.2214 & 0.9911 & 20.0796\\ \hline 
without&5 & 6 & 0.2069 & 0.2188 & 0.8808 & 18.5096\\ 
with&5 & 6 & 0.1127 & 0.2172 & 0.7815 & 16.9045\\ \hline 
without&4 & 11 & 0.1003 & 0.2502 & 0.8708 & 18.3566\\ 
with&4 & 11 & 0.2102 & 0.2502 & 0.9810 & 19.9447\\ \hline 
without&9 & 10 & 0.0952 & 0.2267 & 0.7934 & 17.1059 \\
with&9 & 10 & 0.0831 & 0.2265 & 0.7807 & 16.8899 
 \\\hline
 \end{tabular}
 \end{footnotesize}
 \label{RingNetworkNewPhaseAngleDifference3}
 \end{table}
 
  \begin{table}[h]
\caption{Tail Probability of the ring network according to Table \ref{RingNetworkNewPhaseAngleDifference3}}
\scalebox{0.85}{
  \begin{tabular}{|l|ll|l|l|l|l|}
  \hline
The procdure&i&j&  $P(\omega\in \Omega~|~\theta_{ij}(\omega,t)<-\pi/2)$&$P(\omega\in \Omega~|~\theta_{ij}(\omega,t)>\pi/2)$&\begin{tabular}{l}\mbox{by our criterion}\\ \mbox{(Upper bound)}\end{tabular}\\\hline
without&1 & 12&6.4054e-17 & 8.5576e-04&\\
with&1 & 12&2.7878e-16 & 8.0087e-05 & $2\times 1.9005e-04$\\\hline
 without&4 & 10&1.3186e-19 & 2.3386e-04&\\
with&4 & 10& 11.4193e-19 & 1.7080e-04 &$2\times  1.7128e-04$\\\hline
without& 3 &9& 1.7355e-19 & 1.2905e-04&\\
with& 3 & 9& 1.2915e-19 & 1.7102e-04 &$2\times 1.7102e-04$\\\hline 
 without&2 & 7& 1.3812e-17 & 2.5593e-05&\\ 
 with&2 & 7& 2.5329e-17 & 1.2974e-05 &$2\times 1.7036e-04$\\\hline
  without&1 & 5&1.0250e-19 & 6.4511e-07&\\ 
with&1 & 5&   1.5052e-18 & 3.6604e-08 &$2\times 1.3520e-04$\\\hline 
 without&3 & 8& 4.2129e-15 & 1.0897e-07&\\
with& 3 & 8&   1.8982e-15 & 2.1981e-07 &$2\times 1.6220e-04$\\\hline
 without& 11 & 12& 2.0049e-08 & 1.0636e-12&\\ 
with&11 & 12&   1.1527e-09 & 1.5457e-11 &$2\times 1.6971e-04$\\\hline
without&  7 & 8& 1.1008e-13 & 5.1170e-09&\\ 
 with&7 & 8&   2.2335e-13 & 2.3104e-09 &$2\times 1.5758e-04$\\\hline
  without&2 & 6& 2.0838e-16 & 2.9711e-10&\\ 
 with&2 & 6&  1.0248e-17 & 6.0356e-09 &$2\times 1.3321e-04$\\\hline
without& 5 & 6&  2.2413e-16 & 2.2802e-10&\\ 
with&5 & 6&   4.5620e-15 & 9.5233e-12 &$2\times 1.2766e-04$\\\hline 
 without&4 & 11&  1.2025e-11 & 2.0853e-09&\\ 
 with&4 & 11&  5.4637e-13 & 2.6936e-08 &$2\times 1.7128e-04$\\\hline 
without&  9 & 10& 3.7820e-11 & 9.9920e-14 &\\
with&9 & 10&2.5464e-11 & 1.4179e-13 &$2\times 1.3997e-04$\\\hline
\end{tabular}}
\label{TailProbabilityoftheringnetwork}
\end{table}

\newpage
\section{Tables of a Manhattan-grid Network Example}\label{sec:table3}
\begin{figure}[h]
\centering
\begin{tikzpicture}[tbcircle/.style={circle,minimum size=13pt,inner sep=0pt},
forward tbcircle/.style={tbcircle,fill=blue!70},
backward tbcircle/.style={tbcircle,fill=red!100},
tbsquare/.style={rectangle,rounded corners,minimum width=15pt,minimum height=15pt,fill=orange!50},
tbcarrow/.style={/utils/exec=\tikzset{tbcarrow pars/.cd,#1},
line width=0.1*\pgfkeysvalueof{/tikz/tbcarrow pars/width},
draw=\pgfkeysvalueof{/tikz/tbcarrow pars/draw},
-{Triangle[line width=\pgfkeysvalueof{/tikz/tbcarrow pars/line width},
length=1.1*\pgfkeysvalueof{/tikz/tbcarrow pars/width},
width=1.1*\pgfkeysvalueof{/tikz/tbcarrow pars/width},
fill=\pgfkeysvalueof{/tikz/tbcarrow pars/fill}]},
postaction={draw=\pgfkeysvalueof{/tikz/tbcarrow pars/fill},-,
line width=\pgfkeysvalueof{/tikz/tbcarrow pars/width}-5*\pgfkeysvalueof{/tikz/tbcarrow pars/line width},
shorten <=\pgfkeysvalueof{/tikz/tbcarrow pars/line width},
shorten >=1.25*\pgfkeysvalueof{/tikz/tbcarrow pars/width}-1.5*\pgfkeysvalueof{/tikz/tbcarrow pars/line width}}
},
tbcarrow pars/.cd,
fill/.initial=white,draw/.initial=black,width/.initial=4pt,
line width/.initial=0.4pt,]

\node[forward tbcircle,draw]  at (0,0) {$\textcolor{white}{21}$};
\node[forward tbcircle,draw]  at (1,0) {$\textcolor{white}{22}$};
\node[forward tbcircle,draw]  at (2,0) {$\textcolor{white}{23}$};
\node[forward tbcircle,draw]  at (3,0) {$\textcolor{white}{24}$};
\node[forward tbcircle,draw]  at (4,0) {$\textcolor{white}{25}$};

\node[forward tbcircle,draw]  at (0,1) {$\textcolor{white}{17}$};
\node[forward tbcircle,draw]  at (1,1) {$\textcolor{white}{18}$};
\node[backward tbcircle,draw]  at (2,1) {$\textcolor{white}{4}$};
\node[forward tbcircle,draw]  at (3,1) {$\textcolor{white}{19}$};
\node[forward tbcircle,draw]  at (4,1) {$\textcolor{white}{20}$};

\node[forward tbcircle,draw]  at (0,2) {$\textcolor{white}{14}$};
\node[backward tbcircle,draw]  at (1,2) {$\textcolor{white}{2}$};
\node[forward tbcircle,draw]  at (2,2) {$\textcolor{white}{15}$};
\node[backward tbcircle,draw]  at (3,2) {$\textcolor{white}{3}$};
\node[forward tbcircle,draw]  at (4,2) {$\textcolor{white}{16}$};

\node[forward tbcircle,draw]  at (0,3) {$\textcolor{white}{10}$};
\node[forward tbcircle,draw]  at (1,3) {$\textcolor{white}{11}$};
\node[backward tbcircle,draw]  at (2,3) {$\textcolor{white}{1}$};
\node[forward tbcircle,draw]  at (3,3) {$\textcolor{white}{12}$};
\node[forward tbcircle,draw]  at (4,3) {$\textcolor{white}{13}$};

\node[forward tbcircle,draw]  at (0,4) {$\textcolor{white}{5}$};
\node[forward tbcircle,draw]  at (1,4) {$\textcolor{white}{6}$};
\node[forward tbcircle,draw]  at (2,4) {$\textcolor{white}{7}$};
\node[forward tbcircle,draw]  at (3,4) {$\textcolor{white}{8}$};
\node[forward tbcircle,draw]  at (4,4) {$\textcolor{white}{9}$};

\draw[tbcarrow={fill=black!70}] (0+0.2,0) to (1-0.2,0);
\draw[tbcarrow={fill=black!70}] (1+0.2,0) to (2-0.2,0);
\draw[tbcarrow={fill=black!70}] (2+0.2,0) to (3-0.2,0);
\draw[tbcarrow={fill=black!70}] (3+0.2,0) to (4-0.2,0);

\draw[tbcarrow={fill=black!70}] (0+0.2,1) to (1-0.2,1);
\draw[tbcarrow={fill=black!70}] (2-0.2,1)  to (1+0.2,1);
\draw[tbcarrow={fill=black!70}] (2+0.2,1) to (3-0.2,1);
\draw[tbcarrow={fill=black!70}] (3+0.2,1) to (4-0.2,1);

\draw[tbcarrow={fill=black!70}] (1-0.2,2) to (0+0.2,2);
\draw[tbcarrow={fill=black!70}] (1+0.2,2) to (2-0.2,2);
\draw[tbcarrow={fill=black!70}]  (3-0.2,2) to (2+0.2,2);
\draw[tbcarrow={fill=black!70}] (3+0.2,2) to (4-0.2,2);

\draw[tbcarrow={fill=black!70}] (0+0.2,3) to (1-0.2,3);
\draw[tbcarrow={fill=black!70}] (2-0.2,3) to (1+0.2,3);
\draw[tbcarrow={fill=black!70}] (2+0.2,3) to (3-0.2,3);
\draw[tbcarrow={fill=black!70}] (3+0.2,3) to (4-0.2,3);

\draw[tbcarrow={fill=black!70}] (0+0.2,4) to (1-0.2,4);
\draw[tbcarrow={fill=black!70}] (1+0.2,4) to (2-0.2,4);
\draw[tbcarrow={fill=black!70}] (2+0.2,4) to (3-0.2,4);
\draw[tbcarrow={fill=black!70}] (3+0.2,4) to (4-0.2,4);
\draw[tbcarrow={fill=black!70}] (0,0+0.2) to (0,1-0.2);
\draw[tbcarrow={fill=black!70}] (0,1+0.2) to (0,2-0.2);
\draw[tbcarrow={fill=black!70}] (0,2+0.2) to (0,3-0.2);
\draw[tbcarrow={fill=black!70}] (0,3+0.2) to (0,4-0.2);

\draw[tbcarrow={fill=black!70}] (1,0+0.2) to (1,1-0.2);
\draw[tbcarrow={fill=black!70}] (1,2-0.2) to (1,1+0.2);
\draw[tbcarrow={fill=black!70}] (1,2+0.2) to (1,3-0.2);
\draw[tbcarrow={fill=black!70}] (1,3+0.2) to (1,4-0.2);

\draw[tbcarrow={fill=black!70}] (2,1-0.2) to (2,0+0.2);
\draw[tbcarrow={fill=black!70}] (2,1+0.2) to (2,2-0.2);
\draw[tbcarrow={fill=black!70}] (2,2+0.2) to (2,3-0.2);
\draw[tbcarrow={fill=black!70}] (2,3+0.2) to (2,4-0.2);

\draw[tbcarrow={fill=black!70}] (3,0+0.2) to (3,1-0.2);
\draw[tbcarrow={fill=black!70}] (3,2-0.2) to (3,1+0.2);
\draw[tbcarrow={fill=black!70}] (3,2+0.2) to (3,3-0.2);
\draw[tbcarrow={fill=black!70}] (3,3+0.2) to (3,4-0.2);

\draw[tbcarrow={fill=black!70}] (4,0+0.2) to (4,1-0.2);
\draw[tbcarrow={fill=black!70}] (4,1+0.2) to (4,2-0.2);
\draw[tbcarrow={fill=black!70}] (4,2+0.2) to (4,3-0.2);
\draw[tbcarrow={fill=black!70}] (4,3+0.2) to (4,4-0.2);

\end{tikzpicture}
\caption{A Manhattan-grid  network}
\label{fig:manhattengrid}
\end{figure}
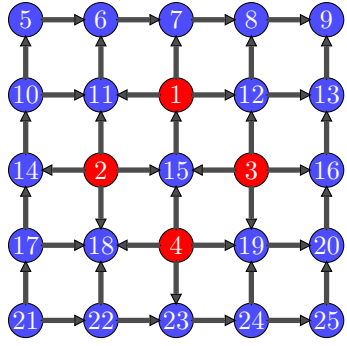

The picture of a Manhattan-grid  power network with 25 nodes is displayed in Fig. \ref{fig:manhattengrid} where the nodes $1,2,3,4$ provide power supply and the other nodes model power demand. Table~\ref{ManhattanGridNetworkParameter} shows the parameters of this network. In this example, the symmetric parameters lead to a symmetric optimal power supply vector which is shown in Table~\ref{ManhattanGridMinimumValue1}. Then, we change the power demand vector from a symmetric one to an asymmetric one, which leads to an asymmetric optimal power supply vector in Table \ref{ManhattanGridMinimumValue2}. What's more, when the vector of disturbance is changed from a symmetric one to an asymmetric one, the power supply also becomes asymmetric, as shown in Table \ref{ManhattanGridMinimumValue3}.

 \par Tables~\ref{ManhattanGridPhaseAngleDifferences1}, \ref{ManhattanGridPhaseAngleDifferences2}, and \ref{ManhattanGridPhaseAngleDifferences3} show the absolute value of mean of phase-angle difference: $|\theta_i-\theta_j|$, the standard deviation of the phase-angle difference of this power line: $\sigma_{ij}$, each component of the control objective vector: $|\theta_i-\theta_j|+3.08\sigma_{ij}$ and line power flow $l_{ij}sin(|\theta_i-\theta_j|+3.08\sigma_{ij})$, and these three tables are ordered by the values of $|\theta_i-\theta_j|+3.08\sigma_{ij}$ in the descending order. As in the particular eight-node academic network, in contrast with the results in Table~\ref{ManhattanGridMinimumValue2}, the computation has been additionally evaluated by discretizing the set of three power supply vectors for each dimension into a grid with 150 steps and the minimum is $1.3416$, which is slightly larger than the value computed by our proposed procedure. Meanwhile, for this power network with large amount of nodes and links, our proposed procedure has a higher efficiency than this partition method.
\par What's more, from Tables~\ref{ManhattanGridPhaseAngleDifferences1}, \ref{ManhattanGridPhaseAngleDifferences2}, and \ref{ManhattanGridPhaseAngleDifferences3}, one can see the total power demand of the neighborhood of a power supply node is much larger than the capacity of the line which connects the neighborhood to this power supply node, e.g.  the neighborhood: node $5,6,7,8,9$ and the connection line $1-7$. If this power network is a tree-like network, then It will experience an immediate power outage. But for this power network, it remains very stable, which is due to its network structure and the number of links. 
\par Furthermore, as for the particular eight-node example, the largest value and the second largest value of the vector $|\theta_i-\theta_j|+3.08\sigma_{ij}$ differ very little. Moreover, the computations from several different initial power supply vectors are considered, and in each case we have computed, the iteration sequence converged to minimizer which is quite close to each other. 
\par
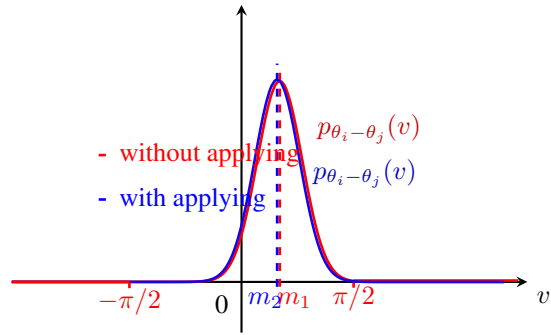
\begin{figure}\label{fig:Manhattannetcontrolwithoutwith}
\begin{center}
\begin{tikzpicture}
 \message{Expected sensitivty & p-value^^J}
  \def\N{100}
  \def\q{0.5363}
   \def\qn{0.4990}
  \def\Bs{2}
  \def\S{\q}
   \def\Sn{\qn}
   \def\Ss{0.3149}
  \def\Ssn{0.3128}   
     \def\xmin{-6*\q}
  \def\xmax{5.2*\q}
  \def\ymin{{-0.16/\Bs}}
  \def\ymax{{0.78/\Bs}}
   \begin{axis}[every axis plot post/.append style={
               mark=none,domain=\xmin:{1.32*\xmax},samples=\N,smooth},
               xmin=\xmin, xmax=1.20*\xmax,
               ymin=\ymin, ymax=\ymax,
               axis lines=middle,
               axis line style=thick,
               enlargelimits=upper, 
               ticks=none,
               xlabel=$v$,
               every axis x label/.style={at={(current axis.right of origin)},anchor=north west},
               y=240pt,
              ]
   \addplot[name path=S,line width=0.35mm,red!90!red ] {gauss(x,\S,\Ss)};
         \addplot[red,dashed,line width=0.35mm]
      coordinates {(\q,{-0.02*exp(-\q/\Bs)/\Bs}) (\q, {1.08*gauss(\q,\S,\Ss)})}
       node[below=-4pt,right=-5pt,pos=-0.06] {$m_1$};
    \addplot[red,line width=0.35mm]
      coordinates {(1.57,{-0.04*exp(-1.57/\Bs)/\Bs}) (1.57, {1.38*gauss(1.57,\S,\Ss)})}
       node[below=-5pt,pos=0] {$\pi/2$};
          \addplot[red,line width=0.35mm]
      coordinates {(1.57, {1.5*gauss(1.57,\S,\Ss)}) (1.39*\xmax, {1.5*gauss(1.57,\S,\Ss)})};
  \addplot[name path=S2,line width=0.35mm,blue!90!red ] {gauss(x,\Sn,\Ssn)};
   \addplot[blue,dashed,line width=0.35mm]
      coordinates {(\qn,{-0.02*exp(-\qn/\Bs)/\Bs}) (\qn, {1.08*gauss(\qn,\Sn,\Ssn)})}
       node[below=1pt,left=-6pt,pos=-0.06] {$m_2$};

       \addplot[red,line width=0.35mm]
      coordinates {(-1.57,{-0.009*exp(1.57/\Bs)/\Bs}) (-1.57, {10.5*gauss(-1.57,\S,\Ss)})}
       node[below=-4pt,pos=0] {$-\pi/2$};    
              \addplot[red,line width=0.35mm]
      coordinates {(\xmin,{2.5*gauss(-1.57,\S,\Ss)}) (-1.57, {2.5*gauss(-1.57,\S,\Ss)})};

    \path[name path=xaxis1]
     (0,0) -- (1.08*\xmax,0);
    \addplot[red!60!blue] fill between[of=xaxis1 and S, soft clip={domain=1.57:1.50*\xmax}];
       \path[name path=xaxis2]
     (-1.08*\xmax,0) -- (0,0);
    \addplot[red!60!blue] fill between[of=xaxis2 and S, soft clip={domain=-1.08*\xmax:-1.57}];
    \node[below left=2pt] at (0,0) {0};

    \node[above right=1.3pt,black!20!blue] at ( 0.42*\Bs,0.13) {$p_{\theta_i - \theta_j}(v)$};
     \node[above right=1.3pt,black!20!red] at ( 0.45*\Bs,0.20) {$p_{\theta_i - \theta_j}(v)$};
     \addplot[red,line width=0.35mm]
      coordinates {(-2.0,0.2) (-1.9, 0.2)}
       node[right=4pt,pos=0] {without applying};    
       \addplot[blue,line width=0.35mm]
      coordinates {(-2.0,0.13) (-1.9, 0.13)}
       node[right=4pt,pos=0] {with applying};

  \end{axis}
  
  \end{tikzpicture}
\end{center}
\caption{Example. Manhattan-grid network, line 1-7, without and with control, disturbance vector asymmetric.}
\end{figure}

Results of with and without applying the proposed procedure are shown in Tables~\ref{ManhattanGridPhaseAngleDifferencessy1} and \ref{ManhattanGridPhaseAngleDifferencessy2} for symmetric power demand vector and symmetric disturbance vector. Then we change the strength of disturbance to
be asymmetric, the results are shown in Tables~\ref{ManhattanGridPhaseAngleDifferences11} and \ref{ManhattanGridPhaseAngleDifferences22}.
These four tables are ordered by the values of $|\theta_i-\theta_j|+3.08*\sigma_{ij}$ in tables of results without applying the proposed procedure. We can see the largest 
and the second largest values of $|\theta_i-\theta_j|+3.08*\sigma_{ij}$ decrease while the other values go up and down.
A figure depicting the probability density function for the cases `without applying' and `with applying' of line $1-7$ where the disturbance vectors are asymmetric is presented, as illustrated in Fig.~\ref{fig:Manhattannetcontrolwithoutwith}. Note that the values are borrowed from in Table~\ref{ManhattanGridPhaseAngleDifferences11}. In this illustration, the reduction in both standard deviation and mean value from the `without applying' to the `with applying' condition is not readily apparent. This is attributed to the small magnitude of the decrease, making it challenging to discern in the visual representation. 
\par
Finally, we show the tail probabilities according to Tables~\ref{ManhattanGridPhaseAngleDifferencessy1} and \ref{ManhattanGridPhaseAngleDifferencessy2} in Tables \ref{TailProbabilityoftheManhattannetworkaccordingtoTablesymmetric1} and \ref{TailProbabilityoftheManhattannetworkaccordingtoTablesymmetric2}, and those according to Tables~\ref{ManhattanGridPhaseAngleDifferences11} and \ref{ManhattanGridPhaseAngleDifferences22} are shown in Tables \ref{TailProbabilityoftheManhattannetworkaccordingtoTableasymmetric1} and \ref{TailProbabilityoftheManhattannetworkaccordingtoTableasymmetric2}. From these tables, one can see the probability of the two most vulnerable lines which go outside the safety region becomes smaller, which means they are more stable than without applying the proposed procedure.
\newpage

\begin{table}[H]
  \caption{The minimum value and optimal power vector starting from the power supply vector [45,53,55] for a symmetric power demand vector, and a symmetric disturbance vector}
  \begin{footnotesize}
\begin{tabular} {|l|l|l|l|l|l|l|l|l|l|l|l|l|} \hline
\text{Minimum}&$p_1$&$p_2$&$p_3$&$p_4$&$p_5$&$p_6$&$p_7$&$p_8$&$p_9$&$p_{10}$&$p_{11}$&$p_{12}$\\
\hline
1.4253 & 45.0000 & 60.0000 & 60.0000 & 45.0000 & -9.5000 & -14 & -8 & -14 & -9.5 & -8 & -8 & -8\\
\hline\hline
$p_{13}$&$p_{14}$&$p_{15}$&$p_{16}$&$p_{17}$&$p_{18}$&$p_{19}$&$p_{20}$&$p_{21}$&$p_{22}$&$p_{23}$&$p_{24}$&$p_{25}$\\
\hline
 -8 & -14 & -8 & -14 & -8 & -8 & -8 & -8 & -9.5 & -14 & -8 & -14 & -9.5 \\
 \hline
\end{tabular}
\label{ManhattanGridMinimumValue1}
\end{footnotesize}
\end{table}

 \begin{table}[H]
  \caption{The minimum value and optimal power vector starting from the power supply vector  [45,53,55] for an asymmetric power demand vector, and a symmetric disturbance vector}
   \begin{footnotesize}
\begin{tabular} {|l|l|l|l|l|l|l|l|l|l|l|l|l|} \hline
\text{Minimum}&$p_1$&$p_2$&$p_3$&$p_4$&$p_5$&$p_6$&$p_7$&$p_8$&$p_9$&$p_{10}$&$p_{11}$&$p_{12}$\\
\hline
1.3413 & 43.1692 & 60.0000 & 60.0000 & 46.8308 & -5 & -8 & -8 & -10 & -8 & -10 & -14 & -16 \\
\hline\hline
$p_{13}$&$p_{14}$&$p_{15}$&$p_{16}$&$p_{17}$&$p_{18}$&$p_{19}$&$p_{20}$&$p_{21}$&$p_{22}$&$p_{23}$&$p_{24}$&$p_{25}$\\
\hline
-8 & -14 & -14 & -14 & -8 & -14 & -14 & -8 & -8 & -8 & -8 & -8 & -5
\\ \hline
\end{tabular}
\label{ManhattanGridMinimumValue2}
\end{footnotesize}
\end{table}

 \begin{table}[H]
  \caption{The minimum value and optimal power vector starting from the power supply vector  [45,53,55] for a symmetric power demand vector, and an asymmetric disturbance vector}
   \begin{footnotesize}
\begin{tabular} {|l|l|l|l|l|l|l|l|l|l|l|l|l|} \hline
\text{Minimum}&$p_1$&$p_2$&$p_3$&$p_4$&$p_5$&$p_6$&$p_7$&$p_8$&$p_9$&$p_{10}$&$p_{11}$&$p_{12}$\\
\hline
1.4625 & 44.2792 & 60.0000 & 60.0000 & 45.7208 & -9.5000 & -14 & -8 & -14 & -9.5000 & -8 & -8 & -8 \\
\hline\hline
$p_{13}$&$p_{14}$&$p_{15}$&$p_{16}$&$p_{17}$&$p_{18}$&$p_{19}$&$p_{20}$&$p_{21}$&$p_{22}$&$p_{23}$&$p_{24}$&$p_{25}$\\
\hline
-8 & -14 & -8 & -14 & -8 & -8 & -8 & -8 & -9.5000 & -14 & -8 & -14 & -9.5000
\\ \hline
\end{tabular}
\label{ManhattanGridMinimumValue3}
\end{footnotesize}
\end{table}

\newpage
 \begin{table}[h]
  \caption{Network parameters in which the line capacity is not listed there, where all line capacities equal 40}
  \begin{small}
\begin{tabular} {|l|l|l|l|l|l|l|}\hline Node &$\begin{array}{l}p^{+,max}\\-p^{-,max}\end{array}$&$\begin{array}{l}p^{+,max}\\-p^{-,max}\end{array}$ &Inertia&Damping&Disturbance &Disturbance \\
$i$&symmetric&asymmetric&$m_i$&$d_i$&$\begin{array}{l}\mbox{symmetric}\\ K_2(i,i)\end{array}$&$\begin{array}{l}\mbox{asymmetric} \\K_2(i,i)\end{array}$
\\\hline1&60&60&10&4&2&1.5682
\\\hline2&60&60&10&4&2& 2.0387
\\\hline3&60&60&10&4&2& 1.3929
\\\hline4&60&60&10&4&2& 1.4432
\\\hline5&-9.5&-5&1&0&2& 1.9966
\\\hline6&-14&-8&1&0&2& 2.7663
\\\hline7&-8&-8&1&0&2& 1.6152
\\\hline8&-14&-10&1&0&2& 2.1857
\\\hline9&-9.5&-8&1&0&2& 1.8872 
\\\hline10&-8&-10&1&1&2& 2.5587
\\\hline11&-8&-14&1&1&2& 1.4555
\\\hline12&-8&-16&1&1&2& 2.0163
\\\hline13&-8&-8&1&1&2& 2.2763
\\\hline14&-14&-14&1&1&2& 2.5503
\\\hline15&-8&-14&1&1&2& 2.7721
\\\hline16&-14&-14&1&1&2& 2.0430
\\\hline17&-8&-8&1&1&2 & 1.2542
\\\hline18&-8&-14&1&1&2& 1.6288
\\\hline19&-8&-14&1&1&2& 1.4692
\\\hline20&-8&-8&1&1&2& 3.1752
\\\hline21&-9.5&-8&1&0&2& 1.6922
\\\hline22&-14&-8&1&0&2& 2.3740
\\\hline23&-8&-8&1&0&2& 1.9038
\\\hline24&-14&-8&1&0&2& 2.4443
\\\hline25&-9.5&-5&1&0&2& 1.6176
\\
\hline
\end{tabular}
\end{small}
\label{ManhattanGridNetworkParameter}
\end{table}

\clearpage
\par
 \begin{table}[H]
 \begin{small}
  \caption{Phase-angle differences and other outputs, power demand vector symmetric, disturbance vector symmetric}
 \begin{tabular}{|ll|l|l|l|l|}\hline  $i$&$j$&$|\theta_i -\theta_j| $&$\sigma_{ij}$&$|\theta_i-\theta_j|+ 3.08\:\sigma_{ij}$&$ l_{ij}\:sin\:(|\theta_i -\theta_j|+ 3.08~\sigma_{ij})$\\\hline 
1 & 7 & 0.5018 & 0.2998 & 1.4253 & 39.5773\\\hline 4 & 23 & 0.5018 & 0.2998 & 1.4253 & 39.5773\\\hline 2 & 14 & 0.5932 & 0.2418 & 1.3379 & 38.9200\\\hline 3 & 16 & 0.5932 & 0.2418 & 1.3379 & 38.9200\\\hline 2 & 18 & 0.4343 & 0.2160 & 1.0997 & 35.6429\\\hline3 & 19 & 0.4343 & 0.2160 & 1.0997 & 35.6429\\\hline 2 & 11 & 0.4343 & 0.2160 & 1.0997 & 35.6429\\\hline 3 & 12 & 0.4343 & 0.2160 & 1.0997 & 35.6429\\\hline 8 & 9 & 0.0904 & 0.3246 & 1.0900 & 35.4652\\\hline 5 & 6 & 0.0904 & 0.3246 & 1.0900 & 35.4652\\\hline 24 & 25 & 0.0904 & 0.3246 & 1.0900 & 35.4652\\\hline 21 & 22 & 0.0904 & 0.3246 & 1.0900 & 35.4652\\\hline 18 & 22 & 0.3044 & 0.2399 & 1.0434 & 34.5643\\\hline 19 & 24 & 0.3044 & 0.2399 & 1.0434 & 34.5643\\\hline 6 & 11 & 0.3044 & 0.2399 & 1.0434 & 34.5643\\\hline 8 & 12 & 0.3044 & 0.2399 & 1.0434 & 34.5643\\\hline 7 & 8 & 0.1410 & 0.2905 & 1.0358 & 34.4114\\\hline 6 & 7 & 0.1410 & 0.2905 & 1.0358 & 34.4114\\\hline 23 & 24 & 0.1410 & 0.2905 & 1.0358 & 34.4114\\\hline 22 & 23 & 0.1410 & 0.2905 & 1.0358 & 34.4114\\\hline 1 & 12 & 0.3276 & 0.2221 & 1.0116 & 33.9063\\\hline 1 & 11 & 0.3276 & 0.2221 & 1.0116 & 33.9063\\\hline 4 & 19 & 0.3276 & 0.2221 & 1.0116 & 33.9063\\\hline4 & 18 & 0.3276 & 0.2221 & 1.0116 & 33.9063\\\hline 12 & 13 & 0.2452 & 0.2468 & 1.0055 & 33.7763\\\hline 10 & 11 & 0.2452 & 0.2468 & 1.0055 & 33.7763\\\hline 19 & 20 & 0.2452 & 0.2468 & 1.0055 & 33.7763\\\hline 17 & 18 & 0.2452 & 0.2468 & 1.0055 & 33.7763\\\hline 17 & 21 & 0.1478 & 0.2699 & 0.9792 & 33.2026\\\hline 20 & 25 & 0.1478 & 0.2699 & 0.9792 & 33.2026\\\hline 5 & 10 & 0.1478 & 0.2699 & 0.9792 & 33.2026\\\hline 9 & 13 & 0.1478 & 0.2699 & 0.9792 & 33.2026\\\hline 14 & 17 & 0.1047 & 0.2622 & 0.9123 & 31.6369\\\hline 16 & 20 & 0.1047 & 0.2622 & 0.9123 & 31.6369\\\hline 10 & 14 & 0.1047 & 0.2622 & 0.9123 & 31.6369\\\hline 13 & 16 & 0.1047 & 0.2622 & 0.9123 & 31.6369\\\hline 2 & 15 & 0.0997 & 0.1506 & 0.5635 & 21.3651\\\hline 3 & 15 & 0.0997 & 0.1506 & 0.5635 & 21.3651\\\hline 1 & 15 & 5.0008e-04 & 0.1625 & 0.5010 & 19.2139\\\hline 4 & 15 & 4.9992e-04 & 0.1625 & 0.5010 & 19.2139\\\hline &  & & & $1.57 \approx \pi/2$ &
 \\\hline
 \end{tabular}
 \label{ManhattanGridPhaseAngleDifferences1}
 \end{small}
 \end{table}

\newpage
 \begin{table}[H]
 \begin{small}
  \caption{Phase-angle differences and other outputs, power demand vector asymmetric, disturbance vector symmetric}
 \begin{tabular}{|ll|l|l|l|l|}\hline  $i$&$j$&$|\theta_i -\theta_j| $&$\sigma_{ij}$&$|\theta_i-\theta_j|+ 3.08~\sigma_{ij}$&$ l_{ij}~sin~(|\theta_i -\theta_j|+ 3.08~\sigma_{ij})$\\\hline 
1 & 7 & 0.4285 & 0.2964 & 1.3413 & 38.9515 \\\hline 4 & 23 & 0.4285 & 0.2964 & 1.3413 & 38.9515 \\\hline 2 & 14 & 0.5614 & 0.2400 & 1.3004 & 38.5470 \\\hline 3 & 16 & 0.5547 & 0.2397 & 1.2929 & 38.4649 \\\hline 3 & 12 & 0.4742 & 0.2169 & 1.1423 & 36.3841 \\\hline 21 & 22 & 0.1100 & 0.3248 & 1.1103 & 35.8330 \\\hline 2 & 11 & 0.4453 & 0.2159 & 1.1102 & 35.8319 \\\hline 8 & 9 & 0.0910 & 0.3239 & 1.0886 & 35.4392 \\\hline 2 & 18 & 0.4155 & 0.2161 & 1.0810 & 35.2966 \\\hline 24 & 25 & 0.0722 & 0.3241 & 1.0704 & 35.0949 \\\hline 5 & 6 & 0.0703 & 0.3235 & 1.0668 & 35.0266 \\\hline 3 & 19 & 0.3984 & 0.2156 & 1.0624 & 34.9418 \\\hline 4 & 18 & 0.3529 & 0.2219 & 1.0363 & 34.4201 \\\hline1 & 12 & 0.3477 & 0.2226 & 1.0333 & 34.3590 \\\hline4 & 19 & 0.3410 & 0.2215 & 1.0231 & 34.1494 \\\hline 7 & 8 & 0.1334 & 0.2885 & 1.0219 & 34.1242 \\\hline 22 & 23 & 0.1203 & 0.2884 & 1.0086 & 33.8443 \\\hline 1 & 11 & 0.3154 & 0.2216 & 0.9980 & 33.6155 \\\hline 23 & 24 & 0.0957 & 0.2880 & 0.9827 & 33.2808 \\\hline17 & 18 & 0.2110 & 0.2461 & 0.9691 & 32.9745 \\\hline 6 & 7 & 0.0826 & 0.2876 & 0.9683 & 32.9573 \\\hline 10 & 11 & 0.2046 & 0.2459 & 0.9619 & 32.8115 \\\hline 19 & 20 & 0.1971 & 0.2457 & 0.9539 & 32.6265 \\\hline 12 & 13 & 0.1907 & 0.2457 & 0.9475 & 32.4785 \\\hline 8 & 12 & 0.2094 & 0.2375 & 0.9407 & 32.3197 \\\hline 9 & 13 & 0.1094 & 0.2690 & 0.9378 & 32.2493 \\\hline 13 & 16 & 0.1199 & 0.2614 & 0.9251 & 31.9470 \\\hline 18 & 22 & 0.1910 & 0.2371 & 0.9213 & 31.8551 \\\hline 6 & 11 & 0.1889 & 0.2372 & 0.9193 & 31.8082 \\\hline 17 & 21 & 0.0903 & 0.2689 & 0.9184 & 31.7852 \\\hline 19 & 24 & 0.1775 & 0.2369 & 0.9071 & 31.5079 \\\hline 10 & 14 & 0.1018 & 0.2614 & 0.9070 & 31.5063 \\\hline 14 & 17 & 0.0808 & 0.2611 & 0.8849 & 30.9547 \\\hline 5 & 10 & 0.0548 & 0.2682 & 0.8809 & 30.8526 \\\hline 20 & 25 & 0.0529 & 0.2683 & 0.8792 & 30.8086 \\\hline 16 & 20 & 0.0571 & 0.2609 & 0.8605 & 30.3279 \\\hline 2 & 15 & 0.1337 & 0.1507 & 0.5978 & 22.5138 \\\hline 3 & 15 & 0.1291 & 0.1507 & 0.5934 & 22.3657 \\\hline 4 & 15 & 0.0753 & 0.1621 & 0.5744 & 21.7340 \\\hline1 & 15 & 0.0128 & 0.1624 & 0.5131 & 19.6341
 \\\hline
 \end{tabular}
 \label{ManhattanGridPhaseAngleDifferences2}
 \end{small}
 \end{table}
 
 \newpage
 \begin{table}[H]
 \begin{small}
  \caption{Phase-angle differences and other outputs, power demand vector symmetric, disturbance vector asymmetric}
 \begin{tabular}{|ll|l|l|l|l|}\hline  $i$&$j$&$|\theta_i -\theta_j| $&$\sigma_{ij}$&$|\theta_i-\theta_j|+ 3.08~\sigma_{ij}$&$ l_{ij}~sin~(|\theta_i -\theta_j|+ 3.08~\sigma_{ij})$\\\hline 
1 & 7 & 0.4990 & 0.3128 & 1.4625 & 39.7656\\\hline  4 & 23 & 0.5046 & 0.3110 & 1.4625 & 39.7656\\\hline  3 & 16 & 0.5932 & 0.2601 & 1.3943 & 39.3787\\\hline  2 & 14 & 0.5932 & 0.2600 & 1.3941 & 39.3770\\\hline  5 & 6 & 0.0896 & 0.3416 & 1.1416 & 36.3717\\\hline  8 & 9 & 0.0896 & 0.3381 & 1.1310 & 36.1934\\\hline  3 & 12 & 0.4379 & 0.2233 & 1.1255 & 36.0993\\\hline  2 & 11 & 0.4379 & 0.2214 & 1.1199 & 36.0019\\\hline  19 & 24 & 0.3040 & 0.2628 & 1.1135 & 35.8899\\\hline  24 & 25 & 0.0912 & 0.3307 & 1.1099 & 35.8256\\\hline  6 & 11 & 0.3049 & 0.2597 & 1.1046 & 35.7320\\\hline  3 & 19 & 0.4307 & 0.2180 & 1.1022 & 35.6882\\\hline  21 & 22 & 0.0912 & 0.3279 & 1.1012 & 35.6696\\\hline  6 & 7 & 0.1397 & 0.3104 & 1.0959 & 35.5730\\\hline  8 & 12 & 0.3049 & 0.2547 & 1.0895 & 35.4555\\\hline  23 & 24 & 0.1422 & 0.3068 & 1.0870 & 35.4102\\\hline  7 & 8 & 0.1397 & 0.3062 & 1.0829 & 35.3328\\\hline  2 & 18 & 0.4307 & 0.2112 & 1.0812 & 35.3018\\\hline  10 & 11 & 0.2439 & 0.2681 & 1.0695 & 35.0792\\\hline  22 & 23 & 0.1422 & 0.3011 & 1.0695 & 35.0788\\\hline  19 & 20 & 0.2465 & 0.2672 & 1.0694 & 35.0757\\\hline  12 & 13 & 0.2439 & 0.2615 & 1.0495 & 34.6861\\\hline  18 & 22 & 0.3040 & 0.2414 & 1.0475 & 34.6473\\\hline  1 & 12 & 0.3233 & 0.2295 & 1.0302 & 34.2956\\\hline  20 & 25 & 0.1470 & 0.2866 & 1.0297 & 34.2848\\\hline  1 & 11 & 0.3233 & 0.2281 & 1.0259 & 34.2074\\\hline  4 & 19 & 0.3319 & 0.2250 & 1.0250 & 34.1878\\\hline  9 & 13 & 0.1486 & 0.2837 & 1.0224 & 34.1340\\\hline  17 & 18 & 0.2465 & 0.2495 & 1.0149 & 33.9772\\\hline  5 & 10 & 0.1486 & 0.2803 & 1.0119 & 33.9140\\\hline  4 & 18 & 0.3319 & 0.2180 & 1.0033 & 33.7302\\\hline  16 & 20 & 0.1026 & 0.2852 & 0.9809 & 33.2407\\\hline  10 & 14 & 0.1068 & 0.2813 & 0.9731 & 33.0644\\\hline  13 & 16 & 0.1068 & 0.2799 & 0.9689 & 32.9717\\\hline  17 & 21 & 0.1470 & 0.2647 & 0.9621 & 32.8154\\\hline  14 & 17 & 0.1026 & 0.2717 & 0.9396 & 32.2931\\\hline  2 & 15 & 0.0997 & 0.1859 & 0.6723 & 24.9118\\\hline  3 & 15 & 0.0997 & 0.1854 & 0.6706 & 24.8577\\\hline  4 & 15 & 0.0079 & 0.1931 & 0.6025 & 22.6696\\\hline  1 & 15 & 0.0069 & 0.1932 & 0.6018 & 22.6465
 \\\hline
 \end{tabular}
 \label{ManhattanGridPhaseAngleDifferences3}
 \end{small}
 \end{table}

\newpage

\begin{table}[h]
\begin{footnotesize}
\caption{Phase-angle differences and other outputs for Manhattan network with and without control, power demand vector symmetric, disturbance vector symmetric}
\begin{tabular}{|l|ll|l|l|l|l|l|}\hline The procedure&$i$&$j$&$|\theta_i -\theta_j| $&$\sigma_{ij}$&\begin{tabular}{l}$|\theta_i-\theta_j|$+ \\~$3.08\:\sigma_{ij}$\end{tabular}&\begin{tabular}{l}$ l_{ij}\:sin\:(|\theta_i -\theta_j|$\\~$+ 3.08~\sigma_{ij})$\end{tabular}\\\hline 
without&1 & 7 & 0.5363 & 0.3017 & 1.4657 & 39.7793\\ 
with&1 & 7 & 0.5018 & 0.2998 & 1.4253 & 39.5773\\\hline
without&4 & 23 & 0.5363 & 0.3017 & 1.4657 & 39.7793\\
 with&4 & 23 & 0.5018 & 0.2998 & 1.4253 & 39.5773\\\hline
without& 2 & 14 & 0.5574 & 0.2408 & 1.2990 & 38.5317\\
with& 2 & 14 & 0.5932 & 0.2418 & 1.3379 & 38.9200\\\hline
without&  3 & 16 & 0.5574 & 0.2408 & 1.2990 & 38.5317\\
 with&3 & 16 & 0.5932 & 0.2418 & 1.3379 & 38.9200\\\hline
without&   5 & 6 & 0.0979 & 0.3242 & 1.0963 & 35.5810\\ 
with&5 & 6 & 0.0904 & 0.3246 & 1.0900 & 35.4652\\\hline
without&   24 & 25 & 0.0979 & 0.3242 & 1.0963 & 35.5810\\ 
with&24 & 25 & 0.0904 & 0.3246 & 1.0900 & 35.4652\\\hline
  without& 8 & 9 & 0.0979 & 0.3242 & 1.0963 & 35.5810\\ 
 with& 8 & 9 & 0.0904 & 0.3246 & 1.0900 & 35.4652\\\hline
without&   21 & 22 & 0.0979 & 0.3242 & 1.0963 & 35.5810\\
 with&21 & 22 & 0.0904 & 0.3246 & 1.0900 & 35.4652\\\hline
 without&   1 & 12 & 0.3836 & 0.2225 & 1.0688 & 35.0655\\ 
with& 1 & 12 & 0.3276 & 0.2221 & 1.0116 & 33.9063\\\hline
 without&   1 & 11 & 0.3836 & 0.2225 & 1.0688 & 35.0655\\ 
 with&1 & 11 & 0.3276 & 0.2221 & 1.0116 & 33.9063\\\hline 
without&    4 & 18 & 0.3836 & 0.2225 & 1.0688 & 35.0655\\ 
with&4 & 18 & 0.3276 & 0.2221 & 1.0116 & 33.9063\\\hline
without&    4 & 19 & 0.3836 & 0.2225 & 1.0688 & 35.0655\\ 
with&4 & 19 & 0.3276 & 0.2221 & 1.0116 & 33.9063\\\hline
 without&   7 & 8 & 0.1561 & 0.2910 & 1.0524 & 34.7451\\ 
 with&7 & 8 & 0.1410 & 0.2905 & 1.0358 & 34.4114\\\hline
 without&   22 & 23 & 0.1561 & 0.2910 & 1.0524 & 34.7451\\ 
  with&22 & 23 & 0.1410 & 0.2905 & 1.0358 & 34.4114\\\hline
  without&  23 & 24 & 0.1561 & 0.2910 & 1.0524 & 34.7451\\
 with& 23 & 24 & 0.1410 & 0.2905 & 1.0358 & 34.4114\\\hline 
   without& 6 & 7 & 0.1561 & 0.2910 & 1.0524 & 34.7451\\ 
   with&6 & 7 & 0.1410 & 0.2905 & 1.0358 & 34.4114\\\hline
  without&  3 & 12 & 0.3771 & 0.2161 & 1.0428 & 34.5527\\ 
  with&3 & 12 & 0.4343 & 0.2160 & 1.0997 & 35.6429\\\hline
without&    2 & 18 & 0.3771 & 0.2161 & 1.0428 & 34.5527\\ 
 with&2 & 18 & 0.4343 & 0.2160 & 1.0997 & 35.6429\\\hline
 without&   3 & 19 & 0.3771 & 0.2161 & 1.0428 & 34.5527\\ 
 with&3 & 19 & 0.4343 & 0.2160 & 1.0997 & 35.6429\\\hline
without&    2 & 11 & 0.3771 & 0.2161 & 1.0428 & 34.5527\\ 
with&2 & 11 & 0.4343 & 0.2160 & 1.0997 & 35.6429\\\hline
 without&   8 & 12 & 0.2966 & 0.2401 & 1.0361 & 34.4178\\ 
 with&8 & 12 & 0.3044 & 0.2399 & 1.0434 & 34.5643\\\hline

 \end{tabular}
 \label{ManhattanGridPhaseAngleDifferencessy1}
 \end{footnotesize}
 \end{table}
 \newpage
\begin{table}[h]
\begin{footnotesize}
\caption{Continue of Table \ref{ManhattanGridPhaseAngleDifferencessy1}}
\begin{tabular}{|l|cc|c|c|c|c|}\hline The procedure&$i$&$j$&$|\theta_i -\theta_j| $&$\sigma_{ij}$&\begin{tabular}{l}$|\theta_i-\theta_j|$+ \\~$3.08\:\sigma_{ij}$\end{tabular}&\begin{tabular}{l}$ l_{ij}\:sin\:(|\theta_i -\theta_j|$\\~$+ 3.08~\sigma_{ij})$\end{tabular}\\\hline 
without&18 & 22 & 0.2966 & 0.2401 & 1.0361 & 34.4178\\ 
with&18 & 22 & 0.3044 & 0.2399 & 1.0434 & 34.5643\\\hline 
without&19 & 24 & 0.2966 & 0.2401 & 1.0361 & 34.4178\\
with&19 & 24 & 0.3044 & 0.2399 & 1.0434 & 34.5643\\\hline
without& 6 & 11 & 0.2966 & 0.2401 & 1.0361 & 34.4178\\ 
with&6 & 11 & 0.3044 & 0.2399 & 1.0434 & 34.5643\\\hline
 without&17 & 18 & 0.2529 & 0.2468 & 1.0132 & 33.9414\\
  with& 17 & 18 & 0.2452 & 0.2468 & 1.0055 & 33.7763\\\hline 
without& 12 & 13 & 0.2529 & 0.2468 & 1.0132 & 33.9414\\
with&12 & 13 & 0.2452 & 0.2468 & 1.0055 & 33.7763\\\hline
without&  10 & 11 & 0.2529 & 0.2468 & 1.0132 & 33.9414\\ 
with&10 & 11 & 0.2452 & 0.2468 & 1.0055 & 33.7763\\\hline
 without& 19 & 20 & 0.2529 & 0.2468 & 1.0132 & 33.9414\\ 
 with&19 & 20 & 0.2452 & 0.2468 & 1.0055 & 33.7763\\\hline
without&  5 & 10 & 0.1402 & 0.2697 & 0.9709 & 33.0156\\
with&5 & 10 & 0.1478 & 0.2699 & 0.9792 & 33.2026\\\hline
without&   9 & 13 & 0.1402 & 0.2697 & 0.9709 & 33.0156\\ 
with&9 & 13 & 0.1478 & 0.2699 & 0.9792 & 33.2026\\\hline
without&   20 & 25 & 0.1402 & 0.2697 & 0.9709 & 33.0156\\ 
with&20 & 25 & 0.1478 & 0.2699 & 0.9792 & 33.2026\\\hline
 without&  17 & 21 & 0.1402 & 0.2697 & 0.9709 & 33.0156\\ 
 with&17 & 21 & 0.1478 & 0.2699 & 0.9792 & 33.2026\\\hline
  without& 14 & 17 & 0.0896 & 0.2618 & 0.8960 & 31.2337\\ 
  with&14 & 17 & 0.1047 & 0.2622 & 0.9123 & 31.6369\\\hline
 without&  13 & 16 & 0.0896 & 0.2618 & 0.8960 & 31.2337\\
 with&13 & 16 & 0.1047 & 0.2622 & 0.9123 & 31.6369\\\hline
 without&   10 & 14 & 0.0896 & 0.2618 & 0.8960 & 31.2337\\ 
 with&10 & 14 & 0.1047 & 0.2622 & 0.9123 & 31.6369\\\hline
  without&  16 & 20 & 0.0896 & 0.2618 & 0.8960 & 31.2337\\
  with&16 & 20 & 0.1047 & 0.2622 & 0.9123 & 31.6369\\\hline 
  without&  1 & 15 & 0.0530 & 0.1622 & 0.5526 & 20.9973\\
  with&1 & 15 & 5.0008e-04 & 0.1625 & 0.5010 & 19.2139\\\hline
   without&  4 & 15 & 0.0530 & 0.1622 & 0.5526 & 20.9973\\ 
  with& 4 & 15 & 4.9992e-04 & 0.1625 & 0.5010 & 19.2139\\\hline
   without&  3 & 15 & 0.0470 & 0.1506 & 0.5110 & 19.5606\\ 
   with&3 & 15 & 0.0997 & 0.1506 & 0.5635 & 21.3651\\\hline
    without& 2 & 15 & 0.0470 & 0.1506 & 0.5110 & 19.5606 \\
    with&2 & 15 & 0.0997 & 0.1506 & 0.5635 & 21.3651\\\hline           
 \end{tabular}
  \label{ManhattanGridPhaseAngleDifferencessy2}
 \end{footnotesize}
 \end{table}
 
 \newpage
 \begin{table}[h]
\caption{Tail Probability of the Manhattan network according to Table \ref{ManhattanGridPhaseAngleDifferencessy1}}
\begin{small}
  \begin{tabular}{|l|ll|l|l|l|}
  \hline
The procdure&i &j&  $P(\omega\in \Omega~|~\theta_{ij}(\omega,t)<-\pi/2)$&$P(\omega\in \Omega~|~\theta_{ij}(\omega,t)>\pi/2)$&\begin{tabular}{l}\mbox{by our criterion}\\ \mbox{(Upper bound)}\end{tabular}\\\hline

without&1 & 7 & 1.4337e-12 & 3.0303e-04 & \\ with& 1 & 7 & 2.3681e-12 & 1.8144e-04 & $2\times 1.8171e-04$\\ \hline
without&4 & 23 & 1.4337e-12 & 3.0303e-04 & \\  with&4 & 23 & 2.3681e-12 & 1.8144e-04 &$2\times 1.8171e-04$\\ \hline
without&2 & 14 & 4.8713e-19 & 1.2856e-05 & \\ with& 2 & 14 & 1.7850e-19 & 2.6386e-05 &$2\times 1.1583e-04$\\ \hline
without&3 & 16 & 4.8713e-19 & 1.2856e-05 & \\  with&3 & 16 & 1.7850e-19 & 2.6386e-05 &$2\times 1.1583e-04$\\ \hline
without&5 & 6 & 2.7707e-06 & 1.3226e-07 & \\  with&5 & 6 & 2.5494e-06 & 1.5466e-07 &$2\times 2.0917e-04$\\ \hline
without&24 & 25 & 1.3226e-07 & 2.7707e-06 & \\  with&24 & 25 & 1.5466e-07 & 2.5494e-06 &$2\times 2.0917e-04$\\ \hline
without&8 & 9 & 1.3226e-07 & 2.7707e-06 & \\  with&8 & 9 & 1.5466e-07 & 2.5494e-06 &$2\times 2.0917e-04$\\\hline
without& 21 & 22 & 2.7707e-06 & 1.3226e-07 & \\  with&21 & 22 & 2.5494e-06 & 1.5466e-07 &$2\times 2.0917e-04$\\\hline 
 without&1 & 12 & 7.9019e-19 & 4.7585e-08 & \\  with&1 & 12 & 6.2900e-18 & 1.0876e-08 &$2\times 9.3823e-05$\\ \hline
 without&1 & 11 & 7.9019e-19 & 4.7585e-08 & \\  with&1 & 11 & 6.2900e-18 & 1.0876e-08 &$2\times 9.3823e-05$\\ \hline
 without&4 & 18 & 7.9019e-19 & 4.7585e-08 & \\  with&4 & 18 & 6.2900e-18 & 1.0876e-08 &$2\times 9.3823e-05$\\\hline
  without&4 & 19 & 7.9019e-19 & 4.7585e-08 & \\  with&4 & 19 & 6.2900e-18 & 1.0876e-08 &$2\times 9.3823e-05$\\ \hline
without&  7 & 8 & 1.4750e-09 & 5.8250e-07 & \\  with&7 & 8 & 1.9010e-09 & 4.2866e-07 &$2\times 1.7124e-04$\\ \hline
  without&22 & 23 & 5.8250e-07 & 1.4750e-09 & \\  with&22 & 23 & 4.2866e-07 & 1.9010e-09 &$2\times 1.7124e-04$\\ \hline
 without& 23 & 24 & 1.4750e-09 & 5.8250e-07 & \\  with&23 & 24 & 1.9010e-09 & 4.2866e-07 &$2\times 1.7124e-04$\\ \hline
without&  6 & 7 & 5.8250e-07 & 1.4750e-09 & \\  with&6 & 7 & 4.2866e-07 & 1.9010e-09 &$2\times 1.7124e-04$\\\hline
  without& 3 & 12 & 9.9459e-20 & 1.6586e-08 & \\  with&3 & 12 & 8.2500e-21 & 7.1420e-08 &$2\times 8.7159e-05$\\\hline
without&    2 & 18 & 9.9459e-20 & 1.6586e-08 & \\  with&2 & 18 & 8.2500e-21 & 7.1420e-08 &$2\times 8.7159e-05$\\ \hline
  without&  3 & 19 & 9.9459e-20 & 1.6586e-08 & \\  with&3 & 19 & 8.2500e-21 & 7.1420e-08 &$2\times 8.7159e-05$\\ \hline
  without&  2 & 11 & 9.9459e-20 & 1.6586e-08 & \\  with&2 & 11 & 8.2500e-21 & 7.1420e-08 &$2\times 8.7159e-05$\\ \hline
  \end{tabular}
  \label{TailProbabilityoftheManhattannetworkaccordingtoTablesymmetric1}
\end{small}
\end{table}
\newpage
 \begin{table}[h]
\caption{Tail Probability of the Manhattan network according to Table \ref{ManhattanGridPhaseAngleDifferencessy2}}
\begin{small}
  \begin{tabular}{|l|ll|l|l|l|}
  \hline
 The procdure&  i&j& $P(\omega\in \Omega~|~\theta_{ij}(\omega,t)<-\pi/2)$&$P(\omega\in \Omega~|~\theta_{ij}(\omega,t)>\pi/2)$&\begin{tabular}{l}\mbox{by our criterion}\\ \mbox{(Upper bound)}\end{tabular}\\\hline
 without&   8 & 12 & 5.5740e-08 & 3.6963e-15 & \\ with& 8 & 12 & 6.4998e-08 & 2.7140e-15 &$2\times 1.1369e-04$\\ \hline
 without&18 & 22 & 3.6963e-15 & 5.5740e-08 & \\ with&18 & 22 & 2.7140e-15 & 6.4998e-08 &$2\times 1.1369e-04$\\ \hline 
without& 19 & 24 & 3.6963e-15 & 5.5740e-08 & \\ with&19 & 24 & 2.7140e-15 & 6.4998e-08 &$2\times 1.1369e-04$\\ \hline 
without&6 & 11 & 5.5740e-08 & 3.6963e-15 & \\ with&6 & 11 & 6.4998e-08 & 2.7140e-15 &$2\times 1.1369e-04$\\ \hline 
without& 17 & 18 & 4.6490e-08 & 7.3763e-14 & \\with& 17 & 18 & 3.9121e-08 & 9.3224e-14 &$2\times 1.2150e-04$\\ \hline 
 without&12 & 13 & 7.3763e-14 & 4.6490e-08 & \\with& 12 & 13 & 9.3224e-14 & 3.9121e-08 &$2\times 1.2150e-04$\\ \hline 
without& 10 & 11 & 4.6490e-08 & 7.3763e-14 & \\ with&10 & 11 & 3.9121e-08 & 9.3224e-14 &$2\times 1.2150e-04$\\\hline 
 without& 19 & 20 & 7.3763e-14 & 4.6490e-08 & \\ with&19 & 20 & 9.3224e-14 & 3.9121e-08 &$2\times 1.2150e-04$\\ \hline 
  without&5 & 10 & 5.6523e-08 & 1.1188e-10 & \\ with&5 & 10 & 6.7359e-08 & 9.6048e-11 &$2\times 1.4783e-04$\\\hline 
 without&  9 & 13 & 5.6523e-08 & 1.1188e-10 & \\ with&9 & 13 & 6.7359e-08 & 9.6048e-11 &$2\times 1.4783e-04$\\ \hline 
  without& 20 & 25 & 1.1188e-10 & 5.6523e-08 & \\with& 20 & 25 & 9.6048e-11 & 6.7359e-08 &$2\times 1.4783e-04$\\ \hline 
  without& 17 & 21 & 1.1188e-10 & 5.6523e-08 & \\ with&17 & 21 & 9.6048e-11 & 6.7359e-08 &$2\times 1.4783e-04$\\ \hline 
  without& 14 & 17 & 1.1323e-10 & 7.6690e-09 & \\with& 14 & 17 & 8.2864e-11 & 1.1255e-08 &$2\times 1.3904e-04$\\ \hline 
   without&13 & 16 & 7.6690e-09 & 1.1323e-10 & \\with& 13 & 16 & 1.1255e-08 & 8.2864e-11 &$2\times 1.3904e-04$\\ \hline 
  without& 10 & 14 & 7.6690e-09 & 1.1323e-10 & \\ with&10 & 14 & 1.1255e-08 & 8.2864e-11 &$2\times 1.3904e-04$\\ \hline 
  without& 16 & 20 & 1.1323e-10 & 7.6690e-09 & \\ with&16 & 20 & 8.2864e-11 & 1.1255e-08 &$2\times 1.3904e-04$\\ \hline 
 without&  1 & 15 & 6.8132e-24 & 4.0799e-21 & \\ with&1 & 15 & 2.0315e-22 & 2.1573e-22 &$2\times 3.5136e-05$\\ \hline 
 without&  4 & 15 & 6.8132e-24 & 4.0799e-21 & \\ with&4 & 15 & 2.0315e-22 & 2.1573e-22 &$2\times 3.5136e-05$\\ \hline 
without&   3 & 15 & 3.2196e-27 & 2.2946e-24 & \\ with&3 & 15 & 6.8384e-29 & 7.7056e-23 &$2\times 2.6038e-05$\\ \hline 
 without&  2 & 15 & 3.2196e-27 & 2.2946e-24 & \\ with&2 & 15 & 6.8384e-29 & 7.7056e-23 &$2\times 2.6038e-05$\\\hline 
 \end{tabular}
   \label{TailProbabilityoftheManhattannetworkaccordingtoTablesymmetric2}
\end{small}
\end{table}

  \newpage
\par 

\begin{table}[h]
\begin{footnotesize}
\caption{Phase-angle differences and other outputs for Manhattan network without and with control,  power demand vector symmetric, disturbance vector asymmetric}
\begin{tabular}{|l|ll|l|l|l|l|}\hline The procedure&$i$&$j$&$|\theta_i -\theta_j| $&$\sigma_{ij}$&\begin{tabular}{l}$|\theta_i-\theta_j|$+ \\~$3.08\:\sigma_{ij}$\end{tabular}&\begin{tabular}{l}$ l_{ij}\:sin\:(|\theta_i -\theta_j|$\\~$+ 3.08~\sigma_{ij})$\end{tabular}\\\hline 
without&1 & 7 & 0.5363 & 0.3149 & 1.5062 & 39.9165\\with& 1 & 7 & 0.4990 & 0.3128 & 1.4625 & 39.7656\\\hline without&4 & 23 & 0.5363 & 0.3128 & 1.4997 & 39.8990\\with&4 & 23 & 0.5046 & 0.3110 & 1.4625 & 39.7656\\\hline without&3 & 16 & 0.5574 & 0.2589 & 1.3550 & 39.0719\\with&3 & 16 & 0.5932 & 0.2601 & 1.3943 & 39.3787\\\hline without& 2 & 14 & 0.5574 & 0.2590 & 1.3550 & 39.0725\\with&2 & 14 & 0.5932 & 0.2600 & 1.3941 & 39.3770\\\hline without&5 & 6 & 0.0979 & 0.3412 & 1.1489 & 36.4920\\with&5 & 6 & 0.0896 & 0.3416 & 1.1416 & 36.3717\\\hline without&8 & 9 & 0.0979 & 0.3377 & 1.1380 & 36.3114\\ with&8 & 9 & 0.0896 & 0.3381 & 1.1310 & 36.1934\\\hline  without&24 & 25 & 0.0979 & 0.3304 & 1.1154 & 35.9243\\with&24 & 25 & 0.0912 & 0.3307 & 1.1099 & 35.8256 \\\hline without&6 & 7 & 0.1561 & 0.3109 & 1.1138 & 35.8953\\with&6 & 7 & 0.1397 & 0.3104 & 1.0959 & 35.5731\\\hline without&19 & 24 & 0.2966 & 0.2631 & 1.1068 & 35.7715\\with&19 & 24 & 0.3040 & 0.2628 & 1.1135 & 35.8899\\\hline without&21 & 22 & 0.0979 & 0.3275 & 1.1065 & 35.7661\\with&21 & 22 & 0.0912 & 0.3279 & 1.1012 & 35.6696\\\hline without&23 & 24 & 0.1561 & 0.3071 & 1.1019 & 35.6832\\with& 23 & 24 & 0.1422 & 0.3068 & 1.0870 & 35.4102\\\hline without&7 & 8 & 0.1561 & 0.3068 & 1.1010 & 35.6670\\ with&7 & 8 & 0.1397 & 0.3062 & 1.0829 & 35.3328\\\hline without&6 & 11 & 0.2966 & 0.2599 & 1.0971 & 35.5958\\with& 6 & 11 & 0.3049 & 0.2597 & 1.1046 & 35.7320\\\hline without&1 & 12 & 0.3836 & 0.2299 & 1.0916 & 35.4955\\with&1 & 12 & 0.3233 & 0.2295 & 1.0302 & 34.2956\\\hline without&1 & 11 & 0.3836 & 0.2285 & 1.0874 & 35.4173\\with&1 & 11 & 0.3233 & 0.2281 & 1.0259 & 34.2074\\\hline without& 22 & 23 & 0.1561 & 0.3016 & 1.0849 & 35.3708\\ with&22 & 23 & 0.1422 & 0.3011 & 1.0695 & 35.0788\\\hline without&8 & 12 & 0.2966 & 0.2549 & 1.0816 & 35.3077\\with&8 & 12 & 0.3049 & 0.2547 & 1.0895 & 35.4555\\\hline  without&10 & 11 & 0.2529 & 0.2681 & 1.0787 & 35.2529\\ with& 10 & 11 & 0.2439 & 0.2681 & 1.0695 & 35.0792\\\hline without&4 & 19 & 0.3836 & 0.2253 & 1.0776 & 35.2324\\ with&4 & 19 & 0.3319 & 0.2250 & 1.0250 & 34.1878\\\hline   
without&19 & 20 & 0.2529 & 0.2672 & 1.0759 & 35.2016\\with&19 & 20 & 0.2465 & 0.2672 & 1.0694 & 35.0757\\\hline
 \end{tabular}
 \label{ManhattanGridPhaseAngleDifferences11}
 \end{footnotesize}
 \end{table}

\newpage
\begin{table}[h]
\begin{footnotesize}
\caption{Continue of Table \ref{ManhattanGridPhaseAngleDifferences11}} 
\begin{tabular}{|l|ll|l|l|l|l|}\hline The procedure&$i$&$j$&$|\theta_i -\theta_j| $&$\sigma_{ij}$&\begin{tabular}{l}$|\theta_i-\theta_j|$+ \\~$3.08\:\sigma_{ij}$\end{tabular}&\begin{tabular}{l}$ l_{ij}\:sin\:(|\theta_i -\theta_j|$\\~$+ 3.08~\sigma_{ij})$\end{tabular}\\\hline 
  without& 3 & 12 & 0.3771 & 0.2233 & 1.0650 & 34.9921\\with&3 & 12 & 0.4379 & 0.2233 & 1.1255 & 36.0993\\\hline without& 2 & 11 & 0.3771 & 0.2218 & 1.0602 & 34.8979\\with&2 & 11 & 0.4379 & 0.2214 & 1.1199 & 36.0019\\\hline    without&12 & 13 & 0.2529 & 0.2614 & 1.0582 & 34.8588\\with&12 & 13 & 0.2439 & 0.2615 & 1.0495 & 34.6861\\\hline  without&4 & 18 & 0.3836 & 0.2184 & 1.0562 & 34.8204\\with&4 & 18 & 0.3319 & 0.2180 & 1.0033 & 33.7302\\\hline without&3 & 19 & 0.3771 & 0.2182 & 1.0493 & 34.6824\\ with&3 & 19 & 0.4307 & 0.2180 & 1.1022 & 35.6882\\\hline  without& 18 & 22 & 0.2966 & 0.2415 & 1.0404 & 34.5038\\with&18 & 22 & 0.3040 & 0.2414 & 1.0475 & 34.6473\\\hline   without&2 & 18 & 0.3771 & 0.2113 & 1.0278 & 34.2468\\with&2 & 18 & 0.4307 & 0.2112 & 1.0812 & 35.3018\\\hline
     without&20 & 25 & 0.1402 & 0.2863 & 1.0220 & 34.1257\\with&20 & 25 & 0.1470 & 0.2866 & 1.0297 & 34.2848\\\hline 
     without&17 & 18 & 0.2529 & 0.2494 & 1.0211 & 34.1071\\with&17 & 18 & 0.2465 & 0.2495 & 1.0149 & 33.9772\\\hline
 without&9 & 13 & 0.1402 & 0.2836 & 1.0137 & 33.9513\\with&9 & 13 & 0.1486 & 0.2837 & 1.0224 & 34.1340\\\hline   without&5 & 10 & 0.1402 & 0.2800 & 1.0026 & 33.7152\\with&5 & 10 & 0.1486 & 0.2803 & 1.0119 & 33.9140\\\hline  without&16 & 20 & 0.0896 & 0.2848 & 0.9667 & 32.9210\\with&16 & 20 & 0.1026 & 0.2852 & 0.9809 & 33.2407\\\hline without&10 & 14 & 0.0896 & 0.2810 & 0.9552 & 32.6562\\ with&10 & 14 & 0.1068 & 0.2813 & 0.9731 & 33.0644\\\hline without&17 & 21 & 0.1402 & 0.2644 & 0.9547 & 32.6449\\with&17 & 21 & 0.1470 & 0.2647 & 0.9621 & 32.8154\\\hline  without&13 & 16 & 0.0896 & 0.2795 & 0.9506 & 32.5503\\with& 13 & 16 & 0.1068 & 0.2799 & 0.9689 & 32.9717\\\hline without& 14 & 17 & 0.0896 & 0.2713 & 0.9252 & 31.9498\\with&14 & 17 & 0.1026 & 0.2717 & 0.9396 & 32.2931\\\hline without&4 & 15 & 0.0530 & 0.1928 & 0.6468 & 24.1064\\with&4 & 15 & 0.0079 & 0.1931 & 0.6025 & 22.6696\\\hline without&1 & 15 & 0.0530 & 0.1928 & 0.6468 & 24.1067\\with&1 & 15 & 0.0069 & 0.1932 & 0.6018 & 22.6465\\\hline without&2 & 15 & 0.0470 & 0.1859 & 0.6197 & 23.2325\\with&2 & 15 & 0.0997 & 0.1859 & 0.6723 & 24.9118\\\hline without&3 & 15 & 0.0470 & 0.1854 & 0.6181 & 23.1804\\with&3 & 15 & 0.0997 & 0.1854 & 0.6706 & 24.8577\\\hline 
 \end{tabular}
  \label{ManhattanGridPhaseAngleDifferences22}
 \end{footnotesize}
 \end{table}
\newpage
 \begin{table}[h]
\caption{Tail Probability of the Manhattan network according to Table \ref{ManhattanGridPhaseAngleDifferences11}}
\begin{small}
  \begin{tabular}{|l|ll|l|l|l|}
  \hline
The procdure&i &j&  $P(\omega\in \Omega~|~\theta_{ij}(\omega,t)<-\pi/2)$&$P(\omega\in \Omega~|~\theta_{ij}(\omega,t)>\pi/2)$&\begin{tabular}{l}\mbox{by our criterion}\\ \mbox{(Upper bound)}\end{tabular}\\\hline
without&1 & 7 & 1.1058e-11 & 5.0963e-04 & \\with& 1 & 7 & 1.8329e-11 & 3.0575e-04 &$2\times 3.0603e-04$\\ \hline 
without&4 & 23 & 8.1268e-12 & 4.7115e-04 & \\ with&4 & 23 & 1.2506e-11 & 3.0370e-04 &$2\times 3.0378e-04$\\ \hline 
without&3 & 16 & 1.0163e-16 & 4.5345e-05 & \\ with&3 & 16 & 4.4033e-17 & 8.5454e-05 &$2\times 2.3582e-04$\\ \hline 
without&2 & 14 & 1.0435e-16 & 4.5630e-05 & \\ with&2 & 14 & 4.2860e-17 & 8.4961e-05 &$2\times 2.3568e-04$\\\hline 
without&5 & 6 & 7.9149e-06 & 5.0247e-07 & \\ with&5 & 6 & 7.2531e-06 & 5.8502e-07 &$2\times 3.4061e-04$\\ \hline 
without& 8 & 9 & 3.8790e-07 & 6.4571e-06 & \\ with&8 & 9 & 4.5315e-07 & 5.9084e-06 &$2\times 3.3655e-04$\\ \hline 
without&24 & 25 & 2.2029e-07 & 4.1380e-06 & \\ with&24 & 25 & 2.5081e-07 & 3.8361e-06 &$2\times 3.2783e-04$\\ \hline 
without& 6 & 7 & 2.6781e-06 & 1.3920e-08 & \\ with&6 & 7 & 2.0086e-06 & 1.7879e-08 &$2\times 3.0302e-04$\\ \hline 
without& 19 & 24 & 6.3440e-13 & 6.3943e-07 & \\ with&19 & 24 & 4.8772e-13 & 7.1642e-07 &$2\times 2.3963e-04$\\ \hline 
without& 21 & 22 & 3.4396e-06 & 1.7414e-07 & \\ with&21 & 22 & 3.2058e-06 & 2.0037e-07 &$2\times 3.2449e-04$\\ \hline 
 without&23 & 24 & 9.3706e-09 & 2.0462e-06 & \\ with&23 & 24 & 1.1791e-08 & 1.6086e-06 &$2\times 2.9849e-04$\\ \hline 
without& 7 & 8 & 9.0767e-09 & 2.0024e-06 & \\with& 7 & 8 & 1.1604e-08 & 1.4789e-06 &$2\times 2.9773e-04$\\ \hline 
without& 6 & 11 & 4.7278e-07 & 3.3590e-13 & \\with& 6 & 11 & 5.4554e-07 & 2.5511e-13 &$2\times 2.3526e-04$\\ \hline 
 without&1 & 12 & 9.3924e-18 & 1.2088e-07 & \\ with&1 & 12 & 7.7141e-17 & 2.7289e-08 &$2\times 1.9125e-04$\\ \hline 
 without&1 & 11 & 5.9884e-18 & 1.0203e-07 & \\ with&1 & 11 & 5.0418e-17 & 2.2619e-08 &$2\times 1.8915e-04$\\ \hline 
without& 22 & 23 & 1.3618e-06 & 5.1479e-09 & \\ with&22 & 23 & 1.0446e-06 & 6.3845e-09 &$2\times 2.9121e-04$\\ \hline 
 without&8 & 12 & 2.8843e-07 & 1.1857e-13 & \\ with&8 & 12 & 3.3451e-07 & 8.9016e-14 &$2\times 2.2814e-04$\\\hline 
 without& 10 & 11 & 4.4235e-07 & 5.1481e-12 & \\with& 10 & 11 & 3.7248e-07 & 6.4959e-12 &$2\times 2.4705e-04$\\ \hline 
 without& 4 & 19 & 2.0743e-18 & 6.8434e-08 & \\with& 4 & 19 & 1.3785e-17 & 1.8332e-08 &$2\times 1.8450e-04$\\ \hline 
 without& 19 & 20 & 4.3898e-12 & 4.0645e-07 & \\ with&19 & 20 & 5.1854e-12 & 3.5943e-07 &$2\times 2.4579e-04$\\ \hline 
  \end{tabular}
    \label{TailProbabilityoftheManhattannetworkaccordingtoTableasymmetric1}
\end{small}
\end{table}
\newpage
 \begin{table}[h]
\caption{Tail Probability of the Manhattan network according to Table \ref{ManhattanGridPhaseAngleDifferences22}}
\begin{small}
  \begin{tabular}{|l|ll|l|l|l|}
  \hline
 The procdure&  i&j& $P(\omega\in \Omega~|~\theta_{ij}(\omega,t)<-\pi/2)$&$P(\omega\in \Omega~|~\theta_{ij}(\omega,t)>\pi/2)$&\begin{tabular}{l}\mbox{by our criterion}\\ \mbox{(Upper bound)}\end{tabular}\\\hline
 without&3 & 12 & 1.3519e-18 & 4.5032e-08 & \\ with&3 & 12 & 1.1757e-19 & 1.9536e-07 & $2\times 1.8194e-04$\\\hline 
without&   2 & 11 & 8.0140e-19 & 3.6860e-08 & \\with&2 & 11 & 5.8050e-20 & 1.5524e-07 &$2\times 1.7908e-04$\\ \hline 
without&   12 & 13 & 1.5115e-12 & 2.3073e-07 & \\with& 12 & 13 & 1.9665e-12 & 1.9459e-07 &$2\times 2.3780e-04$\\ \hline 
without&   4 & 18 & 1.7984e-19 & 2.7263e-08 & \\ with&4 & 18 & 1.2965e-18 & 6.6172e-09 &$2\times 1.7393e-04$\\ \hline 
 without&  3 & 19 & 2.1863e-19 & 2.2419e-08 & \\with&3 & 19 & 2.1320e-20 & 8.4847e-08 &$2\times 1.7393e-04$\\ \hline 
 without&  18 & 22 & 5.2732e-15 & 6.5954e-08 & \\with& 18 & 22 & 4.0391e-15 & 7.7002e-08 &$2\times 2.0887e-04$\\ \hline 
 without&  2 & 18 & 1.5046e-20 & 8.0553e-09 & \\ with&2 & 18 & 1.3112e-21 & 3.3660e-08 &$2\times 1.6359e-04$\\ \hline 
 without&  20 & 25 & 1.1418e-09 & 2.9138e-07 & \\with& 20 & 25 & 1.0256e-09 & 3.3843e-07 &$2\times 2.7224e-04$\\ \hline 
  without& 17 & 18 & 6.3104e-08 & 1.3127e-13 & \\ with&17 & 18 & 5.5478e-08 & 1.6233e-13 &$2\times 2.2066e-04$\\ \hline 
  without& 9 & 13 & 2.2745e-07 & 8.0406e-10 & \\ with&9 & 13 & 2.6790e-07 & 6.7801e-10 &$2\times 2.6836e-04$\\ \hline 
  without& 5 & 10 & 1.6170e-07 & 4.9597e-10 & \\ with&5 & 10 & 1.9494e-07 & 4.2814e-10 &$2\times 2.6378e-04$\\ \hline 
without&   16 & 20 & 2.7707e-09 & 9.9200e-08 & \\with& 16 & 20 & 2.2128e-09 & 1.3167e-07 &$2\times 2.7037e-04$\\\hline 
  without&  10 & 14 & 6.7782e-08 & 1.7222e-09 & \\ with&10 & 14 & 9.7315e-08 & 1.2327e-09 &$2\times 2.6513e-04$\\\hline 
without&     17 & 21 & 4.8600e-11 & 3.1385e-08 & \\ with&17 & 21 & 4.3033e-11 & 3.7470e-08 &$2\times 2.4230e-04$\\ \hline 
without&     13 & 16 & 5.8076e-08 & 1.4199e-09 & \\with& 13 & 16 & 8.4560e-08 & 1.0265e-09 &$2\times 2.6324e-04$\\ \hline 
  without&   14 & 17 & 4.6744e-10 & 2.3857e-08 & \\ with&14 & 17 & 3.6606e-10 & 3.2633e-08 &$2\times 2.5203e-04$\\ \hline 
   without&  4 & 15 & 1.8477e-17 & 1.7397e-15 & \\ with&4 & 15 & 1.4727e-16 & 2.8936e-16 &$2\times 1.3588e-04$\\ \hline 
   without&  1 & 15 & 1.8477e-17 & 1.7397e-15 & \\ with&1 & 15 & 1.5922e-16 & 2.8702e-16 &$2\times 1.3603e-04$\\ \hline 
   without&  2 & 15 & 1.6231e-18 & 1.2337e-16 & \\ with&2 & 15 & 1.2820e-19 & 1.2525e-15 &$2\times 1.2486e-04$\\ \hline 
    without& 3 & 15 & 1.3194e-18 & 1.0263e-16 & \\ with&3 & 15 & 1.0281e-19 & 1.0549e-15 &$2\times 1.2409e-04$ \\\hline 
\end{tabular}
  \label{TailProbabilityoftheManhattannetworkaccordingtoTableasymmetric2}
\end{small}
\end{table}